# Variability in Resistive Memories

Juan B. Roldán,* Enrique Miranda, David Maldonado, Alexey N. Mikhaylov,
Nikolay V. Agudov, Alexander A. Dubkov, Maria N. Koryazhkina, Mireia B. González,
Marco A. Villena, Samuel Poblador, Mercedes Saludes-Tapia, Rodrigo Picos,
Francisco Jiménez-Molinos, Stavros G. Stavrinides, Emili Salvador, Francisco J. Alonso,
Francesca Campabadal, Bernardo Spagnolo, Mario Lanza, and Leon O. Chua

Resistive memories are outstanding electron devices that have displayed a large potential in a plethora of applications such as nonvolatile data storage, neuromorphic computing, hardware cryptography, etc. Their fabrication control and performance have been notably improved in the last few years to cope with the requirements of massive industrial production. However, the most important hurdle to progress in their development is the so-called cycle-to-cycle variability, which is inherently rooted in the resistive switching mechanism behind the operational principle of these devices. In order to achieve the whole picture, variability must be assessed from different viewpoints going from the experimental characterization to the adequation of modeling and simulation techniques. Herein, special emphasis is put on the modeling part because the accurate representation of the phenomenon is critical for circuit designers. In this respect, a number of approaches are used to the date: stochastic, behavioral, mesoscopic..., each of them covering particular aspects of the electron and ion transport mechanisms occurring within the switching material. These subjects are dealt with in this review, with the aim of presenting the most recent advancements in the treatment of variability in resistive memories.

## 1. Introduction

Research and development efforts in the nonvolatile memory arena are focused on a reduced set of innovative components,

among which we can include memristors.[1,2] Memristors are expected to be key players in the electronics landscape of the coming years largely because of the powerful applications that stand upon their unique features.[1,3–5] The switching mechanisms behind memristors differ significantly depending on the physical properties of the structures and the materials involved.[1,3–7] To list some of these mechanisms, we can highlight those devices based on phase-change materials, which can be switched reversibly between amorphous and crystalline phases with different electrical resistivity (phase-change memories, PCMs);[8] devices that take advantage of the magnetic and electrical properties exhibited by some materials with different architectures (magnetic RAMs, MRAMs);[9] also structures where materials with switchable electrical polarization give rise to hysteresis curves of the polarization versus electrical field that can be engineered for storing information (ferroelectric FET, FFET);[10] and, finally, resistive RAMs (RRAMs) where the dielectric conduction properties are altered by means of the internal ion movement and concurrent redox reactions used to generate different resistive states.[1,3,11,12]

J. B. Roldán, D. Maldonado, F. Jiménez-Molinos
Electronics and Computer Technology Department
Science Faculty
Granada University
Avda. Fuente Nueva s/n, 18071 Granada, Spain
E-mail: jroldan@ugr.es

E. Miranda, E. Salvador
Dept. Enginyeria Electrònica
Universitat Autònoma de Barcelona
08193 Cerdanyola del Vallès, Spain

The ORCID identification number(s) for the author(s) of this article can be found under https://doi.org/10.1002/aisy.202200338

A. N. Mikhaylov, N. V. Agudov, A. A. Dubkov, M. N. Koryazhkina, B. Spagnolo
Laboratory of Stochastic Multistable Systems
Lobachevsky University
23 Gagarin prospect, Nizhny Novgorod 603022, Russia

M. B. González, S. Poblador, M. Saludes-Tapia, F. Campabadal
Institut de Microelectrònica de Barcelona
IMB-CNM (CSIC)
Carrer dels Til·lers, s/n. Campus UAB, 08193 Bellaterra, Spain

M. A. Villena, M. Lanza
Physical Science and Engineering Division
King Abdullah University of Science and Technology (KAUST)
Thuwal 23955-6900, Saudi Arabia

R. Picos
Industrial Engineering and Construction Department
University of Balearic Islands
07122 Balearic Islands, Spain

Here, we will focus on the latter devices and associated materials. These devices exhibit good endurance,[13] retention, low-power consumption, full compatibility with the complementary metal-oxide semiconductor (CMOS) technology, and capability of 3D (three-dimensional) stack fabrication.[3–7] Redox-based devices, as nonvolatile memories, have been employed in commercial fabrication processes; for instance, in the 22 nm node we can find companies such as TSMC[14] and INTEL.[15]

The stochastic operation of redox-based devices makes them appropriate for the fabrication of integrated circuits (ICs) devoted to cryptography; in particular, they can be incorporated in the implementation of physical unclonable functions and random number generators.[16–19] These circuits will be essential in the growing context of edge computing, 5G and 6G applications.[1] In addition, the applications linked to neuromorphic computing place these devices (also other memristive devices) in the spotlight of research and development efforts.[20–29] Their inherent features allow them to mimic biological synapses in circuits conceived to replicate mammal neural networks. Considering their low-power consumption[6,28,30] and the possibility of performing analog vector-matrix multiplications in a compute-in-memory context,[1,29,30] their role will be essential to reduce the carbon footprint of developments in new hardware to accelerate artificial intelligence solutions.[31]

There are two main types of redox-based devices: electrochemical metallization cells (also known as conductive bridge RAMs, CBRAMs) and valence change memories (VCMs). The latter devices will be more thoroughly discussed here. In both cases, it is essential to take into consideration the role of variability. Apart from the expected fabrication-dependent device to device variability,[7,32] the inherent stochastic operation of redox-based devices leads to variations in the successive set (transition from the high resistance state into the low resistance state) and reset (transition from the low resistance state into the high resistance state) processes, the well-known cycle-to-cycle (C2C) variability.[32–35] As this is a remarkable feature of these devices that generates a certain concern about control and reproducibility issues, it must be imperatively included in compact modeling to let electronic design automation (EDA) tools be accurate for the massive use needed prior to a commercial step forward in some of the applications described earlier.

S. G. Stavrinides
School of Science and Technology
International Hellenic University
57001 Thessaloniki, Greece

F. J. Alonso
Statistics and Operations Research Department
Science Faculty
Granada University
Avda. Fuente Nueva s/n, 18071 Granada, Spain

B. Spagnolo
Dipartimento di Fisica e Chimica "Emilio Segré"
Group of Interdisciplinary Theoretical Physics
Universitá degli Studi di Palermo and CNISM
Unitá di Palermo, Viale delle Scienze, Edificio 18, I-90128 Palermo, Italy

L. O. Chua
Electrical Engineering and Computer Science Department
University of California
Berkeley, CA 94720-1770, USA

Because of its importance, in this manuscript, we will focus the attention specifically on variability in redox-based devices, both from the experimental and modeling perspectives (thermal aspects of compact modeling have been introduced elsewhere)[36] since, as far as we know, there are no review papers on this issue until now. Before going into the modeling sections, we will describe variability from the experimental viewpoint, making use of structures fabricated with transition metal oxides and the upcoming two-dimensional (2D) materials (Section 2). Subsequently, we will go through the modeling of RRAM devices, highlighting the physical origins and the implementation of variability. We will consider physically based models (Section 3), where the charge transport mechanisms and the conductive filament (CF) formation/destruction processes (in case of filamentary operation) are analytically described; for stochastic models (Section 4), we will consider differential equations with the noise sources inducing random hopping of dielectric structural defects (e.g., oxygen vacancies), thermal and $1/f$ fluctuations of the current to explain the stochastic properties of the memristors; finally, for behavioral models (Section 5), we will consider different modeling strategies to algebraically replicate hysteresis and other characteristic device operation curves, simplifying as much as possible the complexity of the physics behind resistive switching (RS) phenomenon. In addition to presenting the modeling strategies and the subtleties linked to variability, we will discuss the trade-off between accuracy and mathematical complexity to guide the reader through the different modeling options available, considering both pros and cons (Section 6).

## 2. Experimental Evidence of Variability in Resistive Memories

### 2.1. Oxide-Based Devices

#### 2.1.1. Introduction

Resistive memories are mostly built using metal-insulator-metal (MIM) structures that show the RS phenomenon, which consists in a nonvolatile sudden change of the electrical resistance of the structure caused by the application of an electrical stimulus. In filamentary-type devices, the physical mechanism responsible for the RS phenomenon is the formation and partial disruption of a CF in the insulator layer connecting both electrodes. The filamentary RS behavior in VCMs is due to electrochemical reduction–oxidation (redox) processes in the oxide layer and electric-field-assisted oxygen ion migration.[35,37–39] These processes induce a local change in the stoichiometry of the insulating layer and lead to the formation of an oxygen-deficient conductive nanofilament. The applied voltage and the temperature increase induced by Joule heating control the size and stoichiometry of the filamentary path, and therefore, the memristor electrical resistance.

Oxide-based filamentary VCMs exhibit the typical pinched hysteresis current–voltage loop due to the RS phenomenon.[4,39] RS is performed by applying voltages of opposite polarities, or, in other words, in the bipolar operation mode. The most common oxide materials (frequently transition metal oxides) used in these devices are $HfO_2$,[40,41] $Ta_2O_5$,[42–44] $TiO_2$,[45] $ZrO_2$,[46,47] $Al_2O_3$,[48] and nanolaminates of them.[45,49,50] Among the different alternatives, HfO$_2$-based resistive memories have achieved one of the best performances in terms of reproducibility with well-defined resistance states and good device endurance.[40,41,51,52] Furthermore, it has been widely reported that a sub-stoichiometric oxide region of the insulating layer of a MIM-VCM structure may considerably enhance the RS performance of these devices. This can be achieved by depositing a thin oxygen scavenging layer (Ti, Hf, Ta, etc.) on top of the oxide material.[53]

While **Figure 1**a shows a schematic representation of the physical mechanisms responsible for filamentary RS in VCMs, Figure 1b shows the typical current–voltage ($I$–$V$) characteristics during the forming process and the first complete RS cycle for a HfO$_2$-based RRAM. In many cases, after device fabrication (in the pristine state) prior to the RS operation, the devices need to be formed in order to generate an oxygen vacancy ($V_o$) rich filamentary path. The forming process is performed by applying a positive voltage ramp until a sudden and abrupt current increase occurs, leading the device to the low resistance state (LRS) or "on state". In this step, the current is limited with a compliance value to avoid the hard dielectric breakdown of the insulator.[54] After the CF formation, the device can be switched to a high resistance state (HRS) or "off state" by applying a negative voltage ramp that ensures the reset of the device. The reset process is attributed to the drift of oxygen anions caused by the electric field and the recombination of oxygen vacancies at the tip of the CF. Next, a positive voltage ramp is applied during the set process, and the filamentary path is restored, leading the device again to the LRS or "on state."[33] This sequence of positive-negative voltage ramps constitutes a complete RS cycle.

Filamentary-type VCMs are being extensively investigated as promising candidates for a wide variety of potential applications including nonvolatile memories,[51] digital logic circuits,[55] and hardware security systems.[56] In addition, in the last years, intense research is currently ongoing to understand the capability of these devices to store multiple bits to evaluate their potential as synaptic devices in circuits that mimic brain functions, such as reasoning, learning from experience, or decision-making.[1,25,57] This is largely due to the fact that VCMs exhibit excellent characteristics in terms of downscaling within a crossbar array, high-density 3D integration and compatibility with

CMOS manufacturing technology. Furthermore, an analog control of their resistance (synaptic weight) in the context of artificial neural networks has been several times demonstrated.[1,45,49,58,59]

Since the resistance of these devices is mainly determined by the CF state, the stochastic nature of the physical processes responsible for the reversible CF formation induces a significant variability in the electrical characteristics of these devices, consequently reducing their performance. In this context, it is important to investigate the crucial parameters that affect the cycle-to-cycle variability and RS mechanisms governing the growth and shrinkage of CFs.[60] In addition, the device variability must be considered in order to implement memristors in crossbar arrays for emerging applications, where a large number of cells must be integrated. Hence, the analysis of the switching variability in redox-based devices is of great importance in order to achieve an optimal design and modeling for a reliable performance in real applications.

### 2.1.2. Cycle-To-Cycle Variability

In this section, we will characterize the C2C variability in HfO$_2$-based VCMs measured through the sequential use of voltage ramps (**Figure 2**) and pulse schemes (**Figure 3**).

Figure 2a shows the typical bipolar RS behavior for 3000 RS cycles measured by double sweep voltage ramps from 0 to 1.1 V for the set process and from 0 to $-1.4$ V for the reset.[54] To analyze the uniformity of the RS phenomenon during cycling, in Figure 2b, the LRS ($I_{on}$) and HRS ($I_{off}$) currents measured at $-0.1$ V are represented as a function of the number of cycles. In addition, Figure 2d shows the cumulative distribution functions (CDFs) for $I_{on}$ and $I_{off}$. It can be seen a larger C2C spread in the HRS current than in the LRS current.[61] This higher variability is attributed to the variations of the gap distance between the tip of the broken CF and the metal electrode when the reset process is activated. For statistical analysis, set and reset voltages ($V_{set}$ and $V_{reset}$) and currents ($I_{set}$ and $I_{reset}$) are extracted for each RS cycle of the 3000 experimental cycles. In Figure 2c, $V_{set}$ and $V_{reset}$ are represented as a function of the number of cycles, showing a significant cycle-to-cycle variability (for the time-series modeling of these data see Section 5.3). In this study, the reset voltage is extracted as the voltage before the first current drop takes place during the reset transition. This point generally corresponds to the current maximum along the curve obtained for the negative voltage ramps.[62] This result indicates the stochastic nature of the filament formation and dissolution processes due to the field and temperature-assisted oxygen vacancy generation, recombination, and ion migration processes.[35,37,38] In addition, the CDFs for $V_{set}$, $V_{reset}$, and $I_{set}$, $I_{reset}$ are depicted in Figure 2e,f, respectively. Notice that higher C2C variability is observed for $I_{set}$ values compared to $I_{reset}$ values. In addition, the C2C variability of $V_{set}$ is significantly higher than for the $V_{reset}$ case. These results are influenced by the high C2C variability of the HRS prior to the set process and the statistical fluctuations in the HRS current prior to the set transition.

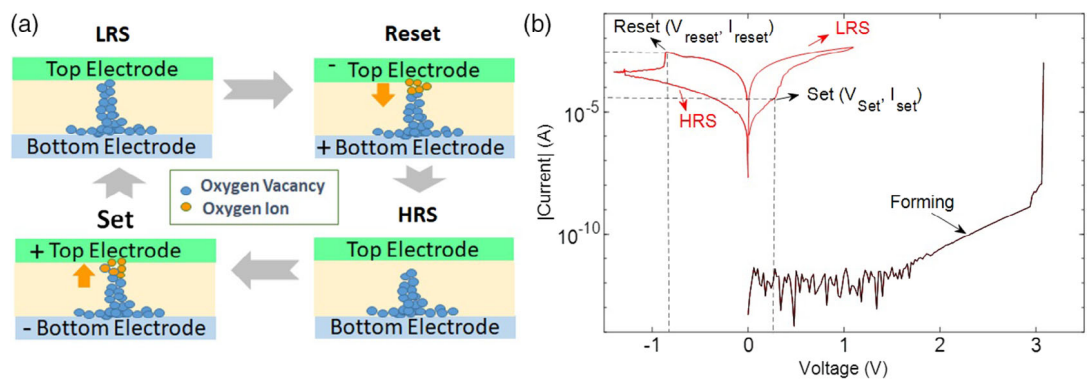

**Figure 1.** a) Schematic representation of the oxide-based filamentary VCM mechanism and b) typical forming process and bipolar resistive switching behavior of HfO$_2$-based RRAMs (including RS parameter definition).

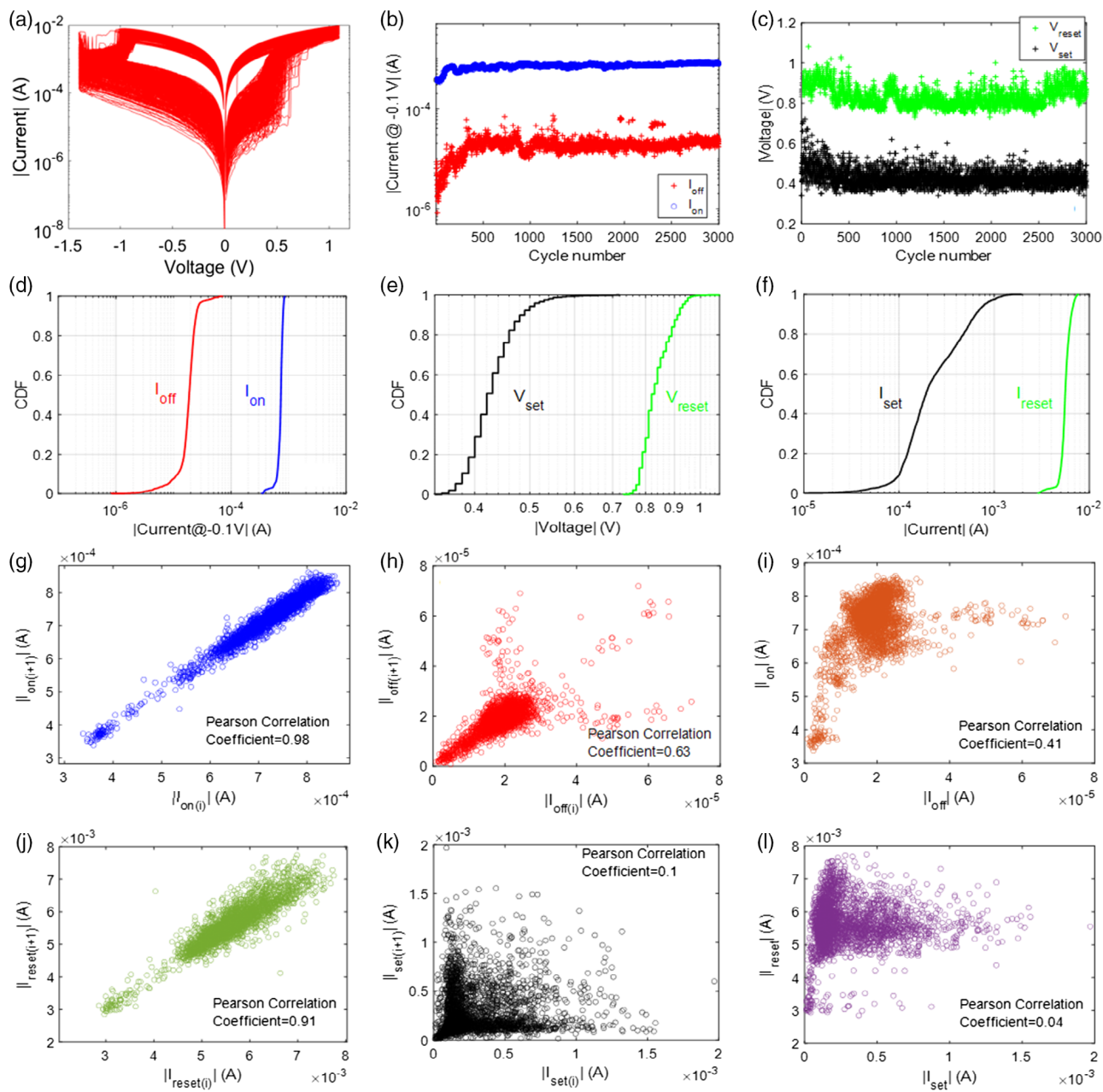

**Figure 2.** a) Typical RS characteristics for 3000 sweep cycles of HfO₂-based RRAMs. b) Evolution of the HRS and LRS current measured at −0.1 V as a function of the number of cycles. c) Evolution of the set and reset voltages as a function of the number of cycles. Cumulative distribution functions of d) $I_{on}$ and $I_{off}$, e) $V_{set}$ and $V_{reset}$, and f) $I_{set}$ and $I_{reset}$. Correlation plots between different RS parameters, g) $I_{on(i+1)}$ versus $I_{on(i)}$, h) $I_{off(i+1)}$ versus $I_{off(i)}$, i) $I_{on}$ versus $I_{off(i)}$, j) $I_{reset(i+1)}$ versus $I_{reset(i)}$, k) $I_{set(i+1)}$ versus $I_{set(i)}$, and l) $I_{reset}$ versus $I_{set(i)}$. The associated Pearson correlation coefficients are indicated in the insets.

To analyze the impact of the stochastic ion migration processes on the HfO₂-based VCM parameters, the connection between the RS parameters in the previous cycle (i) versus the values for the subsequent cycle (i + 1) are represented in Figure 2 for $I_{on}$ (g), $I_{off}$ (h), $I_{reset}$ (j), and $I_{set}$ (k). The associated Pearson correlation coefficients (ρ) are given in the insets.

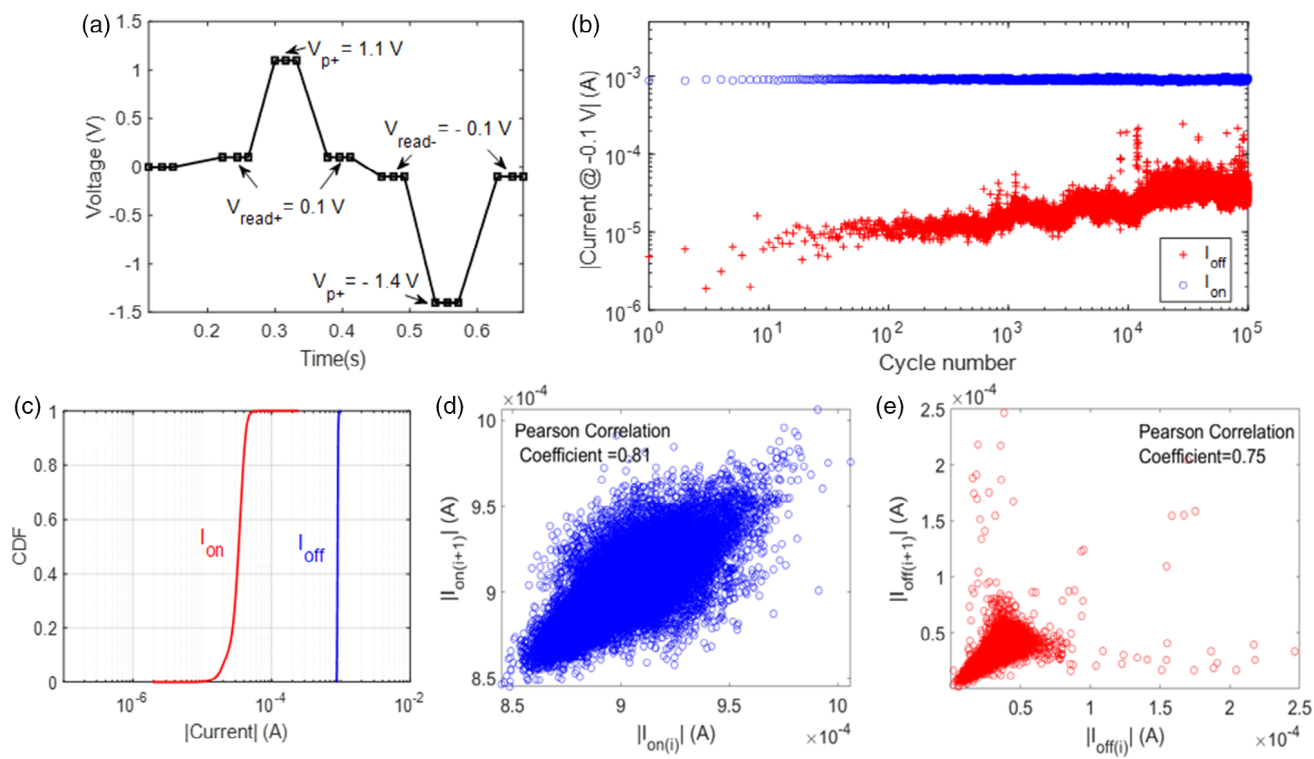

**Figure 3.** a) Pulse scheme applied to the devices. b) Evolution of $I_{on}$ and $I_{off}$ measured at $-0.1\,V$ as a function of the number of pulsed cycles. c) Cumulative distribution functions of $I_{on}$ and $I_{off}$. Correlation plots between different RS parameters: d) $I_{on(i+1)}$ versus $I_{on(i)}$ and e) $I_{off(i+1)}$ versus $I_{off(i)}$. The associated Pearson correlation coefficients are indicated in the insets.

The results reveal a strong correlation of the $I_{on}$ and $I_{reset}$ parameters between two successive cycles, giving $\rho$ values larger than 0.9. This fact can be explained by the relation between the metallic filament strength (size and stoichiometry) for subsequent cycles (in this context the concept of autocorrelation will come up in the time-series analysis described in Section 5.3). In addition, Figure 2h shows that $I_{off(i+1)}$ is statistically correlated to $I_{off(i)}$, with a value of $\rho \approx 0.63$. Notice that there is lower correlation between successive cycles of $I_{off}$ compared to $I_{on}$. This fact is attributed to the higher C2C variability of the off-state current. Furthermore, the correlation plots between $I_{on}$ versus $I_{off}$ and $I_{reset}$ versus $I_{set}$ are depicted in Figure 2i,l, respectively. The results indicate a low correlation between $I_{on}$ versus $I_{off}$, and no association between $I_{reset}$ versus $I_{set}$, corresponding to Pearson correlation coefficients of $\rho \approx 0.41$ and $\rho \approx 0.04$, respectively. This fact may be also influenced by the higher variability of the HRS current and by the noisy set process due to random ion migration events.

For the use of these devices as circuit components, it is interesting to consider the RS behavior when voltage pulses are applied. This behavior is analyzed for the same type of devices henceforth. The pulse sequence scheme is represented in Figure 3a. Notice that the maximum and minimum voltages applied to the device in pulse operation mode are programmed to be the same as for the single sweep cycle in Figure 2a, being in the studied devices $+1.1$ and $-1.4\,V$, respectively. However, it is worth mentioning that the time, and therefore, the energy consumption during sweep and pulse modes differs. Figure 3b shows the evolution of $I_{on}$ and $I_{off}$ at $V_{READ} = -0.1\,V$ for $10^5$ successive pulsed cycles, while Figure 3c shows the CDFs for $I_{on}$ and $I_{off}$. The results show that similar current values were observed in both experiments, obtaining a window of more than one decade between the LRS and the HRS. Furthermore, as in the case of the sweep mode, the C2C variability is also higher in the HRS than in the LRS. However, the $I_{on}/I_{off}$ ratio tends to slightly decrease during cycling. In addition, the first cycles show a significantly larger C2C variability than the following cycles. Finally, the connection between the values of the RS parameters in the previous cycle with respect to the values of the following cycle for $10^5$ pulsed cycles are represented in Figure 3d for $I_{on}$ and in Figure 3e for $I_{off}$, giving values of $\rho \approx 0.81$ and $\rho \approx 0.75$, respectively. This result is in line with the data obtained by the sweep mode, indicating that the correlation is preserved over long series of pulses. More on these issues will be given in Section 5.3 by incorporating the time-series models.

The analysis of the C2C variability of RS parameters of HfO$_2$-based VCMs indicates a highly reproducible bipolar RS behavior, with two well-defined resistance states and good device reliability, which make these devices very competitive for emerging applications.

### 2.1.3. Key Variability Issues

In this subsection, we will present some key variability issues of filamentary RRAMs that are a consequence of the stochastic nature of the physical processes responsible for the reversible

CF formation. It is important to address the main critical aspects inducing device variability to obtain an in-depth understanding of the main variability sources and to assess the electrical behavior of filamentary-based memristors. **Figure 4** shows the evolution of the HRS and LRS currents at −0.1 V versus the number of cycles obtained from voltage ramps measurements. Three different key variability issues are evidenced. In some cases, a higher initial cycle-to-cycle variability for a few cycles is observed until the device stabilizes (Figure 4a). The number of cycles required to achieve the resistive states stabilization may vary depending on the studied device technology and the electrical stimulus applied. Tens to hundreds stabilization cycles may be necessary sometimes.[63] Moreover, it has been reported that multiple discrete current levels can individually last tens or hundreds cycles in the LRS of the devices (Figure 4b), showing a significant C2C variability when a large number of cycles are assessed.[64] This phenomenon is the result of the random nature of the reversible CF path formation through percolation processes, which can be related to a different size, shape, or number of CFs. After some cycles, the same or new CFs could nucleate in the weaker regions of the insulator (thickness or material inhomogeneities, edge of the devices, etc.). It is worth mentioning that this phenomenon mostly occurs in large area devices. The endurance of filamentary RRAMs must also be taken into account[65,66] since a gradual narrowing of the memory window can occur during the degradation phase.[13,67] This phenomenon is associated with a progressive reduction of the $I_{on}/I_{off}$ ratio (Figure 4c).

Finally, it is worth mentioning that filamentary-based memristors have some critical challenges related to major reliability concerns such as retention at high temperatures and the stability of the resistance states.[40] Moreover, the presence of reversible random telegraph noise (RTN) fluctuations can influence the stability of the resistance state. The RTN phenomenon is related to the variations between the electronic states of defects located in the vicinity of the filamentary path.[68,69] Furthermore, irreversible current fluctuations owing to defect density variations inside or near the CF may appear.[69,70] These current variations can be attributed to the field and temperature-assisted oxygen vacancy generation, recombination, or ion migration processes that control the size and stoichiometry of the CF.

## 2.2. Devices Based on Two-Dimensional Materials

### 2.2.1. On the Need for 2D Materials

As highlighted in the previous section, C2C and device-to-device (D2D) variability issues have diminished the initial conception perspectives of RS-based technologies. During the past decade, many groups have suggested alternative techniques and materials configurations focused on mitigating or solving these showstoppers.[71–75]

One of the strategies followed for this purpose is to find the proper material or combination that allows controlling as much as possible the stochastic process associated with the formation of CFs, migrations of ions/vacancies, reduction of the Schottky barrier at the interface, etc., taking into account the materials (dielectrics and electrodes) considered.[39,76] Frequently, researchers have used stack engineering to circumvent the problem of C2C variability. The idea behind this strategy is to try to confine as much as possible all those stochastic mechanisms by splitting the switching process through several material layers. For instance, $HfO_x/TiO_x/HfO_x/TiO_x$ was used as a switching layer showing a promising reduction of C2C and D2D variability.[77] Similarly, Jiang et al.[78] considered $HfO_x/Ti/HfO_x/Ti$ as the switching medium improving the linearity of the electrical response of the device. In the field of neuromorphic applications, the combination $Al_2O_3/TiO_{2−x}$ was successfully used as a switching layer in a neural network to classify grayscale images.[24] As previously mentioned, ion motion and concurrent redox reactions in the dielectric layer play an important role in the variability of the devices. For this reason, another strategy has been employed to control the ions that penetrate the oxide layer by properly designing the electrodes. Yeon et al.[74] employed an Ag–Cu alloy as the top electrode in amorphous silicon (a-Si)-based memristors to improve the uniformity in cycling electrical response. In this configuration, the CF is formed by Ag as a primary mobile metal alloyed with copper. Here, Cu regulates the Ag movement stabilizing the formation and disruption of the filament, i.e., reducing the C2C variability.

In the past decade, bidimensional materials (usually known as 2D materials) have become a good alternative because of the type of defects that appear during the fabrication process (see

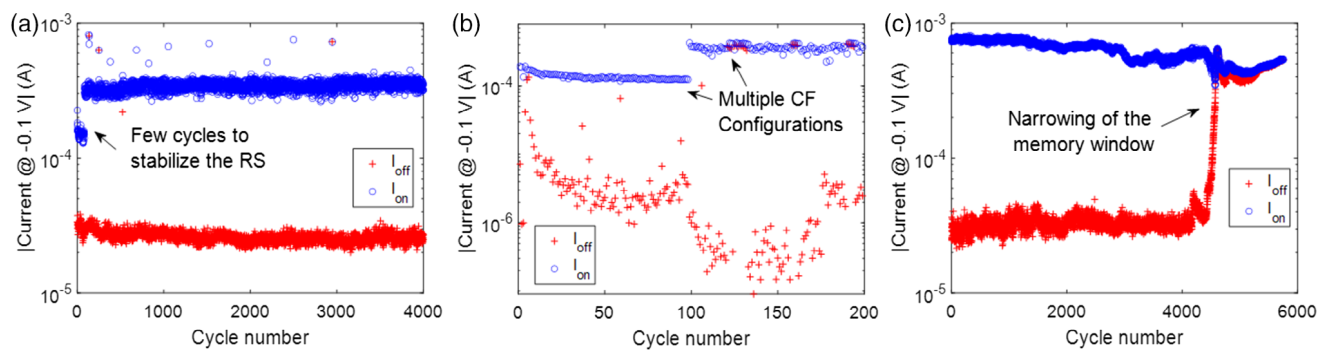

**Figure 4.** Evolution of the HRS and LRS currents at −0.1 V versus the number of cycles measured for a long resistive switching series of voltage ramps that lead to consecutive set and reset processes in filamentary RRAMs. Several cycle-to-cycle key variability issues are evidenced. a) Resistive switching instabilities in the first cycles. b) Formation of well-defined LRS current levels during cycling due to the formation of multiple CF configurations. c) Progressive reduction of the memory window after the degradation of the insulator takes place, for devices fabricated with nonoptimal manufacturing processes.

Section 2.2.2), and their characteristic crystallographic structure helps to confine the RS process. Several studies suggest that the RS process is assisted by the grain boundaries present in these materials.[79–82] In addition, these materials exhibit a wide variety of electronic properties, including semi-metallic, direct/indirect semiconducting, and insulating behavior according to their atomic composition.[83] 2D materials are a large class of materials consisting of stacks of individual layers held together by Van der Waals forces.[84] Each layer is formed by covalently bonded atoms, resulting in a layered structure that is stable even in the form of a single layer. This layered structure allows the accurate control of the thickness of the stack since its variation is discrete, i.e., it is typically defined by the number of layers. This configuration permits memristor fabrication with very thin insulator layers to confine the stochastic processes related to the switching operation. This issue significantly helps to reduce the C2C variability.[85] However, some additional undesired effects related to the nature and fabrication processes of these materials must be taken into consideration.

The second more common strategy followed in order to control the C2C and D2D variability is to connect a transistor in series with the RRAM device (1T1R, see **Figure 5**a) to regulate the switching properties of the device structure as an element in a system.[86–89] The role of the additional transistor is to prevent the sneak-path leakage effect in the memory cell array and force the memristor to work only in a small range of voltages to minimize variability effects. That is, the role of this additional transistor is to connect/disconnect every memristor from the array forcing it to work in the optimum voltage range exclusively. In this line, other works suggested using another RS device as a "selector" substituting the mentioned transistor thus creating a 1S1R memory cell (see Figure 5b).[90–94] In this case, this new selector device controls the operation regime of the memristor. Another proposal that aligns with the philosophy of including additional elements was suggested by Lastras-Montaño et al.[95] Here, one single memory cell contains two serially connected memristors and one transistor (1T2R) as shown in Figure 5c. Using this configuration, the information is stored as the ratio of the resistance between both memristors instead of a single device, decreasing thus the C2C variability effects. Similar to the 1T1R cells, the transistor is in charge of preventing the sneak current.

Here, the 2D materials are also playing an important role since they can also be integrated as a structural part of the transistor as

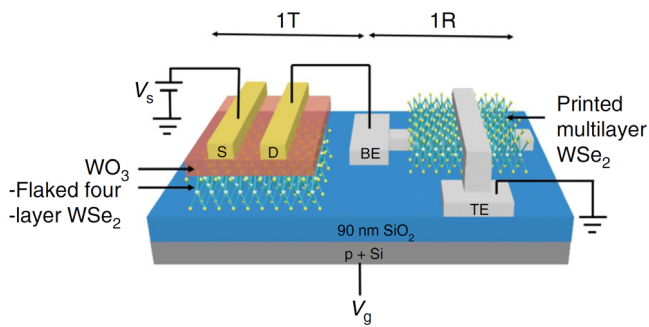

**Figure 6.** Schematic of a 1T1R memory cell with flaked WSe$_2$ transistor and ReRAM and the corresponding circuit representation. Reproduced with permission.[96] Copyright 2019, Springer Nature.

**Figure 6** shows.[96–98] The integration of 2D materials with all electronic components could extend their functionality and help to overcome fundamental device challenges such as the gate leakage current of the transistor, temperature dissipation, and transparency.[99–104] In this application case, 2D materials help to control the RRAM C2C and D2D variability and enhance the transistor properties.

### 2.2.2. Variability Sources on 2D Materials

During the past few years, memristive devices have been realized with exfoliated materials, mainly using h-BN,[105–108] MoS$_2$ nanosheets,[109–113] WS$_2$,[114,115] or WSe$_2$.[116] Unfortunately, their fabrication processes are not highly optimized like in the case of the traditional transition metal oxides and similar materials. For this reason, one of the limitations from the C2C and D2D variability point of view of this type of materials is the fabrication processes available and their limitations. Shen et al. demonstrated[117] that several specific type of defects can appear during the growth and transfer processes of these materials to the device. **Figure 7** shows the types of defects that can appear using this fabrication technique. These defects typically are 1) thicker islands; 2) point defects; 3) contaminant particles; 4) wrinkles; 5) gaps between the h-BN and the substrate; 6) twin boundaries; and 7) grain boundaries. Those defects can potentially induce some parasitic effects such as roughness at the interface between layers, parasitic charge accumulation at the interface or in the impurities, etc. They can critically affect the C2C variability.

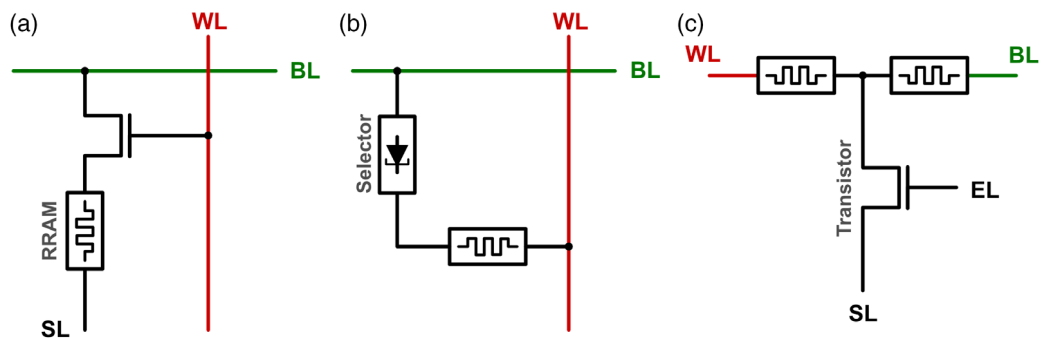

**Figure 5.** Schematic of a) 1T1R, b) 1S1R, and c) 1T2R memory cell. All these types of memory cells are placed in an array where WL is the word-line, BL is the bit-line, SL is the sense-line, and EL is the enable-line.

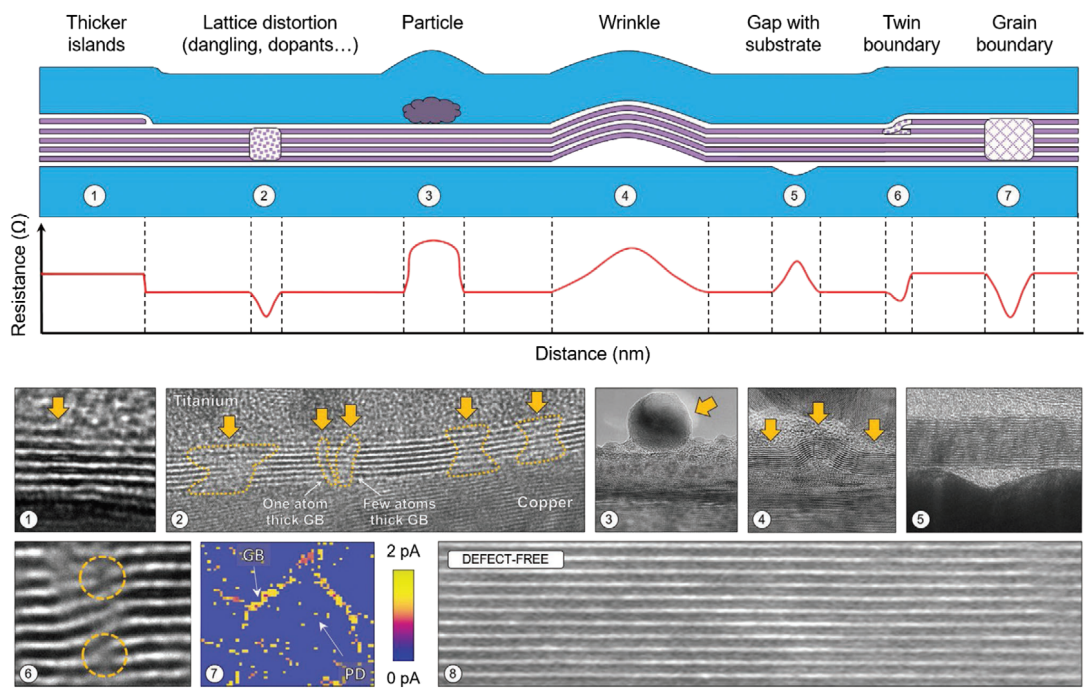

**Figure 7.** Schematic of the main defect types that can be found in any exfoliated 2D material. These defects are (top): 1 thickness fluctuations, 2 defective bonding, 3 rests of polymer particles from the transfer scaffold, 4 wrinkles, 5 suspended 2D materials, 6 twin boundaries, and 7 grain boundaries. At the bottom, cross-sectional TEM images of different samples of layered h-BN are displayed as examples of each type of defect described at the top together with a CAFM current map. Reproduced with permission.[117] Copyright 2021, Wiley-VCH.

On the other hand, the exfoliation technique has two inherent limitations that directly affect the D2D variability. First, it is not possible to fabricate large areas of these materials because the technique implies mechanically transferring the 2D material to the device. In addition, it is not possible to accurately control the thickness (or the number of layers) for large areas.[117–119] These two drawbacks make it very difficult to fabricate many devices while keeping a certain degree of homogeneity of the defect distribution and thickness.

Paradoxically, some of these types of defects can also play a remarkable role in the RS mechanisms and its stability (i.e., low variability) because the RS process usually takes place in the lower resistance regions of the device.[117] That means that if there are thicker islands, lattice distortions, twin boundaries, or mainly grain boundaries in the insulator (which produces a local reduction of the resistance as Figure 7 shows) the RS process will take place in these regions confining the process and reducing the variability.[117] However, an excess of these defects can produce the opposite effect from the variability perspective. Several weaknesses across the device area will produce several CFs, for instance. So, in this case, the behavior of the device will be controlled by the competition between those CFs critically increasing the C2C variability.

The switching behavior has been also observed in 2D materials grown by chemical vapor deposition (CVD) like h-BN,[120–122] MoS$_2$,[123–125] and MoSe$_2$, WS$_2$, WSe$_2$.[123] This technique solves some of the common defects produced during the exfoliation technique like wrinkles. Using this method, memristors show better C2C stability due to the reduction of undesired defects. In addition, CVD technique also allows the growth of larger areas of 2D material increasing the homogeneity between devices, i.e., improving the D2D variability.

## 2.3. Good Practices for the Experimental Data Analysis Focused on Variability

In the current state of maturity of the RS technology, a direct application of these devices is ongoing (TSMC and INTEL have commercial fabrication processes). For this reason, a more exhaustive analysis of experimental results is necessary to provide accurate and meaningful results.

For instance, it is very common to read in many papers a sentence like "Figure X shows a representative *I–V* characteristic of...", and the study of this paper is based on this particular *I–V* curve. At best, only one *I–V* curve is highlighted on top of the shadowed area created by many other curves plotted in the same graph, as the pink dashed line in **Figure 8**a. However, this claim is rarely supported by any statistical evidence. Is this curve statistically relevant or, is it just the curve that is roughly in the center of the shadowed area generated by all *I–V* curves? To avoid considering an edge case as a representative device response, a C2C variability analysis is imperatively required.

Figure 8a shows a typical set of *I–V* curves measured in a standard bipolar memristor, where the red solid lines correspond to the set curves and the blue lines to the reset ones. Checking both shadowed areas, the pink dashed line is one I–V curve placed at the center of the colored area which could be used as a "representative" electrical response of the device under study. Similarly, the median value of all measured I–V curves gives a similar result (green dashed line in Figure 8a), but this curve is not a real measurement in this case. However, before considering the pink or the green line as the best candidate, a better practice is to calculate the probability density over all measured I–V curves in order to identify which electrical response is more frequent (see Figure 8b). This statistical analysis was done in Figure 8b, where the darker regions identify the most probable electrical response of the device. According to this result, it is straightforward to conclude that the highlighted black solid line in Figure 8a is a better representative I–V curve for this device because it overlaps almost all dark regions from the probability density plot. Using the pink or the green line as the reference, we could conclude that the memory window of this device is typically around 7 orders of magnitude, but the most probable memory window is 2 orders of magnitude smaller. In addition, this analysis provides complementary information about the variability of the device by identifying other secondary most probable states of the device. This information is not only important from the variability viewpoint but it is also relevant for understanding the role of the processes that take place during the switching event.

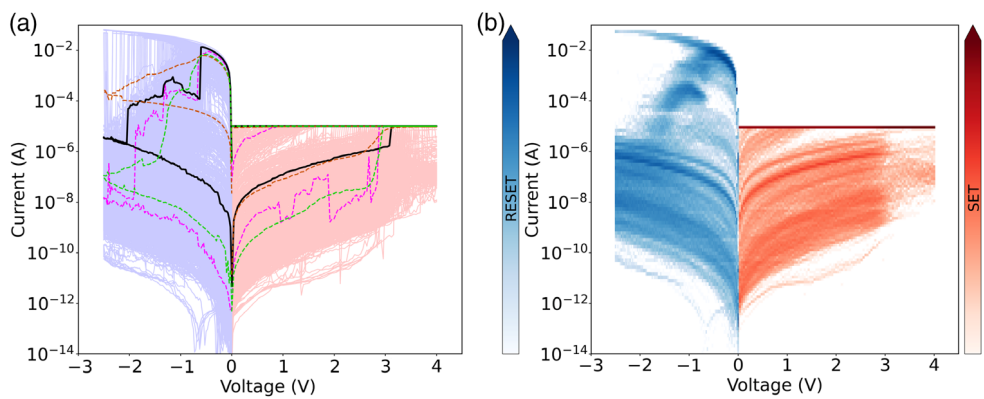

**Figure 8.** a) I–V curves of 1028 cycles of a standard memristor. The pink dashed line is a representative curve of the distribution selected as one I–V curve roughly located in the center of the distribution. The brown dashed line is a representative curve of the distribution calculated as the mean value of all measured I–V curves. The green dashed line is the median value calculated over all measured I–V curves. b) The black solid line is a representative curve estimated from the current density plot.

Another methodology frequently found in the literature is the use of the mean (brown dashed line in Figure 8a) value of all measured I–V curves as the reference curve. Leaving aside that this curve is not a real measurement, this method does not always give a reasonable result. This calculation was also done, and the result is illustrated by the brown dashed line in Figure 8a. This method gives us a good estimate for the set I–V curves according to the criteria defined previously. However, if the electrical response of the device shows a very high current dispersion, the result is completely wrong. This is precisely what occurs in the reset region where there are ≈9 orders of magnitude between the HRS and the LRS. In this case, the brown line is located far from any region of interest concerning the electrical response of the device during the reset process.

In summary, the mean, median, or manually selected I–V curve can be a good representation of the behavior of the device under study according to the context of our research (compact or kinetic Monte Carlo modeling, qualitative descriptions, etc.) However, a statistical study of the measured curves is strongly recommended to avoid misinterpretations.

Another set of parameters widely used for memristor characterization are the set and reset voltages, introduced in Section 2.1. These parameters also need to be treated carefully from the statistical point of view to provide accurate and meaningful results. The first step is to define a clear methodology for the parameter extraction process from the experimental data. It is mandatory to be systematic and consistent to avoid any undesired bias in our data. In this sense, Villena et al.[126] suggested five different extraction methods for the reset voltage (see another paper on parameter extraction techniques in ref. [127]). Once data are collected, their correct treatment and representation are critical for extracting valuable information.[62] A common statistical approach for variability characterization of the set and reset voltages often used by the electron device community is the Weibull distribution (see **Figure 9**). This representation is very useful to understand the variability of the involved parameters. However, it is also highly recommended the inclusion of the distribution parameters in order to correctly justify whether the data follow this distribution or not. Again, the absence of these parameters could lead to a misinterpretation of the obtained results.

**Figure 10** shows other three useful methods for variability representation: histograms, cycle-to-cycle progression, and boxplots. As shown in Figure 10a, it is very common to fit the result of the histogram by a Gaussian distribution. This information is useful to clarify and predict in a rough way the device behavior. However, similarly to what was mentioned previously, it is also important to justify statistically that the experimental data really follow this distribution showing the parameters of the Gaussian distribution. The cycle-to-cycle progression plot (see Figure 10b) is useful to identify any trend in the resistance and variability throughout the device operational life. The progressive device degradation can also affect its variability. This information is critical because one technology can be very stable for a few cycles, but this does not mean that the device will keep this low variability level throughout its entire lifetime.

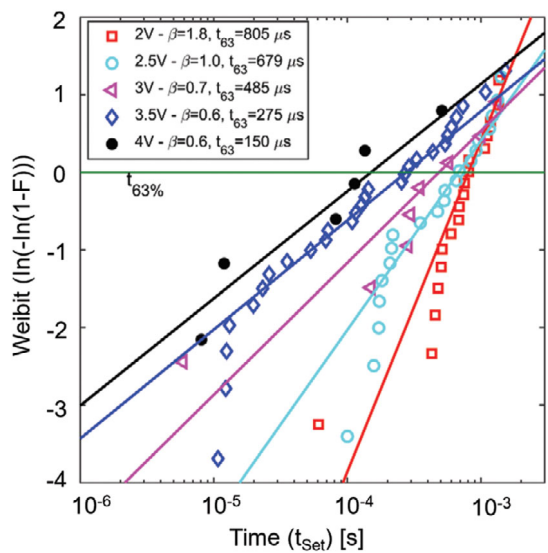

**Figure 9.** Example of a correctly fitted Weibull distribution showing the parameters of the statistical distribution. Reproduced with permission.[250] Copyright 2021, Wiley-VCH.

Finally, the boxplot is one of the most useful representations to understand variability (see Figure 10c,d). Although this graph is not linked to any statistical model and it is not possible to make inferences, it gives a lot of information in a reduced space. For this reason, this kind of representation is also recommended for D2D variability studies, since a complete study will require showing the results for a wide set of devices.

For claiming that the D2D variability of a certain device is high or low, it is necessary to show data related to a statistically significant number of devices. A good example is the work of Chen et al.,[120] where 48 identical devices were studied, and their results were accurately represented, as shown in **Figure 11**. As mentioned above, a boxplot is a good method to show a lot of information in a compact way. In this plot, it is not only possible to see the D2D variability but also the C2C variability. Accordingly, just using this type of plot, it is possible to provide a nice picture of the actual behavior of this technology.

## 3. Variability Physical Models

### 3.1. Stanford-Based Models for Circuit Simulation

In this subsection, we will contextualize the resistive memories study making use of simulations by means of the Stanford model (SM).[128,129] In particular, an extended version of this model has been considered to account for the series resistance.[130] The SM, a compact model developed for circuit simulation, describes the evolution of a CF in the dielectric that shorts the electrodes changing the device resistance (the model is given in Verilog-A version).[131] A gap between the filament tip and the electrode (g) is assumed as the state variable to account for RS operation.[128,129,132] The gap dynamics allows to describe set and reset processes that switch the device resistance between the LRS and HRS. The CF temperature is also calculated to give a highly physical picture of the redox-controlled CF formation and rupture. These characteristics make the SM to be counted as one of the physically based RRAM models; among them the following can be highlighted.[133–137] Other modeling approaches are also possible, see Section 4 and 5 for examples on variability stochastic and behavioral models.

In addition to the SM features explained above, a model to represent the cycle-to-cycle variability has also been implemented.[128,129] As reported in previous sections, variability is always present in RRAMs due to their inherent stochastic behavior; this effect occurs for the different technologies employed to fabricate these devices[3,6,7,34,138–140] and for different operation conditions where temperature and external fields (e.g., magnetic) are changed.[141,142] In this respect, there is a clear need to model variability to make EDA tools suitable for IC design. In this section and the following ones, we will focus on the C2C variability by analyzing RS parameters such as $V_{set}$, $V_{reset}$, and the device resistances at the LRS and HRS. The D2D variability is also a key issue; however, our focus will be mainly on C2C variability.

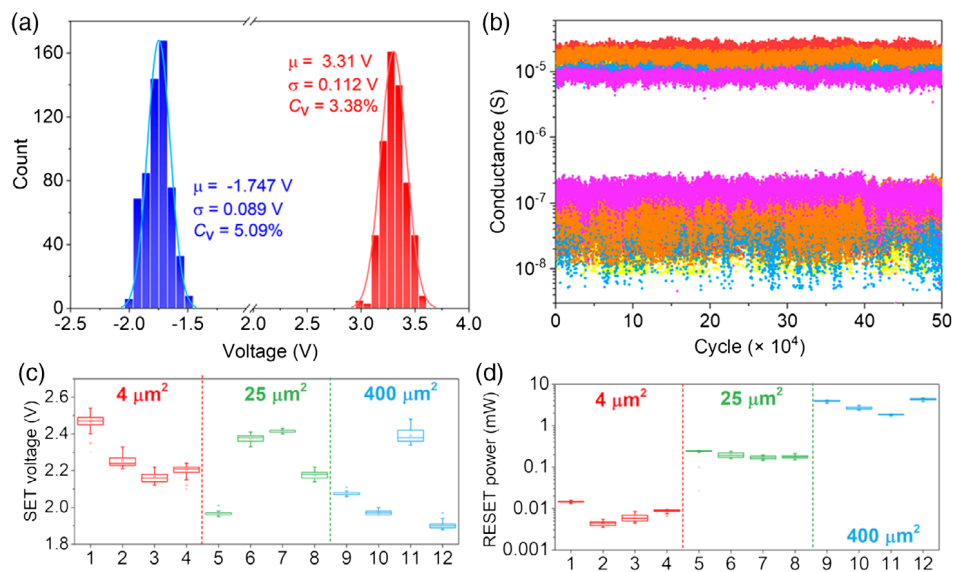

**Figure 10.** Example of different recommended methodologies to estimate and quantify the C2C variability; a) histogram, b) cycle-to-cycle progression, c,d) boxplot. a,b) Reproduced with permission.[251] Copyright 2022, American Chemical Society. c,d) Reproduced with permission.[124] Copyright 2022, American Chemical Society.

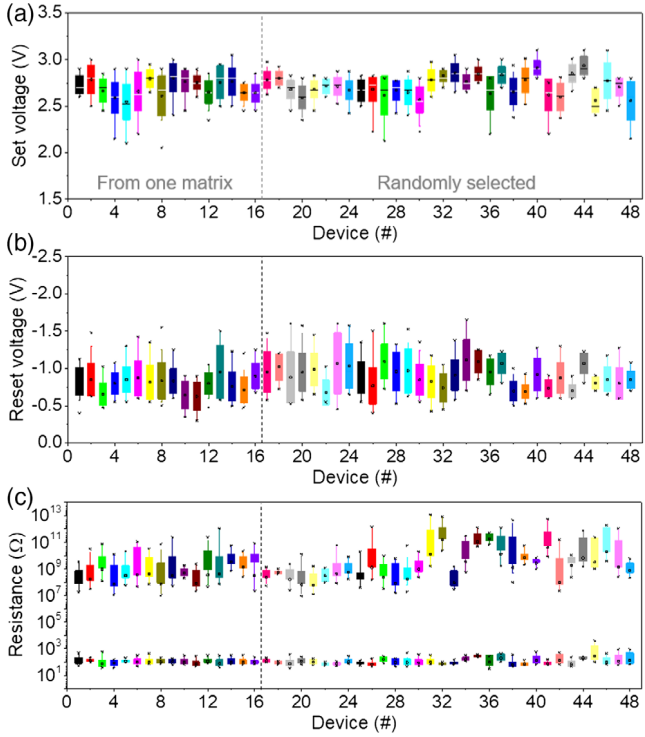

**Figure 11.** Example of a boxplot used to demonstrate the C2C and D2D variability of the device under study. Reproduced with permission.[120] Copyright 2020, Springer Nature.

The use of the SM to fit experimental RRAM I–V curves requires a parameter extraction process; among the parameters, the following can be listed: $I_0$, $g_0$, $V_0$, $\nu_0$, $\gamma_0$, $\alpha$, $\beta$..., see examples in refs. [128–130,132]. In addition, $\delta_g^0$, $T_{crit}$, and $T_{smth}$ are the parameters employed in the variability model, which is implemented as a correction to g, the gap (distance between the CF tip and the electrode), see Equation (1).

$$\delta_g(T) = \frac{\delta_g^0}{\left\{1 + \exp\left[\frac{T_{crit} - T}{T_{smth}}\right]\right\}} \tag{1}$$

where $T$ is the CF temperature.[128] This gap variation is included in the calculation of the state variable, g, as follows (Equation (2))

$$g_{t+\Delta t} = \int \left(\frac{dg}{dt} + \delta_g \times \chi(t)\right) dt \tag{2}$$

where $\delta_g(T)$ models the strength of the random variations, and $\chi(t)$ stands for a zero-mean Gaussian distributed noise sequence with a root mean square of unity[132] that is randomly generated at each time step. The calculation of the g variable is described by Equation (3).

$$\frac{dg}{dt} = -\nu_0 e^{-\frac{E_{g,m}}{k_B T}} \sinh\left(\frac{\gamma(g)a_0 q V_{RRAM}}{t_{ox} k_B T}\right) \tag{3}$$

where $t_{ox}$ is the dielectric thickness, $E_g$ ($E_m$) is the activation energy for vacancy generation (oxygen ion migration) in set (reset) processes, $\nu_0$ is the velocity containing the attempt-to-escape frequency, $a_0$ is the atom spacing and $V_{RRAM}$ the applied voltage, which drops mainly across the gap g; and $\gamma$ is the field local enhancement factor that accounts for the polarizability of the material.[132] It can be calculated as $\gamma = \gamma_0 - \beta g^\alpha$, where $\gamma_0$, $\beta$, and $\alpha$ are fitting parameters. The current was calculated using the voltage and gap information as follows $I = I_0 e^{-\frac{g}{g_0}} \sinh\left(\frac{V_{RRAM}}{V_0}\right)$.[132]

The SM was implemented in Verilog-A. We have reproduced two experimental set and reset curves in **Figure 12** corresponding to RRAM devices based on the Ni/HfO$_2$/Si–n$^+$ stack with a 10 nm-thick oxide grown by atomic layer deposition.[64] Since the mechanisms involved in RS can be modeled, in general, by an Arrhenius-type equation, the fitting of the parameters (mostly the activation energies) allows the simulation of I–Vs from a wide variety of technologies with the SM, including unipolar RRAMs if the absolute value of the applied voltage is considered (Equation (3)) in the calculations (our case for some of the simulations), although the physics might be different from that behind RS in the devices employed to originally develop the SM.

The parameters employed in the simulations are given in **Table 1**. The agreement between experimental and simulated data is reasonably good; see that a few parameters are different for the set and reset processes to improve the fitting, as it was the case in ref. [132], although the technologies are different.

Making use of the parameters reported in Table 1, we have performed simulations accounting for C2C variability. **Figure 13**a shows several experimental RS cycles for the devices described above with unipolar RS. The simulation results are shown in Figure 13b. The variation of the I–V curves is comparable to what is seen experimentally in most aspects, although set and reset voltages variations cannot be fully modeled. Most of the simulated RS parameters spread out as in the experimental case.

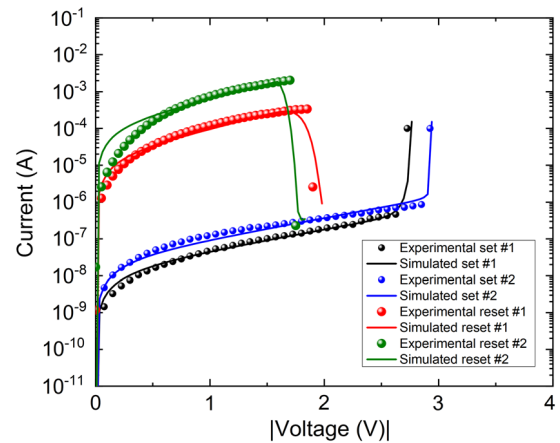

**Figure 12.** Current versus voltage for experimental data (symbols) and simulated data (lines) for the electrochemical metallization cells described in the text. These simulations were obtained by employing the Stanford model using the parameters presented in Table 1 which are in line with those reported in ref. [132].

**Table 1.** Parameters employed in the SM simulation shown in Figure 12.

| Stanford model parameter values | | | | | |
|---|---|---|---|---|---|
| Device parameters | Unit | Cycle #1 | | Cycle #2 | |
| | | Set | Reset | Set | Reset |
| $V_0$ | V | 0.75 | 0.55 | 0.75 | 0.55 |
| $I_0$ | mA | 0.2 | 0.8 | 0.2 | 0.4 |
| $g_0$ | nm | 0.167 | 0.155 | 0.167 | 0.155 |
| $\nu_0$ | m/s | $5 \times 10^6$ | | $5 \times 10^6$ | |
| $\alpha$ | – | 1 | | 1 | |
| $\beta$ | – | 1 | | 1 | |
| $\gamma_0$ | – | 30 | 32 | 25 | 36 |

More than 2800 experimental cycles were measured in a long RS series. The data have been analyzed by calculating the device resistance, $R_{LRS}$, at $V_{RRAM} = -0.1$ V, as presented in **Figure 14**a. The same procedure has been performed to determine the value of the $R_{LRS}$ measured at $-0.1$ V for 23 consecutive simulations performed with the variability model on with the SM, as shown in the inset of Figure 14a. The experimental and simulated $R_{LRS}$ cumulative distribution functions are plotted in Figure 14b; as can be seen, a reasonable agreement is achieved taking into consideration the simplicity of the variability model used (and, in general, of the whole model since several conduction mechanisms could be present along an I–V curve and, also, RS could be influenced by several effects at once). At the physical level, variability is linked to different random processes that lead to RS, in this respect, an accurate modeling would imply a very complex modeling process. This issue will be tackled from different viewpoints in the following sections. At this compact modeling level, the model can generate variability in line with experimental data in some qualitatively aspects.

As described in Equation (1), the variability model makes use of 3 parameters: $\delta_g^0$, $T_{crit}$ and $T_{smth}$.[128,143] It is interesting to study the influence of each parameter on its own; to do so, we simulated I–V curves changing just one of these three parameters at a time, see **Figure 15**. Five different cycles are employed in the simulations, and, to quantify the level of influence of the variability parameters, the current variation (see Figure 15) is calculated at a voltage of 1.5 V for the sets and 1 V for the reset transitions.

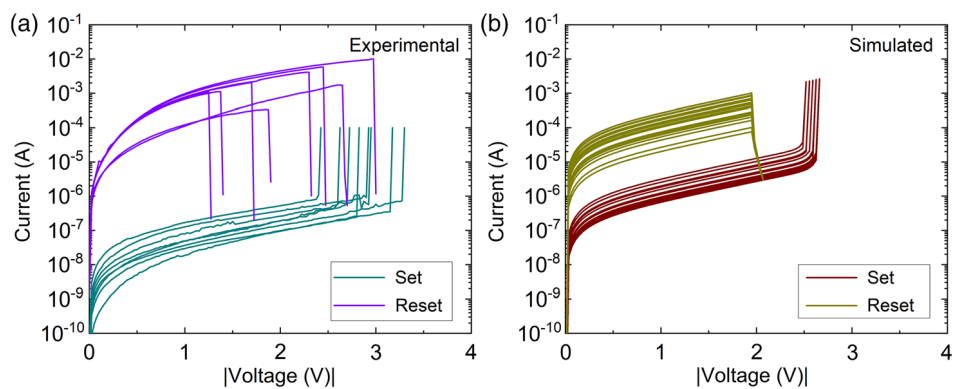

**Figure 13.** Current versus voltage set and reset cycles for a) experimental, b) simulated data for the investigated devices with unipolar RS. The SM was employed including the variability module using Equation (1) with the following parameters $V_0 = 0.75$ V, $I_0 = 3$ mA, $g_0 = 0.167$ nm, $\nu_0 = 5 \times 10^6$ m s$^{-1}$, $\alpha = 1$, $\beta = 1$, $\gamma_0 = 30$, $\delta_g^0 = 0.5$ nm, $T_{crit} = 450$ K, $T_{smth} = 400$ K.

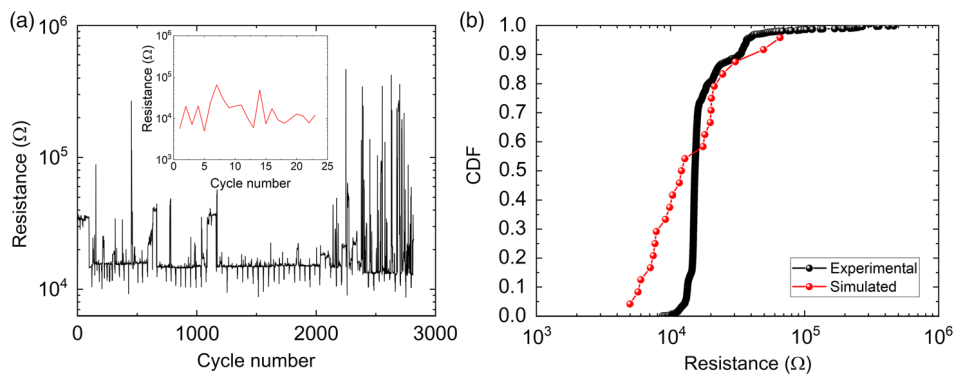

**Figure 14.** a) $R_{LRS}$ measured at $-0.1$ V versus cycle number for the experimental and simulated (inset) curves in the whole RS series, b) cumulative distribution functions for the values plotted in (a).

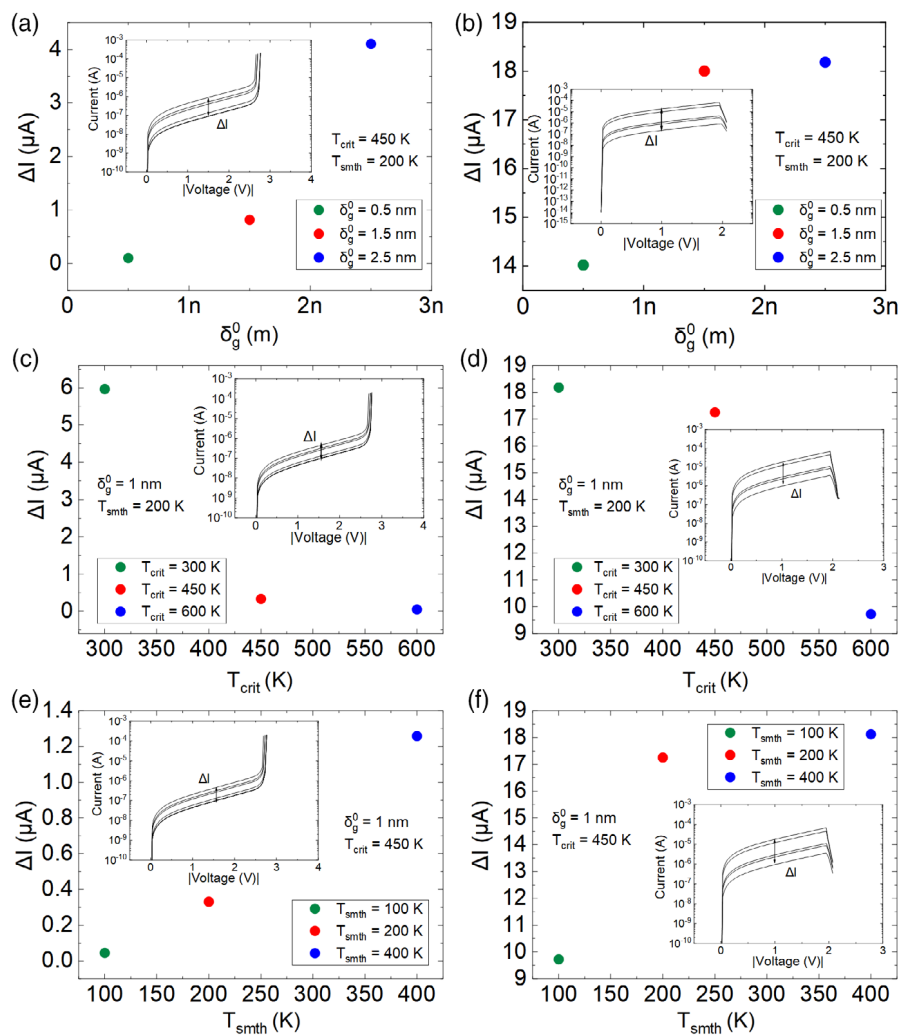

**Figure 15.** Current variation calculated at a fixed voltage of 1.5 V (1 V) for 5 set (reset) different cycles versus $\delta_g^0$, $T_{crit}$ and $T_{smth}$ variability parameters. Inset: current versus voltage for SM simulated data. These simulations were obtained by enabling the variability module and isolating some parameter oscillations: a) $\delta_g^0$ (set), b) $\delta_g^0$ (reset), c) $T_{crit}$ (set), d) $T_{crit}$ (reset), e) $T_{smth}$ (set), f) $T_{smth}$ (reset).

The influence of $\delta_g^0$ on the set (reset) curves is described in Figure 15a (15b). As the $\delta_g^0$ parameter rises, the current variations increase too. This fact arises since $\delta_g^0$ stands for the gap deviation, which affects the final g value, that is responsible for the total current flowing across the device. In Figure 15c, the role of $T_{crit}$ is explained in set curves (Figure 15d for the reset); the higher this parameter, the lower the current variation. In this case, $T_{crit}$ works as a threshold temperature. The last parameter, $T_{smth}$, is a parameter that helps building the smoothing function that is employed in the variability model. Its variation works as $\delta_g^0$; when it is increased, the current variations rise too, as shown in Figure 15e (15f) for a set (reset) process.

The series resistance ($R_{series}$) is also a key parameter for modeling purposes.[130] An accurate determination of this parameter allows the correct fitting of $I$–$V$ curves. The series resistance particularly influences the $I$–$V$ curve shape of some bipolar VCMs. Thus, in some technologies it is mandatory to include a proper value of the series resistance to obtain a good fitting. See **Figure 16**a for an example of SM simulation of VCM devices based on the TiN/Ti/HfO$_2$/W stack.[130] The $R_{series}$ increase shifts the reset curve maximum to higher voltages (in absolute value) and makes que current drop after the current maximum (the numerical condition to fix the reset voltage)[62] more abrupt. The set $I$–$V$ curve also changes with the inclusion of the series resistance. The simulated $I$–$V$ curve without series resistance (in orange color) cannot fit the experimental data (in blue symbols), neither in the set nor in the reset processes. However, the needed $I$–$V$ curve shape change produced by the $R_{series}$ inclusion (in particular, for this case, 22.3 Ω) allows the fitting. These effects cannot be obtained by changing other model parameters, as shown in Figure 11 in ref. [130].

In this case, the SM parameters are given in **Table 2**.[130] The procedure to extract $R_{series}$ is given in this reference. A step-by-step technique allows the extraction by means of an automatic numerical process. A normalized voltage is defined as $V_N = V_{Applied} - I_{Measured} \times R_{series}$ and the $I_{Measured}$ versus $V_N$ is plotted for several $R_{series}$ till a vertical I–V set curve is obtained; at this point the $R_{series}$ parameter is selected. The extraction procedure was applied on a data set of 1000 consecutive experimental cycles in a long RS series, see in Figure 16b the corresponding $R_{series}$ histogram.

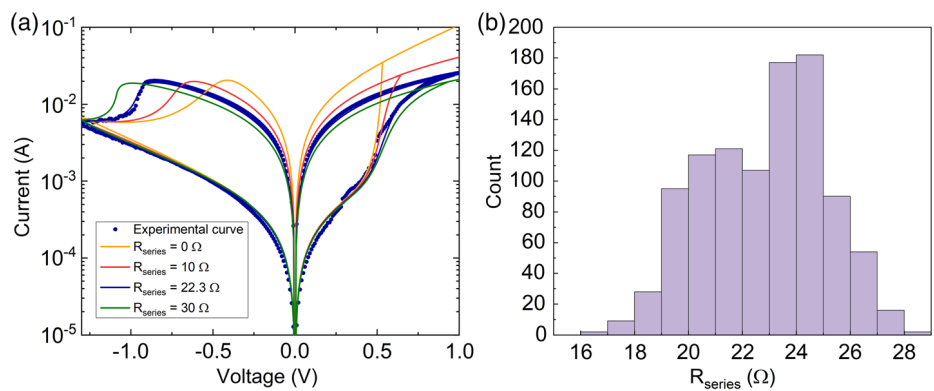

**Figure 16.** a) Current versus voltage for experimental (symbols) and simulated data (lines) to reproduce the typical curve shape of valence change memory cells. Each color line represents a simulation obtained for a different series resistance; the selected value to reproduce the experimental data is 22.3 Ω. These simulations were obtained by employing the SM using the parameters presented in Table 2 which are in line with those given in ref. [132]. b) Histogram for the experimental series resistances extracted for 1000 I–V cycles in a long series employing the method presented in ref. [130].

**Table 2.** Parameters employed in the SM simulation to fit I–V curves of valence change memory cells.

| Stanford model parameter values | | | |
|---|---|---|---|
| Device parameters | Unit | Cycle #1 | |
| | | Set | Reset |
| $V_0$ | V | 0.45 | |
| $I_0$ | mA | 48 | |
| $g_0$ | nm | 0.35 | |
| $\nu_0$ | m/s | $5 \times 10^6$ | |
| $\alpha$ | – | 1 | 1.1 |
| $\beta$ | – | 1 | 15 |
| $\gamma_0$ | – | 20 | |

It is clear that the series resistance also influences the final device C2C variability. Therefore, in addition to the model explained in Equation (1), the inclusion of the series resistance might be needed with the corresponding distribution function to correctly account for the variability of a certain technology. At this point, it is also necessary to comment on time series (the subject of Section 5.3) since in most cases[144] the RS parameter variation is self-correlated along a series of cycles. This issue can be included in the SM for circuit simulation. The development of the time series that includes the dependencies of a certain parameter on regressive values must be done in the context explained in Section 5.3. However, these time series can be incorporated in the SM, as performed in ref. [145].

### 3.2. Quantum Approach to Variability in RRAMs

C2C variability in RRAMs is well known to be the consequence of structural modifications of the atomic bridge (metal ions or oxygen vacancies) that constitutes the switching CFs. Even though a microscopic approach seems a priori an unquestionable requisite to simulate the effects of such discrete atomic rearrangements on the electron transport characteristics, an alternative description, suitable for compact modeling and circuit simulators, arises from the so-called physics of mesoscopic conductors.[146–150] Briefly, mesoscopic physics deals with the conduction properties of systems whose size is in between the macroscopic (bulk material) and the microscopic (atomic or molecular) worlds. Since the electron wavelength in comparison with the size of the confining structure plays a fundamental role within this framework, quantum mechanics is inexorably linked to its foundations. Mesoscopic theory describes electron transport in terms of charge reservoirs, quasi-Fermi levels, electrochemical potentials, potential barriers, transmission probabilities, etc., without making specific reference to the underlying microscopic nature of the filamentary conduction problem. More advanced approaches such as the nonequilibrium Green's function method allows considering a detailed representation of the atomic network in which electrons move, but this tool requires a large computational effort that inhibits its use in circuit simulation environments.[151]

The observation of experimental conductance values around integer and half-integer multiples of the quantum conductance unit $G_0 = 2e^2/h$, where e is the electron charge and h the Planck's constant,[152,153] is considered very often the signature of mesoscopic conduction. Conductance values measured at a fixed bias, obtained from a number of consecutive cycles or conductance values measured during a constant voltage/current stress or bias sweep, are represented using histogram plots with the x-axis normalized to $G_0$. In many cases, these histograms reveal peaks that are interpreted as an indicator of the number of channels available for conduction or as the occurrence of preferred atomic configurations for the CF.[154] Many experimental results on RRAMs have been described in terms of filamentary conduction through atom-sized structures.[119] This is the case of a wide variety of binary oxides such as $SiO_x$, $CeO_x$, $HfO_2$, $Ta_2O_5$, NiO, ZnO, a-Si:H, $TiO_2$, $V_2O_5$, and $YO_x$.[155] Even though the observation of the peaks in the histogram is often straightforwardly associated with quantum point-contact conduction, caution should be exercised with this interpretation: only for simple s-electron metals, the transmission probability for the conductance channels is expected to be close to integer values.[149] For that reason, observations in the RRAM field should be more appropriately considered to be in the quantum (rather than in the quantized) conductance regime.[154] In addition, it is worth emphasizing that measurements obtained from RRAMs are not as clean as those obtained from split gate-based quantum point contacts in which the size of the constriction can be externally modulated by the voltage applied to the control electrode. Measurements in RRAMs are largely affected by a number of nonideal factors such as the existence of multiple conduction paths, internal series resistance, scattering caused by the granularity of matter, and nonadiabaticity of the CF confining walls.

According to the mesoscopic theory viewpoint for RRAMs,[156,157] the LRS is associated with a linear I–V curve with conductance values in the regime $G \geq G_0$. In this case, the device conductance very often reaches values in the range from 10 to 100 times $G_0$ pointing out the large number of atoms that constitute the CF.[158] Because of that, variability in LRS, though always present, is rarely considered a major technological issue. On the other hand, the HRS is often associated with a nonlinear I–V curve with conductance values in the regime $G < G_0$. However, because of the low number of conducting sites involved in this regime, the transport characteristics are extremely sensitive to structural modifications of the atomic bridge becoming a serious concern in terms of leakage current variation and memory state stability. As revealed from first-principle studies,[159] HRS is characterized by the existence of a gap or potential barrier along the filament that performs as a blocking element for the electron flow. Depending on the transmission properties of this barrier, the CF is able to conduct more or less current. At the end of the day, this is a matter of the size of the constriction's bottleneck and how the atomic valence orbitals in the CF couple: while narrow constrictions lead to HRS, wide constrictions are associated with LRS. As the starting point for the inclusion of variability effects in RRAMs, the

Büttiker–Landauer approach for quantum point contacts is considered here.[160] The analysis does not discriminate between CBRAM and VCM devices, so that these structures are treated on equal grounds; only the confining effect is the relevant feature here. Importantly, the mesoscopic model described below should only be considered as a framework for the understanding of the phenomenology associated with memristors in the quantum regime and not as a complete simulation tool.

According to the finite-bias Landauer's approach,[161] the I–V characteristic of a mesoscopic conductor with a gap or barrier along its length reads

$$I = \frac{2e}{h} \int T(E)[f(E - \beta eV) - f(E + (1 - \beta)eV)]dE \quad (4)$$

where $E$ is the energy, $T(E)$ is the transmission probability, $f(E)$ is the Fermi–Dirac distribution function, and $0 \leq \beta \leq 1$ is the fraction of the applied voltage that drops at the source side of the constriction (see **Figure 17**). For a symmetrical structure, $\beta = 1/2$. Assuming that an inverse parabolic potential barrier for the constriction's bottleneck, $T(E)$, is given by the expression[156]

$$T(E) = \{1 + \exp[-\alpha * (E - \varphi)]\}^{-1} \quad (5)$$

where $\alpha$ is a coefficient related to the longitudinal curvature of the potential barrier and $\varphi$ is the height of the potential barrier that represents the confinement effect. Notice that the kind of barrier we are discussing here is not material-related, but the bottom of the first quantized sub-band within the CF. In the zero-temperature limit, (4) and (5) yield for a monomode conductor

$$I(V) = G_0 \left\{ V + \frac{1}{e\alpha} \ln \left[ \frac{1 + \exp[\alpha(\varphi - \beta eV)]}{1 + \exp[\alpha(\varphi + (1 - \beta)eV)]} \right] \right\} \quad (6)$$

which is represented in **Figure 18**a for different values of $\varphi$. Variability is introduced here by assuming a stochastic model for the confinement barrier $\varphi$. For the sake of simplicity, a normal-distributed barrier height $\varphi \approx N(\varphi_{0i}, \sigma_i^2)$ was considered in ref. [162], but more complex dynamics such as an Ornstein–Uhlenbeck process with self-correlation effects may be considered as well. $\varphi$, $\varphi_{0i}$, and $\sigma_i^2$ are the barrier height for cycle $i$, the average barrier height (which can include a trend as a function of the cycle number, see Section 5.3 in relation to time-series modeling), and its dispersion (which can also include a trend), respectively. While the value of $\varphi$ is irrelevant for LRS ($E \gg \varphi$) because the current in (6) reduces to the low-voltage form of the Landauer's approach $I = G_0 V$, it has a strong impact on HRS as it will be demonstrated next. For $E \ll \varphi$, i.e., the tunneling regime, (6) reads[163]

$$I(V) \approx \frac{2e}{\alpha h} \exp(-\alpha\varphi)\{\exp[\alpha\beta eV] - \exp[-\alpha(1-\beta)eV]\} \quad (7)$$

which can be regarded from the circuital viewpoint as the current flowing through two opposite-biased ideal diodes. This is the fundamental equation for the electron transport in the memdiode model as described in Section 5.2 since for a symmetric structure (7) reduces to the expression

$$I(V) \approx \frac{4e}{\alpha h} \exp(-\alpha\varphi)\sinh\left(\frac{\alpha e}{2}V\right) = I_0(\lambda)\sinh[\alpha(\lambda)V] \quad (8)$$

where $I_0$ and $\alpha$ are assumed to be linear functions of the memory state of the device $\lambda$.[164,165] In addition, notice that for $\alpha \to 0$ (collapse of the confining barrier), (8) reduces again to $I = G_0 V$. Since $\alpha$ is assumed to be normally distributed, the current amplitude factor $I_0$ in (8) is lognormally distributed. This observation is consistent with recent reports on the subject.[162,166]

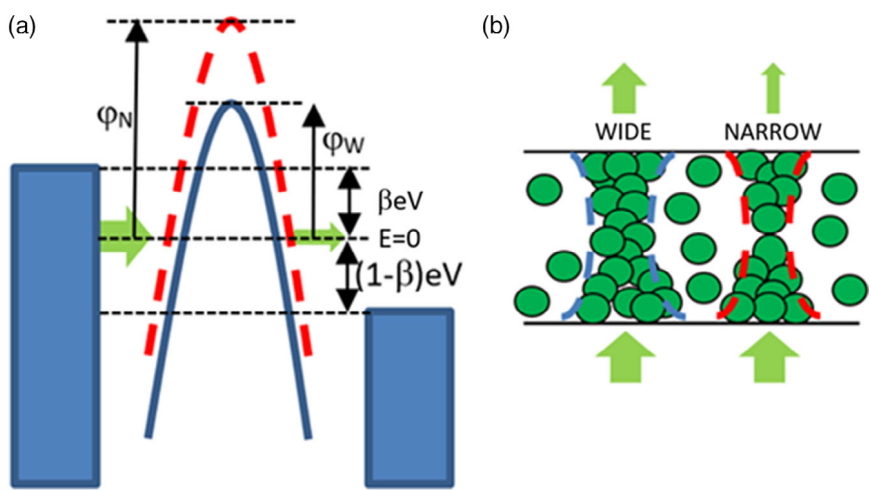

**Figure 17.** a) Schematic of the confinement potential barrier model for HRS. The dashed red solid line corresponds to a narrow ($\varphi_N$) constriction, while the blue solid line corresponds to a wide ($\varphi_W$) constriction. The blue boxes at both sides of the barrier are the cathode and anode injecting electrodes. For LRS, the top of the barrier would be below the anodic quasi-Fermi level. b) Schematic of a narrow and a wide atomic bridge with their corresponding transmitted currents.

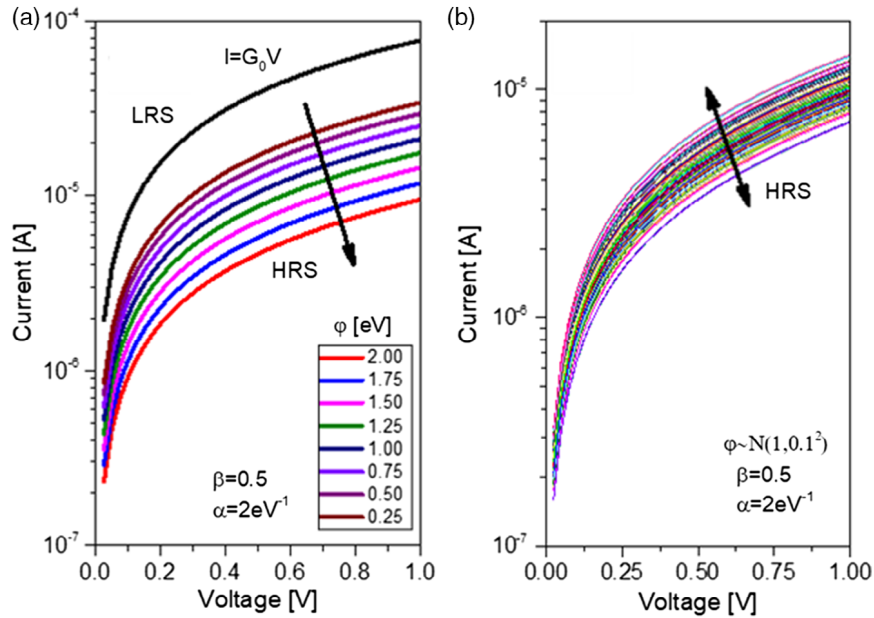

**Figure 18.** a) Model curves for HRS and LRS using expression 6. As the confinement barrier height increases, the current that flows through the device decreases. $I = G_0 V$ corresponds to the complete absence of a tunneling barrier. b) Distribution of the $I$–$V$ curves ($N = 100$) for a normal-distributed barrier height. The dispersion of the curves represents the C2C variability in the HRS regime.

In practice, the HRS barrier fluctuations can be calculated using the expressions

$$\Delta\varphi_V \approx -\frac{1}{\alpha}\Delta\ln(I_V) \quad (9)$$

$$\Delta\varphi_I \approx \beta e\Delta V_I \quad (10)$$

where $\Delta\ln(I_V)$ is the C2C current variation in log scale measured at a fixed bias $V$ (**Figure 19**a) and $\Delta V_I$ is the C2C voltage variation measured at a fixed current $I$ (Figure 19b). The discrepancy between the standard deviations $\sigma_V$ and $\sigma_I$ comes from the difference in using the exact current computation (6) and the approximate voltage computation (8), respectively. Expressions 9 and 10 assume that the barrier profile $\alpha$ (slope of the $I$–$V$ curve in log-linear axis) and the asymmetry of the curves $\beta$ do not change too much with the current variation, which is a reasonable approximation for HRS (based on observations). Importantly, deviations from the normal distribution of the $\Delta\varphi$ histogram could indicate that jumps (RTN, Levy flights, etc.) or strong correlation effects are more relevant for the C2C stochastic process than initially expected. In this regard, a thorough characterization of the Quantum Point Contact model parameters and their correlation for C2C variability can be found in ref. [158].

## 4. Variability Stochastic Models

In this section, we will focus on the role of noise and fluctuations in RS and on how they can be reflected in the stochastic modeling approach. This modeling viewpoint complements the other sections since it inherently includes stochasticity, making use of a statistical mechanics framework. After the classification of different modeling perspectives in Section 4.1, we introduce a stochastic modeling approach to wrap up this subsection.

**ADVANCED
INTELLIGENT
SYSTEMS**

www.advintellsyst.com

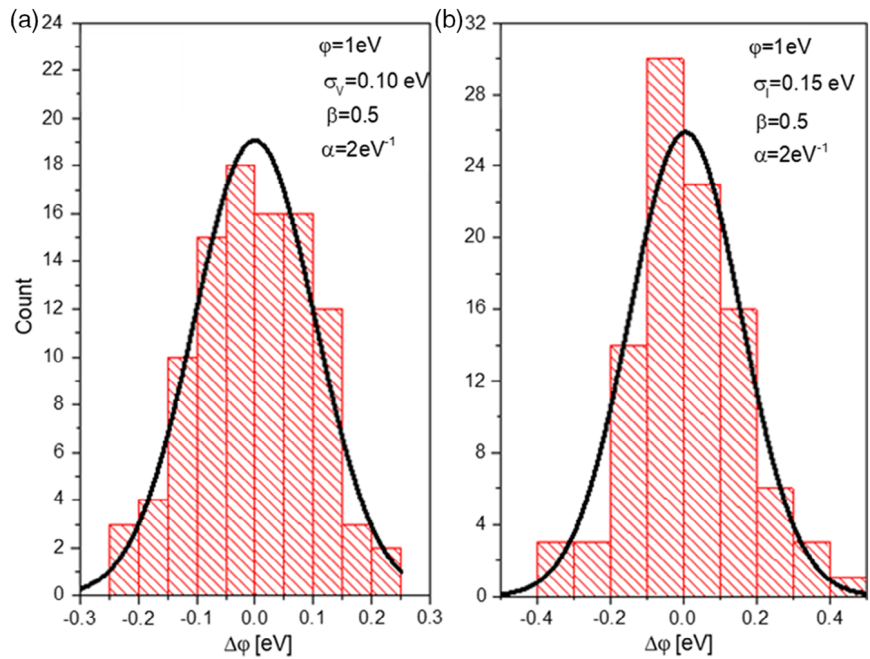

**Figure 19.** a) Evaluation of the barrier height fluctuation using expression 9 for the current variation represented in Figure 18b. b) Evaluation of the barrier height fluctuation using expression 10. The difference between the two histograms arises because the data sets are not the same and because different expressions were used for the computing the barrier height fluctuations.

Then, we provide a compact stochastic model that allows an adequate framework for the simulation of RS regularities and important stochastic phenomena based on the model parameters estimated from experimental RS data (Section 4.2 and 4.3).

### 4.1. Variability and Noise in Memristor Models

The stochasticity of memristive devices observed in many experiments manifests itself as an important intrinsic property of all memristors. An adequate model of a memristive device must necessarily take into account its stochastic properties, such as variability and fluctuations. By D2D variability, we mean random changes in device parameters that occur during the fabrication process, as well as after the electroforming process is completed (if necessary). In addition to the D2D variability, as already highlighted in previous sections, fluctuations are observed during operation, which are C2C random changes in the device resistance and other RS parameters, and fluctuations in the response to a deterministic action.

To take into account stochastic properties and understand how they affect the dynamics of a system (a memristive device in our case), it is important to build an appropriate theoretical model.[167] This model should reflect the memristor general properties and include information about its stochastic properties. Various theoretical models of memristive systems have been proposed in the literature. We can distinguish four main approaches to build such models: dynamical, microstructural, thermodynamical, and stochastic (**Figure 20**).

The dynamical approach is based on simple equations that reflect the general physical properties of the memristor as a dynamical system.[2] This approach includes the models presented in refs. [168–172] and the CF growth model,[173] among others. These models usually include at least two equations: one is an ohmic-type relationship between voltage and current and the other is a first-order differential equation for the state variable. Stochasticity is not covered by dynamical models, although it inevitably manifests itself in the system due to many reasons, including the uncertainty of the model itself. Model uncertainty arises because the dynamical approach catches only the basic properties of the memristive system to build the model while other details are omitted. The chosen basic properties are mainly determined by the selection of internal state variables, which often cannot be directly observed in the experiments (e.g., the gap size $g$ or height of potential barrier $\varphi$ in Section 3). There are models of memristive systems described by various state variables, such as doping ratio,[171] doped region size,[170] CF diameter,[173] oxygen vacancy concentration in the gap region,[174] and tunneling barrier width.[175]

The microstructural approach provides greater accuracy in describing all physical processes occurring at the microscopic level to describe RS.[176–178] While the dynamical approach gives a practical correspondence of the abstract mathematical model to the generalized experimental data of the RS processes, microstructural models aim to obtain an exact correspondence to the physical dynamics of the fabricated devices. In the latter case, the mathematical complexity of the description increases significantly, since such a model must include a large number of different algebraic relations and differential equations. This leads to the fact that microstructural models allow only numerical simulation with high computational resources. At the same time, the simulated values sometimes give only a qualitative agreement with the experimental data.[178] Indeed, to simulate a microstructural model, dozens of additional physical parameters may be required, each of which may contain some errors. As a result, significant uncertainty remains in the description of the system (that represents the memristive device) as a whole. In addition, due to the increased complexity of this approach, the resulting model does not allow efficient application of analytical methods to describe the system and leads to time-consuming computational procedures.

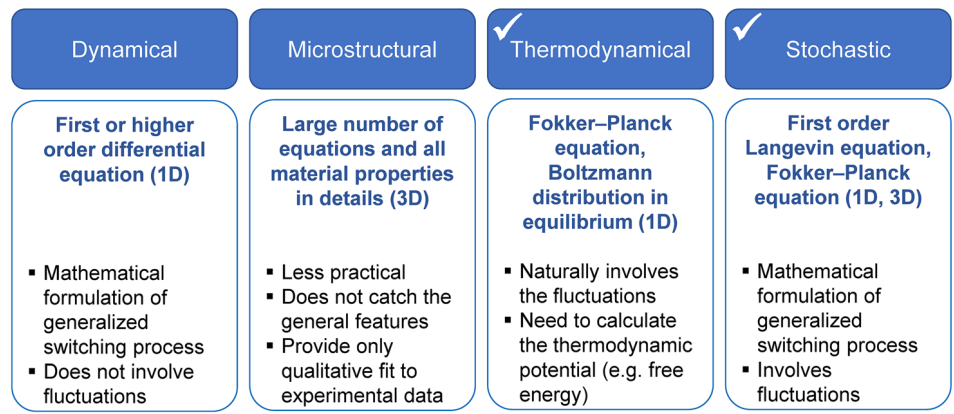

**Figure 20.** Theoretical approaches to modeling memristive devices.

Thermodynamical models take into account fluctuations in a natural way. As in the dynamical approach, in this case the state of a memristive system is described by an internal state variable, a system parameter, a representative or configuration coordinate, and so on. The system tends to a state with a minimum value of the thermodynamical potential, e.g., the free energy $F$ as a function of the internal state variable. According to ref. [179], the free energy of a memristive system can have three local minima separated by energy barriers of different heights. A thermodynamical system can change a locally stable state under the influence of fluctuations or due to a change in external parameters (e.g., under the action of an external electric field). In equilibrium, thermodynamical models lead to a Boltzmann probability distribution for an internal state variable, which may be, for instance, the CF radius or length. The evolution of a nonequilibrium thermodynamical system is usually described by the Fokker–Planck equation (FPE) for the probabilistic distribution of the state variable in the field of a thermodynamical force determined by the free energy profile.[179,180] FPE takes fluctuations into account and describes a nonlinear relaxation from an initial nonequilibrium distribution toward equilibrium with a gradual decrease in thermodynamic flows to zero. If there is a constant thermodynamic flow different from zero, then the system can come to a nonequilibrium steady state.[181–183]

Within the framework of the stochastic approach, random variables are used in the mathematical model. Similar to the dynamical approach case, stochastic models are based on at least two equations: an ohmic-type relation and a first-order Langevin equation (the differential equation with a noise source).[184–186] Langevin equations describe how a system evolves when subjected to a combination of deterministic and random forces. Historically, it was first proposed for the description of Brownian motion.[187] The movements of a Brownian particle in different time intervals are considered as statistically independent processes. The first-order Langevin equation used for the memristor model in ref. [188] is known as the model of over-damped Brownian motion in a field of force, which is often used for description of diffusion in solids. The Langevin equation corresponds to the FPE for the concentration of Brownian particles or for their probability distribution (if the normalization condition is satisfied). The Langevin equation and FPE are equivalent in the sense that they both describe the same dynamics, but in different ways: based on a stochastic equation describing the random trajectory of a Brownian particle or using a dynamical equation describing the evolution of particle concentration.[189] Therefore, the thermodynamical and stochastic approaches have a common base and can be considered as variations of the same mathematical model. On the other hand, a stochastic model can also be built on the basis of a generalization of dynamical equations that describe the key physical properties of memristive systems. Fluctuations are not taken into account within the framework of dynamical approach, but, nevertheless, they inevitably arise in the system for many reasons, including insufficient accuracy of the dynamical equations themselves, which assume some idealization or simplification of real physical microprocesses occurring inside the memristive device and omitting less essential details. The influence of the latter can be taken into account in the form of noise sources correctly included in the Langevin equation.

Further in this section, based on a generalization of theoretical and experimental data, distributed and lumped stochastic models of a voltage-controlled RS memristive system (a VCM device) are considered. It is shown that both models adequately describe bipolar RS and the relationship between them is addressed. Under certain assumptions, stochastic models become quite simple and can be described by Brownian diffusion equations, which do not require significant computational resources for numerical simulation, and in some cases lead to exact analytical solutions. The lumped model incorporates the properties of widely used SPICE (Simulation Program with Integrated Circuit Emphasis) dynamical models and fluctuations. The latter is especially important in the case of memristive devices because of their variability. Below, the compact lumped model proposed is shown to be close to the SM described in Section 3.1, but with a different noise source formulation, which should be consistent with fluctuation–dissipation relations. In addition, the lumped model allows the application of the mathematical apparatus that implements the first passage time of the boundaries by a Markovian process to analyze the RS random time characteristics.[190] On the basis of stochastic approaches to model memristive devices, stochastic resonance phenomena[63] and transient bimodality[190] induced by external noise that occur during RS were theoretically identified and then confirmed and investigated experimentally.

## 4.2. Stochastic Approach to Memristor Modeling

Stochastic properties are inherently linked to memristive devices; consequently, they should be correctly included to the device and circuit simulation models in computer-aided design tools for ICs. The widespread use of these tools is essential for the development and commercialization of emerging memristive devices for the different key applications under consideration in the industry.[1] As noted above, a stochastic model can be built on the basis of stochastic concepts of diffusion processes underlying RS mechanisms[188] or by generalizing dynamical equations that describe the basic physical properties of memristive systems.[2] This subsection discusses the first approach and then shows the assumptions required to make it equivalent to adding a noise source to the equations describing dynamical models.

### 4.2.1. Selection of Adequate Variables to Describe Memristor Stochastic Behavior: Distributed and Lumped Stochastic Models

The main physical quantity describing a memristor state is its resistance $R$. As already stated, a particular set of memristors is linked to resistive memories. For these devices, operation under filamentary conduction, RS is based on the forming and rupture of conductive regions (CFs) in the dielectric layer enclosed between two metal electrodes, changing the resistance of the sample as a whole. As commented above, the creation of conducting regions is due to an increase in the concentration of metal ions or microstructural defects (oxygen vacancies in the case of the VCM mechanism) in the dielectric layer. The diffusion process behind RS dynamics could be based on random transitions of positively charged particles (metal ions or microstructural defects) between stable states in the structure of a dielectric material. Using the electric field created by the potential difference $V$ at the electrode contacts, it is possible to shape the positively charged diffusing particles distribution and, thereby, control memristive device behavior.

The voltage $V$ can serve as a control parameter, then the device is called the voltage-controlled memristor, and the current $I$ flowing through it is a random variable. If the control parameter is the current $I$, the device is called the current-controlled memristor, and the voltage drop $V$ across the device is a random variable. In any case, fluctuations of the random variable will be due to variations in the memristor resistance $R$. Henceforth, we will consider voltage-controlled memristors.

The resistance $R$ is a macroscopic parameter. To establish the relationship between $R$ and the microscopic processes occurring in the dielectric layer, it is convenient to introduce a state variable $\gamma$, which determines the device resistance

$$R = g(\gamma) \tag{11}$$

A memristive system is usually described by at least two more dynamical equations (a first-order equation for the state variable and an equation relating current, voltage, and temperature)

$$\frac{d\gamma}{dt} = f(\gamma, V, T) \tag{12}$$

$$I = I(V, \gamma, T) \tag{13}$$

where $T$ is the temperature of the sample and $t$ is the time. The current voltage dependence $I(V)$ can be linear or nonlinear, symmetric, or asymmetric depending on the materials properties but in any case $I(0) = 0$. The resistance is the ratio

$$R = V/I(V, \gamma, T) \tag{14}$$

and, in general case, it is an ambiguous nonlinear function of $V$ and other parameters. Memristor can heat up under the influence of an electric field and current flowing through by means of Joule heating.[36] Therefore, the temperature is also an important variable describing the state of the memristive system. In general, the differential equation for the temperature can be written as follows

$$\frac{\partial T}{\partial t} = Q(\gamma, V, I, T) \tag{15}$$

The state variable can be a function of coordinates $r$ and time: $\gamma = \gamma(r, t)$. Then, we can speak about a distributed model of a memristive system. For example, this is the case if $\gamma(r, t)$ is the concentration of particles diffusing in the dielectric. If the state variable is a function of time only $\gamma(t)$, then such a model can be called lumped.

The distributed model is more correct, since it can more accurately reflect the main microprocesses occurring inside the dielectric, taking into account its geometry. The lumped model allows more uncertainty, but, due to its simplicity, it is more convenient for numerical modeling, and, therefore, for practical use.[190]

Fluctuations in the memristive device are primarily due to the random nature of the transition of each diffusing particle (metal ion or microstructural defect) from one stable state to another due to activation processes. The stochastic equation for the coordinate vector $r$ of such a particle will include the noise source

$$\mu\dot{r} = F(r, E, \xi(t)) \tag{16}$$

where $E$ is the vector of strength of the external electric field arising in the dielectric layer when voltage is applied to the electrodes $E = E(r, V)$, $\mu$ is the viscosity coefficient, and $\xi(t)$ is the noise source. The right-hand side of Equation (16) stands for the force acting on the particle. In the left-hand side, we neglected the particle mass, i.e., the term with the second time derivative, as is usually done when considering diffusion processes. The force includes a regular part and random components, described by one or more noise sources. For its description, it is convenient to introduce the potential profile $U(r, V)$

$$F(r, E, \xi(t)) = -\nabla U(r, V) + F_N(\xi(t)) \quad (17)$$

where $F_N(\xi(t))$ is the random term. The regular forces acting on the diffusing particle are primarily due to the dielectric material internal structure where the diffusion occurs. Due to the crystal structure of solids, it can be assumed that the corresponding regular potential profile $\Phi(r)$ is a function with a periodic arrangement of local minima—potential wells separated from each other by potential barriers with the height equal to the activation energy $E_a$. Deviations from the ideal periodic structure, which occurs in real materials, can be taken into account by a spatio-temporal noise source $\eta(r,t)$. For definiteness, we can separate the periodic and random components of potential profile as follows

$$U = \Phi(r) + \Phi_\eta(r)\eta(r, t) \quad (18)$$

where $\Phi(r)$ is a periodical function and the random term $\Phi_\eta(r)\eta(r,t)$ with deterministic function $\Phi_\eta(r)$ models the deviations from the ideal periodic structure $\Phi(r)$ in real materials including the D2D variability. It also provides a possibility to consider variability of the energy profile over time, for example, fluctuations of the activation energy $E_a$ in time (e.g., fluctuating barriers as in ref. [191]).

The regular force should also include the effect of external electric field $E(r,V)$ on positively charged diffusing particles. To define the vector $E(r,V)$ at each point of the dielectric layer, it is necessary to rely on the distribution of resistivity inside the sample $\rho(r,t)$. The resistance of memristive device as a whole reads

$$R(t) = \iiint \rho(r, t)\mathrm{d}r \quad (19)$$

The distribution of the strength vector can be defined from the Maxwell's equations (Equation (20))

$$\nabla E = \frac{4\pi\sigma(r,t)}{\varepsilon} \quad (20)$$

where $\sigma(r,t)$ is the density of charged particles (defects and free electrons) inside the memristor, and $\varepsilon$ is the dielectric material constant.

The force caused by the external electric field $E(r,V)$ can be considered as an additive component and represented by potential $\varphi(r,V)$, which is constant under $V = 0$. Then the potential $U$ includes the three terms as follows

$$U(r, V, \eta(r, t)) = \Phi(r) + \Phi_\eta(r)\eta(r, t) - \varphi(r, V) \quad (21)$$

The noise source $\xi(t)$ is involved in the random term $F_N(\xi(t))$ of the force $F$ in (17), where $F_N$ is a deterministic function. The diffusing particles random hopping from one potential well to another occurs due to the action of this component. It has a thermal nature and the intensity of noise source $\xi(t)$ depends on the temperature $T$.

In addition to the fundamentally irremovable thermal noise in semiconductor and other solid-state structures (including memristive structures), there is also low-frequency $1/f$ noise that occurs when current passes, and it is often interpreted as "noise of dirt". Its origin is attributed to the presence of defects and impurities that form local two-level systems (fluctuators) with atoms of the crystal lattice. Each fluctuator provides random micro-switchings of resistance between two metastable states.[192–194] Resistance fluctuations under an applied voltage are transformed into current fluctuations appearing in the form of random telegraph signal. Sometimes, $1/f$ noise is also associated with fluctuations in the number of carriers in the bulk of the sample. It should be noted that there is a number of experiments and theoretical concepts where different authors argue for the natural origin of $1/f$ noise.[195,196] We can take into account $1/f$ noise and thermal fluctuations of charge carriers by adding the noise source $\zeta(t)$ into (13)

$$I = I(V, \gamma, T, \zeta(t)) \quad (22)$$

For the distributed model, it is convenient to choose the defect concentration spatial distribution function as the state function $\gamma(r,t)$. The defect concentration at a given point $r$ in space and at a given time $t$ is the limit of the ratio between the number of defects in a small subdomain $\Omega$ of the sample, enclosing this point, and the volume of the subdomain when the subdomain shrinks to the point $r$. In this case, the resistivity is determined by the concentration $\gamma(r,t)$ in a given area

$$\rho(r, t) = G(\gamma(r, t)) \quad (23)$$

The function $G$ is known to be nonlinear (see e.g.,[197]): the resistivity sharply decreases with increasing concentration $\gamma$ when a certain threshold value is exceeded. Equation (19) and (23) define the function $g(y)$ in (11).

The distributed model with the state function $\gamma(r,t)$ may take into account the material temperature rise in accordance with (15) where in this case $T = T(r,t)$.

The equation with random sources (16), describing the movement of each charged diffusing particle, corresponds to the following equation for concentration of the particles (it is a continuity equation)

$$\frac{\partial \gamma}{\partial t} + \nabla j = 0 \quad (24)$$

where $j$ is the particle concentration flux density. The phenomena of the birth and death of particles in the dielectric layer can also be taken into account, then the corresponding terms must be added to Equation (24). The boundary conditions for (24) are determined by the physical configuration of the memristive devices (dielectric and electrode materials, interfaces, etc.). In particular, the distributed models were considered in refs. [188,198] where the state function was the defect concentration.

The memristor lumped model is based on additional assumptions on the features of charged diffusion particles during the device operation. The electrode material (or the boundary conditions for Equation (24)) can be chosen so that the high (above-threshold) values of the particle concentration will be localized in a certain conductive region. Let one side of this region always adjoin one of the electrodes and let the size of the region change under the influence of the electric field and diffusion processes. This area may have a filamentary structure, including one or more CFs. The size of this region, in the simplest case, is its length (see **Figure 21**) or CF length (the length of the longest filament if there are many) can be chosen as a lumped state variable $\gamma(t)$. Then $\gamma(t)$ can vary from 0 to some maximum value, limited by the size of the device itself. For example, for the memristive device shown in Figure 21, the state variable may vary in the interval $[0, L]$, corresponding to the oxide film thickness.

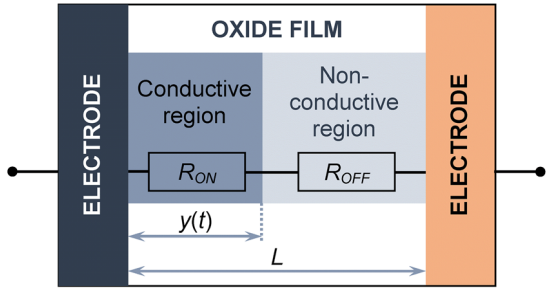

**Figure 21.** Schematic representation of a memristive device with a conductive region with $R_{ON} \ll R_{OFF}$. The conductive region length can be selected as a lumped state variable $\gamma(t)$.

The conductive region resistivity can be considered equally low, $\rho = \rho_{ON}$, and equally high, $\rho = \rho_{OFF}$, outside it. Thus, knowing the size of the conductive region, it is possible to define the memristive device resistance according to (19).

In particular, for the memristive device shown schematically in Figure 21, a linear dependence was assumed in ref. [168]

$$R(\gamma) = R_{ON}\frac{\gamma(t)}{L} + R_{OFF}\left(1 - \frac{\gamma(t)}{L}\right) \tag{25}$$

as well as the validity of Ohm's law

$$I = V/R \tag{26}$$

Later, in a number of papers, it was shown that for lumped models, the dependences (11) and (13) are more complex and, in the general case, must be described by nonlinear functions. For example, in ref. [199] it was assumed that

$$I = \left(\frac{\gamma(t)}{L}\right)^n \lambda \sinh \alpha V + \chi(\exp \gamma V - 1) \tag{27}$$

where $n$, $\lambda$, $\alpha$, and $\chi$ are some fitting parameters. Various abstract nonlinear and nonseparable functions (11) and (13) were considered in ref. [200]. However, in the general case, these functions should be determined by specific mechanisms and parameters of electron transport in the corresponding memristor states[201] (see also Equations (6)–(8) in Section 3.2).

Thus, when constructing a lumped model, the specific dependence (13) (or (22) with the noise source), as well as the specific temperature dependence (15), must be chosen for each particular memristive device in accordance with the properties of the materials from which it is made.

The memristor stochastic lumped model with the thermal noise source was proposed in ref. [190]. For a particular kind of the function $F$ in (17), it was shown that the stochastic equation for the lumped state variable $\gamma(t)$, which describes the conducting region length, can coincide with the equation of motion of an individual charged particle in the distributed model. This is because among all charged particles diffusing in the sample, only those that are near the boundary of the conductive region have the greatest activity (and, therefore, the greatest influence on RS process). Indeed, inside the conducting region, the magnitude of the electric field is much smaller than at the boundary. Therefore, the movement of particles inside the region is much slower than at the boundary or outside it. On the other hand, the electric field outside the conductive region is quite large, but the charged particles concentration far from the conductive region is low. Thus, all important changes that significantly affect the conducting region size occur only at its boundary. The importance of the particle motion near the boundary of the conducting region was also noted in ref. [202], where it was shown that stochastic processes at the tip of the conducting filament play a key role in the RS dynamics of memristive devices. Generalizing this result for an arbitrary $F$, we write the general form of the equation for $\gamma(t)$ as a one-dimensional version of (16)

$$\dot{\gamma} = F(\gamma, E, \xi(t)) \tag{28}$$

Similar to the case of the distributed model (17), the right-hand side of Equation (28) includes regular and random components, and it can be represented by a one-dimensional potential profile $U = U(\gamma, E)$

$$F = -\frac{\partial U(\gamma, V, \eta(\gamma, t))}{\partial \gamma} + F_N(\xi(t)) \tag{29}$$

The potential profile can be described by three terms

$$U(\gamma, V, \eta(\gamma, t)) = \Phi(\gamma) + \Phi_\eta(\gamma)\eta(\gamma, t) - \varphi(\gamma, V) \tag{30}$$

The term $\Phi(\gamma)$ has periodically positioned local minima separated by potential barriers with the height $E_a$, reflecting the ideal internal structure of a dielectric material

$$\Phi(\gamma + \ell) = \Phi(\gamma) \tag{31}$$

where $\ell$ is a period of the crystal structure. The term $\varphi(\gamma, V)$ describes the effect of external electric field under $V \neq 0$. The fluctuating part is represented by noise sources. The deviations of the internal periodic structure from the ideal, as well as variations in the structure itself with time, can be represented by the spatio-temporal noise source $\eta(\gamma, t)$ and a deterministic function $\Phi_\eta(\gamma)$. The noise source $\xi(t)$ is of a thermal nature and reflects the action of a random force leading to the diffusion of particles at the conducting region boundary. Also, the $1/f$ noise source $\zeta(t)$ can be added to (26) or (27).[192]

The random variable $\gamma(t)$ can be described by the probability density function $P(\gamma, t)$, which is proportional to the probability that the random process $\gamma(t)$ is located in the interval $[\gamma, \gamma + \Delta\gamma]$ at time $t$. The probability density function satisfies the normalization condition

$$\int_0^L P(\gamma, t)d\gamma = 1 \tag{32}$$

The stochastic Equation (28) with the noise sources corresponds to the equation for the probability density function $P(\gamma, t)$.

$$\frac{\partial P}{\partial t} = \hat{L}P \tag{33}$$

where $\widehat{L}$ is a kinetic operator. The boundary conditions for Equation (33) are determined by the properties of the electrode materials.

Local heating of the active region under the influence of an electric field $E(V)$ can be taken into account according (15). In particular, the following heat conduction equation can be considered

$$\frac{\partial T}{\partial t} = \alpha E^2 - \kappa(T - T_0) \tag{34}$$

where $T_0$ is the ambient temperature, $\alpha$ and $\kappa$ are fitting parameters.

Thus, a distributed stochastic model can be described by Equation (15–17, 21, 22, 24) of which the Equation (16) describing the random motion of particles diffusing in a dielectric layer is stochastic. The appropriate noise source is added also in (22) for the description of $1/f$ noise and thermal fluctuations of the current. Equation (16) corresponds to the Equation (24) for the concentration $\psi(\mathbf{r},t)$ of diffusing particles, which is the state function of distributed model. A lumped stochastic model can be described by Equation (15, 22, 28–30, 33). The stochastic Langevin Equation (28) corresponds to the Equation (33) for probability density function $P(y,t)$ of the state variable $y(t)$.

### 4.2.2. On the Correct Introduction of a Thermal Noise Source

Let us now consider the problem of the correct introduction of a thermal noise source into the stochastic equations of the memristive device. As it is known, all macroscopic physical systems are subject to thermal fluctuations. One of the most fruitful approaches in modern statistical physics is the one where consideration of a complex thermodynamically nonequilibrium system is reduced to studying the equivalent open subsystem with a few degrees of freedom. The excluded degrees of freedom are considered to be the external environment (thermostat) that has a random effect on the selected subsystem. As a result, by introducing random forces, the behavior of a dynamical subsystem can be described in terms of stochastic differential equations: microscopic or macroscopic (phenomenological). At the same time, the construction of a model stochastic equation of Langevin type for macroscopic variables has to be thermodynamically correct. Wherein, the basic principles of statistical physics should not be violated: the Gibbs' law for an equilibrium statistical ensemble and the reversibility in time of microscopic equations of motion. The consequences of these fundamental first principles are the so-called fluctuation-dissipation relations and theorems[203–208] connecting fluctuation and dissipation characteristics of a subsystem. The validity of these relationships serves as a criterion for the correct construction of a phenomenological model for a subsystem.

Let us demonstrate the above procedure using the example of the stochastic Langevin equation for the coordinate $x(t)$ of a Brownian particle of mass $m$ moving in a one-dimensional potential profile $U(x)$ and interacting with a thermostat of temperature $T$

$$m\ddot{x} + \mu\dot{x} + U'(x) = \xi(t) \tag{35}$$

Here, the random force (heat source) $\xi(t)$ is white Gaussian noise with $<\xi(t)> = 0$ and $<\xi(t)\xi(t + \tau)> = 2D\delta(\tau)$. Noise intensity $2D$ is "consistent" with the linear dissipation parameter $\mu$ by the Sutherland–Einstein fluctuation–dissipation relation

$$D = k_B T \mu \tag{36}$$

where $k_B$ is the Boltzmann constant. The Langevin Equation (35) corresponds to the following equation for the joint probability density function $P(x,v,t)$ of the coordinate and velocity $v$ of a Brownian particle

$$\frac{\partial P}{\partial t} = -v\frac{\partial P}{\partial x} + \frac{U'(x)}{m}\frac{\partial P}{\partial v} + \frac{\mu}{m}\frac{\partial}{\partial v}\left(vP + \frac{kT}{m}\frac{\partial P}{\partial v}\right) \tag{37}$$

It is easy to check that the steady-state solution of Equation (37) is the equilibrium Gibbs distribution

$$P_{st}(x,v) = c_0 \exp\left\{-\frac{H_0(x,v)}{k_B T}\right\} \tag{38}$$

where $H_0(x,v) = U(x) + mv^2/2$ is the Hamiltonian of the considered subsystem and $c_0$ is the normalization constant. Relation (38) confirms the thermodynamic correctness of the stochastic model (35). This approach can be generalized for a three-dimensional case.

The first-order Langevin equation can be obtained from (35) for case of overdamped Brownian motion, when we can neglect by inertia of the Brownian particle and consider the limit $m \to 0$

$$\mu\dot{x} + U'(x) = \xi(t) \tag{39}$$

where the random force $\xi(t)$ is the same as in (35) with noise intensity (36). The equation for the probability density function $P(x,t)$ corresponding to Langevin Equation (39) known also as FPE and reads

$$\frac{\partial P}{\partial t} = \frac{1}{\mu}\frac{\partial}{\partial x}[PU'(x)] + \frac{D}{\mu^2}\frac{\partial^2}{\partial x^2}P \tag{40}$$

The steady-state solution of FPE (40) is the Boltzmann distribution

$$P_{st}(x) = c_0 \exp\left\{-\frac{U(x)}{k_B T}\right\} \tag{41}$$

which confirms the correct formulation of stochastic model of overdamped Brownian motion (39) and (36).

However, everything becomes less obvious when the thermal noise source $\xi(t)$ in equation (35) is replaced by non-Gaussian white noise, or "colored" Gaussian noise. In this case, the Langevin equation itself should be changed. For an additive Gaussian thermal noise source with zero mean and correlation function $K_\xi(\tau) = <\xi(t)\xi(t + \tau)>$, the stochastic model (35) with $U(x) = 0$ becomes the generalized Langevin equation in the form Kubo-Mori[209] for the particle velocity

$$m\dot{v} + \int_0^t \mu(t - \tau)v(\tau)d\tau = \xi(t) \tag{42}$$

where the nonlocal dissipation $\mu(\tau)$ and the correlation function $K_\xi(\tau)$ of thermal noise are connected in accordance with the fluctuation–dissipation relation

$$K_\xi(\tau) = k_B T \mu(\tau) \tag{43}$$

which is a generalization of (36).

By virtue of the central limit theorem, the Gaussian thermostat in many cases can be a good approximation, although it is an idealized physical situation. Even in the simple case of a Brownian particle interacting with solution molecules, random collisions can be described using Poisson rather than Gaussian statistics. Only in the case of frequent and small energy changes during collisions (limit of weak collisions), the central limit theorem guarantees Gaussian statistics. On the other hand, rare and strong collisions require a description in terms of the basic Boltzmann kinetic equation with a collision kernel. The two typical cases indicated refer to the situation with a gas in the Rayleigh limit, i.e., a heavy particle colliding with a thermal reservoir of light particles. A similar situation with nonlinear non-Gaussian thermostat is ubiquitous for scenarios of chemical reactions, i.e., for the theory of rates of monomolecular reactions.

Some first steps toward solving the problem of non-Gaussian thermal fluctuations in nonlinear systems by introducing multiplicative white Gaussian noise into the Langevin equation, along with nonlinear friction, were taken within the framework of the theory of Markovian processes (see, for example.[210]). Another example is the description of electrical circuits with a capacitor and a nonlinear diode using the Langevin equation in the linear-quadratic approximation.[211] Despite all previous works, the extension of the phenomenological Langevin method to stochastic nonlinear dynamical systems containing non-Gaussian thermal noise is still an unsolved problem.

From the general ideology of introducing thermal noise sources, let us turn to electrical circuits and particularize to memristors. Thermal noise occurs in any electrical conductor with active resistance, and it is associated with the chaotic movement of mobile charge carriers. In the case of memristors, the charge carriers are free electrons. If we are dealing with a linear resistance $R$, then, according to the Nyquist formula, a random voltage source in the form of white Gaussian noise with an intensity of $2k_B TR$ has to be included as additive source into the corresponding circuit, that is, into the Langevin equation for voltage.

However, first, the memristor is not only an ordinary resistor but also has a capacitive (reactive) component, which cannot be a source of thermal noise. Second, the memristor has a nonlinear current–voltage characteristic in the form of hysteresis that can be switched by an external voltage into states with low resistance $R_{ON}$ and high resistance $R_{OFF}$. Moreover, even in the simplest Chua's model of an ideal memristor,[212] the memristor resistance (understood in the sense of the ratio of applied voltage to the current flowing through it at a fixed time) is considered as a function of charge, i.e., as a nonlinear functional of the current flowing through it

$$R_m = R(Q) = R\left(\int_0^t I(\tau)d\tau\right) \tag{44}$$

In accordance with the nonlinear fluctuation–dissipation theorems mentioned earlier, the presence of a nonlinear resistor in an electric circuit implies the existence of higher order correlations in equilibrium current fluctuations.

Let us demonstrate the above-mentioned procedure to introduce the thermal noise source on the example of a nonlinear electric circuit containing an ideal memristor with resistance $R$ (44) and a standard capacitor with capacitance $C$ (see **Figure 22**). This is a lumped model where the cumulative charge $Q(t)$ crossing the device is the state variable. This model with an additive Gaussian noise voltage $V(t)$ was analyzed in ref. [213].

Since the memristance $R$ in (44) is a nonlinear function of the charge $Q$, the intensity of a random voltage $V(t)$ should depend on the state variable $Q$: $V(t) = \Psi(Q)\xi(t)$. As a result, using Kirchhoff's laws for the electrical circuit shown in Figure 22 one can write the Langevin equation for the charge $Q$ with a multiplicative thermal noise source in the following form

$$R(Q)\frac{dQ}{dt} + \frac{Q}{C} = \Psi(Q)\xi(t) \tag{45}$$

where $\xi(t)$ is white Gaussian noise with zero mean and unit intensity. FPE for the probability density function of charge $P(Q,t)$ corresponding to (45) reads

$$\frac{\partial P}{\partial t} = \frac{1}{C}\frac{\partial}{\partial Q}\frac{Q}{R(Q)}P + \frac{1}{2}\frac{\partial}{\partial Q}\frac{\Psi(Q)}{R(Q)}\frac{\partial}{\partial Q}\frac{\Psi(Q)}{R(Q)}P \tag{46}$$

Without external forces, the stationary solution of Equation (46) has to be an equilibrium Gibbs distribution

$$P_{st}(Q) = c_0 \exp\left\{-\frac{Q^2}{2k_B TC}\right\} \tag{47}$$

On the other hand, the steady-state solution of the Equation (46) is

$$P_{st}(Q) = B\frac{R(Q)}{\Psi(Q)}\exp\left\{-\frac{2}{C}\int\frac{QR(Q)}{\Psi^2(Q)}dQ\right\} \tag{48}$$

Comparing Equation (47) and (48), we obtain the following first-order differential equation for the unknown function $\Psi^2(Q)$

$$\frac{d\Psi^2(Q)}{dQ} - 2\left[\frac{Q}{k_B TC} + \frac{1}{R(Q)}\frac{dR(Q)}{dQ}\right]\Psi^2(Q) = -\frac{4QR(Q)}{C} \tag{49}$$

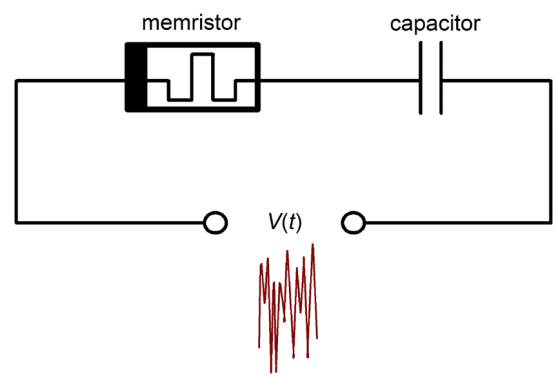

**Figure 22.** The electric circuit with memristor and capacitor.

The finite solution of Equation (49) can be written in quadrature as follows

$$\Psi^2(Q) = \frac{4R^2(Q)}{C} \int_Q^\infty \frac{q}{R(q)} \exp\left\{\frac{Q^2 - q^2}{k_B T C}\right\} dq \tag{50}$$

Equation (50) is exact and determines the intensity of thermal multiplicative noise in the Langevin Equation (45).

If we insert standard resistance $R = const$ in the electric circuit in Figure 22 instead of a memristor, then Equation (50) gives an obvious result

$$\Psi^2 = 2k_B T R \tag{51}$$

In such a case, the white Gaussian noise source in (45) becomes additive, and its intensity satisfies the well-known Nyquist formula.

In the distributed and lumped stochastic models introduced in the Section 4.2.1, the state function $\gamma(r,t)$ or the state variable $\gamma(t)$ are not directly related to the current carriers, as in the model of an ideal memristor (44), but related to concentration of metal ions or defects diffusing in dielectric layer. This concentration or the CF length in turn affects the memristance according to (11), (19) and (23). In particular cases, the dependence of memristance on CF length can be linear (25) or nonlinear (27). Therefore, the random variables in these Langevin equations are not the electric current or the carriers charge like in (45), but the coordinates of a single defect hopping between metastable states in a solid-state structure (16), or the coordinates of the tip of CF formed by a large number of hopping defects (28). The model of overdamped Brownian motion in potential field of force (39) is the most suitable for correct description of diffusing particles in solids. If we follow it, we should rewrite (16) and (17) with a linear additive source of the thermal noise

$$\mu \dot{r} = -\nabla U(r, V, \eta(r)) + \xi(t) \tag{52}$$

where the potential profile $U(r,V,\eta(r))$ is defined according to (21) and contains only spatial noise $\eta(r,t) = \eta(r)$ with zero mean, which describes D2D variability. The thermal noise $\xi(t)$ is the Gaussian vector process with zero mean, statistically independent, and identically distributed components having correlation functions $<\xi(t)\xi(t + \tau)> = 2D\delta(\tau)$.

If we neglect the D2D variability and take $\eta(r) = 0$, then Langevin Equation (52) corresponds to the following FPE for the concentration $\gamma(r,t)$

$$\frac{\partial \gamma(r,t)}{\partial t} = \frac{1}{\mu} \nabla[\gamma(r,t)\nabla U(r, V)] + \frac{D}{\mu^2}\Delta\gamma(r,t) \tag{53}$$

In a similar way, we can introduce the additive linear thermal noise into the lumped model (28), (29), then

$$\dot{\gamma} = -\frac{\partial U(\gamma,V,\eta(\gamma))}{\partial \gamma} + \xi(t) \tag{54}$$

and potential profile $U(\gamma,V,\eta(\gamma))$ is defined according to (30) where the spatial noise $\eta(\gamma)$ with zero mean describes D2D variability. The thermal noise $\xi(t)$ is the white Gaussian noise with $<\xi(t)> = 0$ and $<\xi(t)\xi(t + \tau)> = 2D\delta(\tau)$, where $D$ is defined according (36).

Langevin Equation (54) with $\eta(\gamma) = 0$ corresponds to the FPE for probability density function $P(\gamma,t)$

$$\frac{\partial P(\gamma,t)}{\partial t} = \frac{\partial}{\partial \gamma}\left[\frac{\partial U(\gamma,V)}{\partial \gamma}P(\gamma,t)\right] + D\frac{\partial^2 P(\gamma,t)}{\partial \gamma^2} \tag{55}$$

with the following boundary conditions

$$j(0,t) = j(L,t) = 0 \tag{56}$$

where $j(\gamma,t)$ is the probability flow

$$j(\gamma,t) = -\frac{\partial U(\gamma,V)}{\partial \gamma}P(\gamma,t) - D\frac{\partial P(\gamma,t)}{\partial \gamma} \tag{57}$$

### 4.2.3. A Memristor Compact Stochastic Model

Besides the elaboration of general universal approaches to the construction of stochastic models and the correct choice of noise sources, the development of compact models for describing real memristive systems is essential as well. Compact models are used in EDA tools for IC design. Until now, for this purpose, many dynamical models have been used, although they do not reflect accurately the inherent fluctuation processes in memristor operation. In Section 3.1, the Stanford-based simulation model with a noise source was considered. In this subsection, we consider the construction of a simple and compact stochastic model based on the lumped model introduced above with the consistent thermal noise source. The compact model is linked to the SM, but the noise source should be different to satisfy the fluctuation–dissipation relations considered in Section 4.2.2.

Let us consider the lumped stochastic model where the state variable $\gamma$ is the conduction region length in the direction from one electrode to another (see Figure 21). If the conduction channel consists of one or more CFs, then the length of the longest CF is $\gamma$. We also assume that the electroforming process and switching from the HRS to the LRS occur at a voltage $V > 0$. Accordingly, the switching from LRS to HRS takes place for negative voltages (assuming bipolar operation). In the stochastic equation for the lumped state variable (54), we neglect the D2D variability assuming $\eta(\gamma) = 0$. Then, the potential profile (30) has only two terms

$$U(\gamma, V) = \Phi(\gamma) - \varphi(\gamma, V) \tag{58}$$

where $\Phi(\gamma)$ is periodical function (31). The probability density function $P(\gamma,t)$ obeys the FPE (55).

The term $\varphi$ describes the effect of the external electric field $E(V)$ under $V \neq 0$. For the compact model, we assume that the field near CF boundary, $\gamma$, increases with the voltage linearly, $E(V) = BV$ and appropriately

$$\varphi(\gamma, V) = BV\gamma \tag{59}$$

in which parameter $B$ depends on the distance between electrodes $L$, charge of diffusing particle $q$ and the dielectric constant $\varepsilon$

$$B = \frac{q\ell}{\varepsilon L} \quad (60)$$

It is difficult to measure or calculate the values $q$ and $\varepsilon$ for real materials. Therefore, $B$ can be considered as a fitting parameter, which is estimated from experiments (for more details see Section 4.3). Notice that there are different electrical properties on the two sides of the boundary $\gamma$: to the left there is a conduction region with low resistivity $\rho_{ON}$ and to the right there is a region with a high resistivity $\rho_{OFF}$. Taking this fact into account, we should expect different values of $B$ for the set and reset processes

$$\varphi(\gamma, V) = \begin{cases} B_{SET}V\gamma, & V > 0 \\ B_{RES}V\gamma, & V < 0 \end{cases} \quad (61)$$

where $B_{SET} > B_{RES}$. The values of parameters $B_{SET}$ and $B_{RES}$ for various memristive systems, obtained on the basis of experimental data, are given in Section 4.3.

Equations (54), (55), (58), (61) along with equation for the current as a function of $V$ and $\gamma$ (for example, (27)) describe a lumped stochastic model that can be implemented numerically. One can add the Equation (34) to describe Joule heating.

We can determine the view of the probability density function $P(\gamma, t)$ and the main properties of the random process $\gamma(t)$ from a general analysis of these equations. Since the potential profile (58) contains a periodical term (31), the distribution $P(\gamma, t)$ will have many peaks corresponding to the local minima of $U(\gamma, V)$. Under the condition

$$E_a - B|V| \gg k_B T \quad (62)$$

(which is typical for solids) the probability for a diffusing particle to be near the barrier top is very low. In this case, the transitions through potential barriers will be relatively rare random events with probability described by Poisson's law. The mean passage time over such a barrier is known as Kramer's time

$$\tau = \tau_0 e^{\frac{E_a - B|V|}{k_B T}} \quad (63)$$

The lumped model described above becomes even simpler, if we transfer from the serrated distribution $P(\gamma, t)$ consisting of many minima and maxima corresponding to potential barriers and wells to a coarse grained model with distribution $P_1(\gamma, t)$, averaged over these small-scale inhomogeneities. According to ref. [190], in this case, the equation for the state variable $\gamma(t)$ reads

$$\dot{\gamma} = -\frac{\partial U_{eff}(\gamma, V)}{\partial \gamma} + \xi(t) \quad (64)$$

where $\xi(t)$ is white Gaussian noise with $<\xi(t)> = 0$ and $<\xi(t)\xi(t+\tau)> = 2D_{eff}\delta(\tau)$. The noise intensity is defined by the effective diffusion coefficient $D_{eff}$. $U_{eff}(\gamma, V)$ is the effective potential which is a nonlinear function of $V$ and linear one of $\gamma$

$$U_{eff}(\gamma, V) = v_{eff}(V)\gamma \quad (65)$$

where $v_{eff}$ is effective drift coefficient. The effective coefficients of coarse-grained model are nonlinear functions of the driving voltage and the temperature

$$v_{eff}(V) = \frac{2\ell}{\tau_{kr}} \sinh \frac{BV}{k_B T} \quad (66)$$

$$D_{eff}(V) = \frac{2\ell}{\tau_{kr}} \cosh \frac{BV}{k_B T} \quad (67)$$

The value $\tau_{kr}$ is Kramer's time (63) at $V = 0$

$$\tau_{kr} = \tau_0 e^{\frac{E_a}{k_B T}} \quad (68)$$

Expressions of the compact model described by Equation (64)–(68) are valid under condition (62).

According to (61), the coefficient $B$ is asymmetric and can be different for positive and negative values of the driving voltage. Therefore, we should use $B = B_{SET}$ for the voltage values at which the set process occurs (for the structure shown in Figure 21 at $V > 0$) and $B = B_{RES}$ for the voltage values at which the reset process takes place (for the structure shown in Figure 21 at $V < 0$).

The compact model (64)–(68) is relatively close to the widely known dynamical models used for modeling in SPICE applications with various window functions;[55,170,175,214] but, unlike the latter, it takes into account the influence of C2C fluctuations modeled by the noise source $\xi(t)$. A more detailed comparison of dynamical and stochastic models is given in ref. [190]. The compact model is also close to SM described in Section 3.1, where the state variable $g$ is the size of the gap between CF tip and electrode: $g = L - \gamma$. However, the noise intensity in (2) is different from $D_{eff}$ (67) and the consistency with fluctuation–dissipation relations is not achieved in SM. The gap variation (1) in the SM is selected only as a fitting function without significant physical explanation.

Stochastic Equation (64) corresponds to the FPE for coarse grained probability density $P_1(\gamma, t)$

$$\frac{\partial P_1(\gamma, t)}{\partial t} = \frac{\partial}{\partial \gamma}\left[P_1(\gamma, t)\frac{\partial U_{eff}(\gamma, V)}{\partial \gamma}\right] + D_{eff}(V)\frac{\partial^2}{\partial \gamma^2}P_1(\gamma, t) \quad (69)$$

with the following boundary conditions

$$j_1(0, t) = j_1(L, t) = 0 \quad (70)$$

$$j_1(\gamma, t) = -P_1(\gamma, t)\frac{\partial U_{eff}(\gamma, V)}{\partial \gamma} - D_{eff}(V)\frac{\partial P_1(\gamma, t)}{\partial \gamma} \quad (71)$$

Stochastic Equation (64) models a single realization $\gamma(t)$ of random RS process in a memristor. The FPE (69) describes evolution of probability density $P_1(\gamma, t)$ obtained after averaging over and an infinite ensemble of memristors: $P_1(\gamma, t) = <\delta(\gamma - \gamma(t))>$.

In ref. [190], an exact analytical solution of Equation (69) was obtained, which can be represented as an infinite series of functions decreasing exponentially in time. Under $t \to \infty$ and fixed value $V = V_0$, the probability distribution tends to the stationary Boltzmann distribution $P_1(\gamma, t) \to P_1^{st}(\gamma)$

$$P_1^{st}(\gamma) = Ae^{-\frac{U_{eff}(\gamma, V_0)}{D_{eff}}} \quad (72)$$

After forming the memristor device works with the CF tip moving near the boundary $L$: $\gamma_{HRS} < \gamma < L$. The gap $g = L - \gamma_{HRS}$ is supposed to be a small value because the CF is not destroyed fully under operation of memristor. In this case, when we apply $V_0 > 0$, the probability $P_1(\gamma, t)$ tends to the stationary LRS distribution (72) with the maximum at $\gamma = L$, corresponding to the potential minimum $U_{eff}(\gamma, V_0)$. When we apply $V_0 < 0$, the CF decreases and $\gamma(t)$ tends to the opposite boundary $\gamma = 0$, but we stop the transition process somewhere in an intermediate stage before any stationary distribution is formed. At this time, the distribution $P_1(\gamma, t)$ is not stationary, it is rather flat and it does not have its maximum value near the HRS state. That is why we usually observe a greater variability in the HRS current rather than in the LRS, as it is shown in Figure 2b and 3b.

The switching time from one state to another (e.g., from HRS to LRS) can be estimated as the relaxation time to the stationary distribution, or as the mean first passage time (MFPT) $\Theta(\gamma_0 \rightarrow \gamma_1)$ of the boundary $\gamma = \gamma_1$ by the random process $\gamma(t)$ if initially it was in point $\gamma = \gamma_0$. According to ref. [190], for $\gamma_0 < \gamma_1$ the exact expression for MFPT is as follows

$$\Theta(\gamma_0 \rightarrow \gamma_1) = \frac{\gamma_1 - \gamma_0}{v_{eff}} - \frac{D_{eff}}{v_{eff}^2}\left(e^{-\frac{v_{eff}}{D_{eff}}\gamma_0} - e^{-\frac{v_{eff}}{D_{eff}}\gamma_1}\right) \quad (73)$$

As an example, to describe switching from HRS to LRS, we can take $\gamma_0 = L/2$ and $\gamma_1 = L$. Then according to (73)

$$\Theta\left(\frac{L}{2} \rightarrow L\right) = \frac{L}{2v_{eff}} - \frac{D_{eff}}{v_{eff}^2}\left(e^{\frac{v_{eff}L}{D_{eff}2}} - e^{-\frac{v_{eff}}{D_{eff}}L}\right) \quad (74)$$

At a low intensity of the thermal noise compared to the height of potential barriers (62) and at sufficiently high switching voltages $V_0 \gg k_B T/B$, expression (74) can be simplified

$$\Theta\left(\frac{L}{2} \rightarrow L\right) = \frac{L}{2\ell}\tau_0 e^{\frac{E_a - B|V_0|}{k_B T}} \quad (75)$$

The approximate expression (75) shows that under the above conditions, the memristor switching time is defined mainly by the Kramer's time (63), which is the mean transition time of a Brownian particle over potential barrier with height $E' = E_a - B|V_0|$. The coefficient $L/2\ell$ corresponds to the quantity of such barriers at the interval from $L/2$ to $L$.

Table 3 summarizes the main features of distributed, lumped, and lumped compact stochastic models. The distributed model contains less idealization, and it is based on Langevin equation describing a random motion of microstructural defects or metal ions in dielectric layer. On the other hand, this model is the most complex from the mathematical point of view especially in 3D case.

The lumped models assume a simple structure of the conductive region, when there are always one or more CFs in contact to one of the electrodes. The CF lengths change randomly under the influence of thermal noise and the external voltage. The lumped models are less complex and more useful from the circuit simulation viewpoint, but sometimes they cannot reflect some real

**Table 3.** Summary of the stochastic models.

| Features | Type of stochastic model | | |
|---|---|---|---|
| | Distributed | Lumped | Lumped compact |
| State variable $\gamma$ | Time and space dependent $\gamma = \gamma(\mathbf{r}, t)$. For example, $\gamma(\mathbf{r}, t)$ can be the concentration of diffusing defects. | Time dependent $\gamma = \gamma(t)$. For example, $\gamma(t)$ can be the CF length. | |
| First order Langevin equation | Equation (52) describing random motion of microstructural defects or metal ions in the dielectric layer. | Equation (54) describing random variation of CF length. | Equation (64) describing random variation of CF length. |
| Potential profile | Tilted, quasi-periodical. It can be 1D, 2D, or 3D. An example of 3D is (21). | Tilted, quasi-periodical, 1D. An example is (30). | Tilted 1D effective potential. An example of linear one is (65). |
| Conductive region structure | An arbitrary structure of conductive and nonconductive regions. | Two regions: CF and nonconductive region. | |
| $I$–$V$ relation | Depends on spatial distribution of state variable $I = I(V, \gamma(\mathbf{r}))$. | Depends on the value of state variable $I = I(V, \gamma)$. Examples are (25) and (27). See also comments after (27). | |
| Thermal noise | It can be 1D, 2D, or 3D vector value. Equation (52) with an additive white Gaussian vector noise source is an example of thermal noise source consistent with fluctuation–dissipation relations. | An additive 1D white Gaussian noise source in (54) is an example of thermal noise source consistent with fluctuation–dissipation relations. | An additive white Gaussian thermal noise in (64) is consistent with fluctuation-dissipation relations. |
| 1/f noise | Possible to add 1/f noise source into $I$–$V$ relation (22). | | |
| Spatial noise for D2D variability modeling | Spatial noise $\eta = \eta(\mathbf{r})$ in the expression for potential (21). | Spatial noise $\eta = \eta(\gamma)$ in expression for potential (30). | Does not reflect D2D variability. |
| Temperature variations | Equation (15). | Equation (15) or, in particular, (34). | |
| Mathematical complexity of the model | It is high for 3D and 2D cases, where only numerical simulation is possible. An analytic solution is possible for 1D case under some additional assumptions.[188] | Medium complexity. | Least complexity. Analytic solutions are possible and exact expression for MFPT (73) and for variance of FPT can be obtained.[190] |

phenomena. For example, when there are two CFs in the dielectric layer connected to opposite electrodes.[188]

The simplest lumped model is the compact one. It involves additional assumptions on ideal periodical microstructure of dielectric material (61) and low intensity of the thermal noise comparing to the activation energy (62). In fact, various lumped compact models without noise sources are widely used as SPICE dynamic models in EDA design tools.[190] Among the variability models considered in this manuscript, from the perspective given in Figure 20, in line with Table 3, the Stanford model in Section 3.1 and the memdiode model in Section 5.2 could be classified as lumped compact stochastic models. The quantum transport variability model introduced in Section 3.2 could also be regarded as a lumped compact model, as well as the model depicted in Section 5.2.

### 4.3. Estimation of Stochastic Model Parameters from Experimental Resistive Switching Data

In the compact model described by Equation (64)–(68), there is a parameter set that is determined by the device dimensions and material properties. The parameters are the distance between the electrodes $L$, the period of the dielectric material structure $\ell$, the height of potential barriers $E_a$ or activation energy of diffusing particles (metal ions, or oxygen ions and vacancies in the case of VCMs), prefactor $\tau_0$ of Kramer's time (63) or period of thermal vibrations of dielectric lattice atoms, the coefficients $B_{SET}$ and $B_{RES}$ in (61) for positive and negative voltages $V$, as well as the active layer temperature $T$ during the switching process.

Parameter $L$ is known for each device, the parameters $\ell$, $E_a$, and $\tau_0$ can be found for each dielectric material. According to many estimates, the temperature $T$ in the active zone inside the sample could be much higher than the ambient temperature, but accurate measurements are not available. Finally, the parameters $B_{SET}$ and $B_{RES}$ cause the greatest difficulty in the determination process.

According to (75), at a constant temperature, the dependence of the logarithm of the average switching time $\Theta$ on voltage $V = V_0$ is a straight line

$$\lg \Theta = a - \frac{BV_0}{k_B T} \lg e \tag{76}$$

where $a = \lg(L/2\ell) + \lg \tau_0 + (E_a/k_B T) \lg e$ is a constant independent of $V_0$ if the temperature is independent of $V_0$. In the general case, it can be assumed that the sample temperature does not remain constant and is a function of the control voltage, $T = T(V_0)$. For example, the temperature may rise as $V_0$ increases. In this case, the $\lg(\Theta(V_0))$ plot deviates from a straight line.

Let us first consider the case when the temperature does not depend on the voltage and the $\lg(\Theta(V_0))$ plot is a straight line, as shown in **Figure 23**. Such dependence was observed experimentally in refs. [178,184,215]. The description of the experiment for measuring $\Theta$ is given below.

Notice that for the use of the compact model proposed above, there is no need to determine the absolute values of parameters $\ell$, $L$, $E_a$, $B_{SET}$, $B_{RES}$, and $T$, but only the ratios $\ell/L$, $\beta = E_a/k_B T$, $B_{SET}/k_B T$, and $B_{RES}/k_B T$. The ratios $B_{SET}/k_B T$ and $B_{RES}/k_B T$ can be determined from the slope of the experimental straight line $\lg(\Theta(V_0))$.

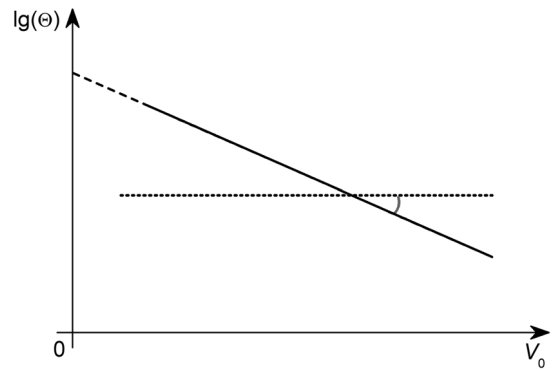

**Figure 23.** Linear part of the dependence of the logarithm of the average switching time $\lg \Theta$ of a memristive device (in the transition from LRS to HRS) on the switching voltage $V_0$ at a constant temperature ($k_B T = $ const).

If we write the equation of the straight line shown in Figure 23 in the following form

$$\lg \Theta = a - bV_0 \tag{77}$$

the parameter $b$ can be used to estimate the ratio of the parameters $B_{SET}/k_B T$ and $B_{RES}/k_B T$ of the compact model. Comparing (76) and (77), we get

$$\frac{B}{k_B T} = \frac{b}{\lg e} \tag{78}$$

Let us turn to the experimental data. **Table 4** provides specific $B_{SET}/k_B T$ and $B_{RES}/k_B T$ values determined from published data. The dependencies of the switching times on voltage drop for various materials and structures are obtained. It can be seen that these parameters depend on specific materials of the device structure and the type of diffusing particles.

Similar measurements were also carried out in the framework of this work for the structure Au(20 nm)/Ta(40 nm)/ZrO$_2$(Y)(20 nm)/Pt(20 nm) with an active area of $20 \times 20 \, \mu m^2$. The fabrication details of these devices are described in ref. [216]. It was shown that this technology exhibits stable bipolar RS associated with oxygen ion/vacancy migration and formation of regions of different species concentration.

The choice of ZrO$_2$(Y) as the functional dielectric material in these devices was due to the following. For RS in VCMs with transition metal oxides, a sufficient concentration of oxygen vacancies is required. Usually, the required vacancy concentration is achieved by the deposition of nonstoichiometric oxides. This is obtained by oxide film annealing in vacuum after the layer deposition, or as a result of electrochemical oxidation/reduction reactions at the interfaces of a functional oxide with reactive metal electrodes.

In ZrO$_2$(Y), oxygen vacancies are the elements of the crystal structure, and the vacancy concentration in equilibrium is determined by the Y concentration (the number of oxygen vacancies is 1/2 of the number of Y atoms). Thus, by varying the Y fraction, one can control the equilibrium concentration of oxygen vacancies in ZrO$_2$(Y). It should be emphasized that the oxygen vacancy concentration in ZrO$_2$(Y) is determined solely by the Y concentration and almost does not depend on temperature, environment, etc. This, along with the high mobility of anions, makes ZrO$_2$(Y) a promising material for memristive applications.[63,217–220]

**Table 4.** Parameters $B_{SET}/k_BT$ and $B_{RES}/k_BT$ determined from published experimental data for various materials and structures.

| The structure of the memristive device | Diffusing particles | $B_{SET}/k_BT$ [$V^{-1}$] | $B_{RES}/k_BT$ [$V^{-1}$] | References |
|---|---|---|---|---|
| Pt/TiO$_2$–TiO$_{2-x}$/Pt | Oxygen vacancies | 5.1 | 3.3 | [184] |
| Ag/a–Si/p$^+$–Si/Pt | Silver ions | – | 6.14 | [215] |
| Pt/Ta/Ta$_2$O$_5$/Pt | Oxygen vacancies | – | 8.9 | [178] |
| Pt/W/Ta$_2$O$_5$/Pt | Oxygen vacancies | 18.6 | – | [252] |
| Pt/Ta/Ta$_2$O$_5$/Pt | Oxygen vacancies | 28.8 | – | [252] |
| Au/Ta/ZrO$_2$(Y)/Pt | Oxygen vacancies | – | 30.5 | This work |

The electrical contacts to the device pads were provided using the Everbeing EB-6 probe station. The sign of the voltage across the device corresponds to the potential of the bottom electrode relative to the potential of the top electrode. The studies were carried out using the USB-6341 data acquisition system (National Instruments). All the measurements were carried out on the same memristive device.

In this work, we analyzed the change in the device resistive state from the initial LRS to the HRS under the influence of positive pulses of different amplitudes. For this purpose, an input signal was applied to the device, which was an alternating sequence of positive and negative pulses with a duration of 0.1 s. The amplitude of the positive pulses $V_0$ varied in the 1.0–1.2 V range. To ensure the invariance of the initial resistive state, the amplitude of the negative pulses was constant ($-2$ V). The input signal was applied to the device with a sampling frequency of 2 MHz. The device response was recorded at a sampling rate of 2 MHz. The experiment was repeated 114 times for each positive pulse amplitude.

The change in the resistive state occurred at different rates depending on the value of the positive pulses amplitude (see **Figure 24**). We used the threshold level of the current to calculate the first passage time $\theta$ for each current waveform and then averaged the obtained values within one value of $V_0$.

**Figure 25** shows the dependence of the MFPT $\theta$ on the constant voltage $V_0$ (at 300 K).

The dependence of $\lg(\theta(V_0))$ in the range $V_0 = 1.0$–1.2 V can be approximated by a straight line; therefore, we can conclude that the temperature $T$ in this range does not depend on the voltage $V_0$. According to these data, the ratio $B_{RES}/k_BT = 30.5 \pm 0.6$ V$^{-1}$. This value is in qualitative agreement with microscopic parameters according to (60).

Let us now consider the case when the temperature goes up with an increasing voltage. According to the results of the study of memristive devices using microstructural models (see, for example,[178]) thermal equilibrium is established in memristors in a fast manner, compared to the switching time. Therefore, to determine the temperature in the active region of the device, one can use the stationary solution of the equation for temperature (15) or (34). According to (34), the stationary temperature value as a function of voltage $V_0$ has the form

$$T = T_0 + AV_0^2 \tag{79}$$

where $A$ is some coefficient. Substituting (79) into (76) we get

$$\lg \Theta = a_1 + \frac{E_a}{k_B(T_0 + AV_0^2)} \lg e - \frac{BV_0}{k_B(T_0 + AV_0^2)} \lg e \tag{80}$$

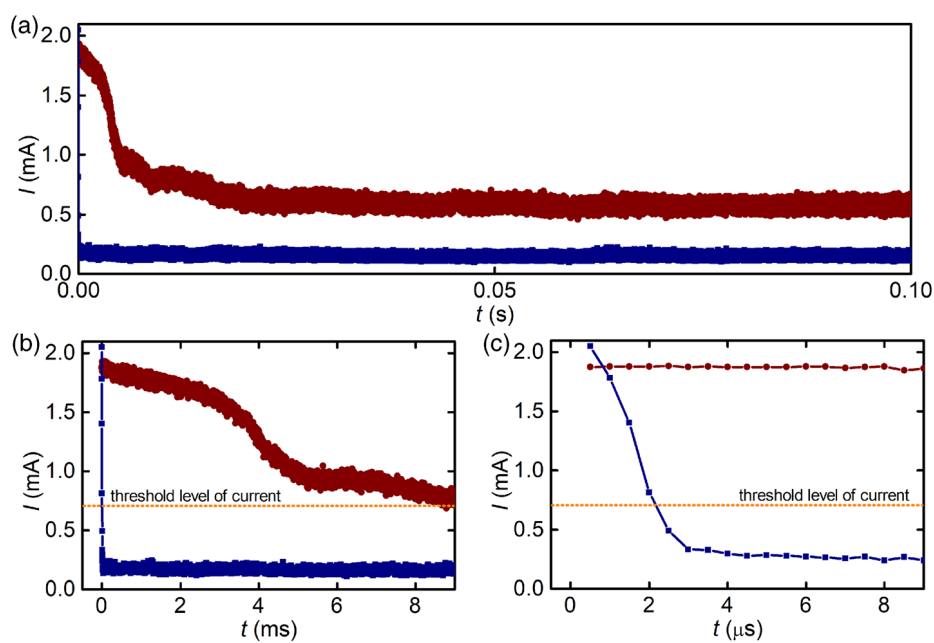

**Figure 24.** Examples of an $I(t)$ waveform obtained under the influence of a constant bias voltage of +1.0 V (red circles) and +1.2 V (blue squares) for different time scales: a) 0–0.1 s, b) 0–9 ms, and c) 0–9 μs.

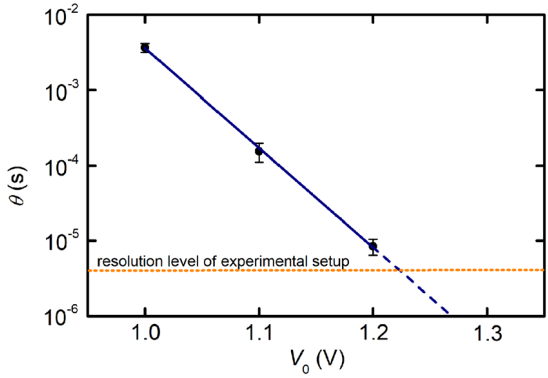

**Figure 25.** Dependence of the mean first passage time $\theta$ on the value of the constant voltage $V_0$ (at 300 K).

where $a_1 = \lg(L/2\ell) + \lg \tau_0$. The dependence $\lg(\Theta(V_0))$ is shown in **Figure 26**.

From the results presented in Figure 26, it follows that, given the increase in temperature, the switching time decreases as $V_0$ rises in faster way than at a constant temperature. According to the proposed model, temperature determines the noise intensity —a random force due to which diffusion particles can overcome potential barriers. As the noise intensity increases, such transitions become more frequent and the average switching time decreases. From the results presented in Figure 26, it follows that with $A = 400$, the switching time can decrease by 1 orders of magnitude. For $A < 100$, the RS acceleration is much smaller, and the $\lg(\Theta(V_0))$ plot is difficult to distinguish from a straight line. The measurement results for various samples, taken into account in Table 4, do not have significant deviations from a straight line. Therefore, considering the error of these measurements, we can conclude that for these memristive devices, the effect of temperature increase with increasing voltage $V_0$ is insignificant.

The developed stochastic approach to model memristive systems was successfully tested on devices based on $ZrO_2(Y)$. Experimentally determined parameters made it possible to qualitatively, or even quantitatively, describe such classical phenomena as stochastic resonance,[34] resonant activation,[221] as well as the phenomenon of transient bimodality, which is characteristic for stochastic systems.[190] The manifestation of these and other phenomena in memristive systems indicate that the memristor is a complex stochastic system, and it allows the successful application of an arsenal of modern statistical methods not only to describe and predict the rich response of memristic devices but also to control their behavior using noise.

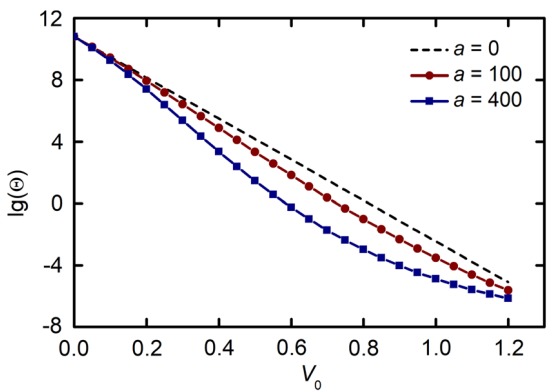

**Figure 26.** The dependence $\lg(\Theta(V_0))$ at temperature changing according to the law described in Equation (80), and $A = 100$ (red circles) and $A = 400$ (blue squares). The dashed line denotes the Arrhenius dependence (Equation (76)) at a constant temperature $T = T_0$. Parameters $E_a/k_B T_0 = 40.3$ and $B/k_B T_0 = 30.5 \, V^{-1}$.

## 5. Variability Behavioral Models

### 5.1. Variability under the General Memristor Model Formalism

In this section, we present an approach tackling with the variability of the parameters employed in a memristor model, the relation between these parameters and the consequences on the model implementation. Specifically, we consider a memristor representation described in the charge and flux domain, and we describe how we have implemented a Monte Carlo framework on the basis of this model to consider C2C variability. The results are obtained using a $HfO_2$-based memristor device (in fact a RRAM device), forced to commute between high and low resistance states for several tens of cycles.

#### 5.1.1. Memristor Model Description

In this section, we refer to a simple model describing the operation of a memristor device, based on the charge and flux approach, as defined in ref. [222]. According to this framework, flux and charge can be obtained from the time integrals of voltage and current as follows

$$\phi(t) = \int_{-\infty}^{t} V(\tau) d\tau \tag{81}$$

and

$$Q(t) = \int_{-\infty}^{t} I(\tau) d\tau \tag{82}$$

In the case studied here, a train of identical pulses to program the device, i.e., to define its state of operation. Under this circumstance, from Equation (81), the flux for a train of identical pulses can be expressed as

$$\phi = m \cdot \Delta V \cdot \Delta t \tag{83}$$

where $m$ is the pulse number, $\Delta V$ is the voltage amplitude over a zero volts reference, and $\Delta t$ IS the temporal width of an individual pulse. It is apparent that we have considered an ideal flux-controlled memristor, defined by a simple relation between charge and flux[223–226]

$$Q = Q_0 \left( \frac{\phi}{\phi_0} \right)^{1+n} \tag{84}$$

In the above equation, $Q_0$, $\phi_0$, and $n$ are fitting parameters. $n$ is defined such that when $n = 0$, a linear relation between charge and flux exists. By rewriting Equation (84) in a more compact way, it transforms into the following expression

$$Q = G_0 \cdot \phi^{1+n} \tag{85}$$

See that in this model (corresponding to Section 5.1) $G_0$ is a model parameter and should not be confused with the quantum conductance that has a fixed value. Then, the current can be simply obtained by deriving the charge in the model Equation (85)

$$I = Q \cdot \frac{V}{\phi} \cdot (n+1) = G_0 \cdot (1+n) \cdot \phi^n \cdot V \tag{86}$$

and as a result, the (mem)conductance $G$ could be then calculated as

$$G = \frac{dQ}{d\phi} = (n+1) \cdot \frac{Q}{\phi} = (n+1) \cdot G_0 \cdot \phi^n \tag{87}$$

assuming that $n$ is a constant. This can be used for our convenience as

$$\frac{dn}{dt} = 0 \tag{88}$$

### 5.1.2. Statistical Analysis

We have applied the model described above (see also[227]) to fit two sets of experimental cycles; the first with $\Delta V = +0.9/-0.7$ V and $\Delta t = 10\,\mu s$, the second with $\Delta V = +0.75/-0.6$ V and $\Delta t = 100$ ns; where each data set consists of ten cycles, each comprising 500 identical pulses of consecutive potentiation and depression processes. Both the experimental and modeled conductance versus the number of applied pulses are shown in **Figure 27**. It should be noted that in this device both potentiation and depression operations are considered to be in the LRS.[228]

The model, as described by Equation (85), needs in principle only two parameters (i.e., $G_0$ and $n$) to describe the device behavior during the depression or the potentiation phases. Since we are considering the device operation in the context of a potential neuromorphic application, we focus on the relation between the two parameters in consecutive potentiation and depression events. The Monte Carlo model implementation (see the details below) provides a reasonable guess on the parameter values from one cycle to the next. In brief, we need to determine if there exists any relation between $G_0$ and $n$ in the potentiation (or depression), as well as in the values corresponding to the next depression (or potentiation) steps.

In the first step, the values extracted from the model fitting to the experimental measurements are plotted as illustrated in Figure 27a. This fitting provides us with a list of values for each potentiation/depression step. We first analyze the possible correlation of each parameter separately. To do so, in Figure 27b, we plot $n$ in one potentiation (depression) cycle versus the following depression (potentiation) cycle, for all the series of cycles appearing in Figure 27a. Then, we can model the value of the parameters in the next cycle $(t+1)$ as a function of the values of the parameters in cycle $t$ plus a random error

$$n(t+1) = \alpha_n n(t) + \varepsilon_n(n(t)) \tag{89}$$

$$G_0(t+1) = \alpha_g G_0(t) + \varepsilon_g(G_0(t)) \tag{90}$$

where $\alpha_n$ and $\alpha_g$ are constants, and the error $\varepsilon_x$ ($x$ could be $n$ or $g$, depending on the parameter selected) can be written as a Gaussian error as follows

$$\varepsilon_x(p) = P_x e^{-p^2/\sigma_x^2} \tag{91}$$

where in the second step, it is verified whether a relationship between $n$ and $G_0$ exits or not. To do so, $n$ versus $G_0$ is represented and the correlation calculated. The connection between them can be written as follows

$$G_0 = G_1 e^{n/n_0} + \varepsilon_{p,d}(n) \tag{92}$$

In the above equation, two indexes $\{p,d\}$ where used to indicate that the error could be different in the transition from potentiation to depression or the other way around. Although, for small

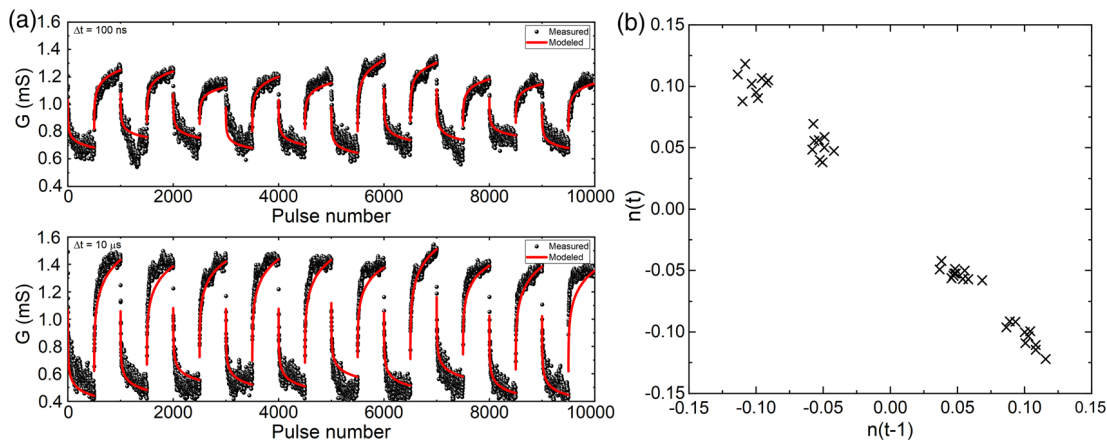

**Figure 27.** a) Evolution of the experimental (symbols) and modeled (line) conductance versus the number of applied pulses. This plot shows the conductance for two different pulse widths. b) The value of $n$ at timestep $t$ versus the value of $n$ in the previous step t-1. Notice the clear linear relation between the potentiation ($n > 0$) and the next depression cycle ($n < 0$).

values of the parameter n, the last relation could be also proposed in a linear manner as follows

$$G_0 = G_2 + G_1 n + \varepsilon_{p,d}(n) \tag{93}$$

Thus, the Monte Carlo model has two parts: first, an estimation of the next value of $n$ is calculated from the previous value using Equation (89), and second, the conductance is obtained from Equation (93).

### 5.1.3. Device Description and Measurement Procedure

The analyzed nanodevice is a four-layer stack 40 nm(TiN)/ 5 nm(HfO$_2$)/10 nm(Ti)/40 nm(TiN), with an area of 40 × 40 μm². Deposition of the Ti and TiN layers was performed using magnetron sputtering, while the deposition of the active memory layer, made of HfO$_2$, was performed using atomic layer deposition.[229] These devices demonstrate filamentary-type memristive behavior described by.[228] The switching is initiated by means of a forming process that is not shown here.

Starting from the initial state defined by a DC voltage, within the resistance distribution of either LRS or HRS, a pulsed characterization was carried out with a sequence of 300 identical pulses with an 80 ms period and a reading operation at 100 mV after each pulse. Potentiation curves were obtained, starting from the HRS, while depression curves were started from the LRS. In order to evaluate the effect of the programming conditions, i.e., $\Delta t$ and $\Delta V$, after each sequence the cell was re-initialized by DC operations and a new sequence was performed with slightly modified programming parameters. The voltage variation, $\Delta V$, of the applied pulses was span between 0.3 (−0.3 V) and +1 V (−1 V) with a 50 mV step for depression (potentiation) operations. Having in mind to avoid any irreversible damage to the memory cell, the maximum applied voltage during this step was limited to the maximum value applied during previous DC characterization. The $\Delta t$ was varied between 100 ns and 300 μs, with two series per decade for both types of operation, with a fixed pulse rise and fall time of 40 ns. It is worth to emphasize that no external current limitation was applied during the pulsed characterization.

An important issue for characterizing memristor devices is the potentiation dynamics (conductance increase) and depression dynamics (conductance decrease). These could be characterized and assessed by pulsed measurements. Of course, those measurements should include a broad spectrum of parameters values, varying with the pulse height ($\Delta V$) and its duty cycle, i.e., the pulses time duration ($\Delta t$), while the memory-cell's initial state should be kept constant. Speaking in terms of neuromorphic-oriented applications, connectivity strength is a function of the incoming pulses; something that is subject to proper programming. As a result, identification of the operating regions of increasing and decreasing connectivity strength could be identified.

In this study, we applied a series of identical pulses to the memristor device, while monitoring its conductance. In addition, a read operation at the end of each pulse was performed. Additional information on how the original experiment was performed can be found in ref. [228]. An example of the conductance sequences resulting from 500 identical pulses, repeated with increasing

$\Delta V$ and a fixed duration of $\Delta t = 10$ μs and 100 ns, is given in Figure 27a for both potentiation and depression processes.

### 5.1.4. Obtained Results and Discussion

It is worth calling the reader's attention to the fact that in the studied devices, the transition process from HRS to LRS is faster than that from LRS to HRS. Accordingly, the reinforcement process of the CF creation is faster than the dissolution process on average. This result reflects on the potentiation and depression operations though both processes occur in the LRS. This would lead only to a change in the model parameters, but the model equations would still be able to reproduce measurements in a satisfactory way.

As far as the Monte Carlo approach is concerned, as discussed above, two different steps are needed: first, we need to relate $n$ variation to $G_0$ variation; second, we also need a model for the variation of any of those variables. As a first step, data from Figure 27 were used to extract the n and $G_0$ parameters for each cycle shown in **Figure 28**. Then, the values of $G_1$ and $G_2$ were obtained by fitting Equation (93) to these n and $G_0$ values. In this way, the Monte Carlo model reduces to a single parameter variation, thus simplifying the implementation. This order reduction has already been observed in other devices and seems to be caused by the underlying physical origin of the values used by this behavioral model.[230]

From the above process, we get the following calculated values $G_1 = 14.31$ S and $G_2 = −7.03$ S. The error $\varepsilon$ can be modeled as a Gaussian distribution in Equation (91), with a standard deviation $\sigma = 152 \times 10^{-6}$ S and mean $\mu = −22.4 \times 10^{-6}$ S. In a second step, $n$ is obtained. In this case, we can see in **Figure 27b** and **29** that there is a clear correlation between $n$ in the potentiation part of the cycle and $n$ in the depression region. No such relation is observed in the opposite case, or even in between the values of $n$ in the same region (potentiation or depression) for consecutive cycles, as seen in Figure 29. Thus, $n$ must be modeled in two different ways: one for the transition between potentiation and depression (a linear relation works well), and another one for the opposite transition.

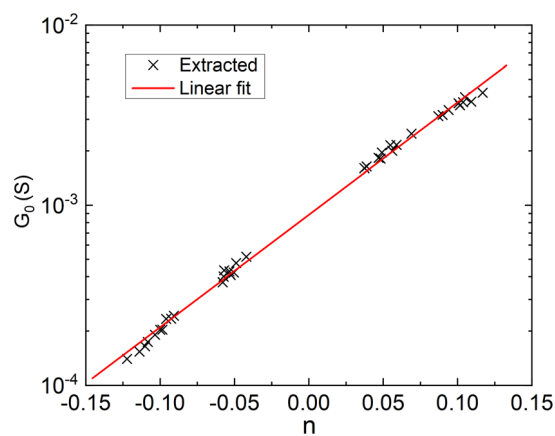

**Figure 28.** Experimental values of $n$ and $G_0$ for each potentiation and depression set of cycles. Notice that $n > 0$ corresponds to potentiation, while $n < 0$ corresponds to the depression part.

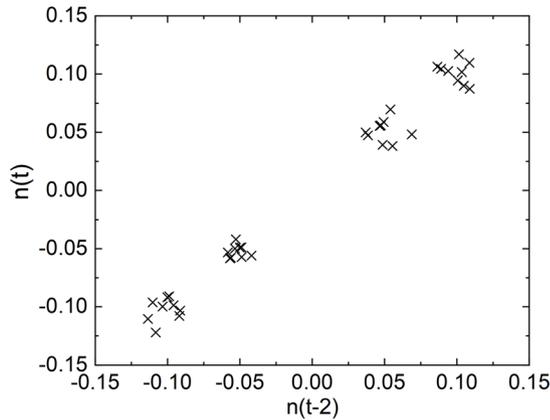

**Figure 29.** Parameter $n$ at transition $t$ versus $n$ at a transition of the same type ($t$-2).

In the first case, i.e., the potentiation to depression transition, the relationship appears to be simple

$$n(t) = -n(t-1) \tag{94}$$

In the case for the depression to potentiation transition, this can be modeled by a Gaussian distribution, with standard deviation $\sigma = 0.07$ and mean $\mu = 0.10$. Notice that the difference is very clear, since in this occasion the relations $n(t-1) < 0$ and $n(t) > 0$ hold true, while the opposite relations hold true for the potentiation to depression transition. A modeling scheme based on time series that complements the analysis unfolded in this section is given in Section 5.3.

## 5.2. Application of the Dynamic Memdiode Model to C2C Variability

As highlighted in the introduction and previous sections, variability in resistive memories in general and in bipolar-type filamentary RS devices in particular are hot research topics. Variability was shown to be the result of morphological changes in the CF at the atomic scale.[5,231] Since this kind of variability is inherent to RRAMs operation, any compact model intended to reproduce as realistically as possible their behavior in the circuit simulation landscape must contemplate this phenomenon. In particular, this section focuses on the inclusion of noncorrelated C2C variability in the Dynamic Memdiode Model (DMM) for RRAM devices.[164,165] The DMM describes the conduction characteristics of RS devices under the application of arbitrary input signals. The origin of the switching behavior is related to the creation and destruction of the CF that spans across the dielectric layer. This in turn is linked to the metal ions or oxygen vacancies displacement depending on the device type (CBRAM or OxRAM). In bipolar devices, the transition between HRS and LRS (CF formation) and vice versa (CF rupture) takes place at opposite voltages.[232,233] This is the case we are going to analyze next.

The DMM reproduces the hysteretic behavior of the memristive device by means of two nonlinear coupled equations: one for the electron transport (Equation (95)) and one for the memory state (Equation (96)). According to the DMM, the $I$–$V$ characteristic reads (see expression (8))

$$I(V) = I_0(\lambda) \sinh \{\alpha(\lambda)[V - (R_c(\lambda) + R_i)I]\} \tag{95}$$

where $V$ is the applied voltage, $I_0$ is the current amplitude factor, $R_c$ is a variable series resistance, and $\alpha$ is a variable model parameter. $R_i$ refers to a fixed resistance related to the snapback voltage correction ($V_{sb} = V - R_i \cdot I$).[234] The $R_i$ value corresponds to a vertical current increase at the transition voltage $V_T$. Notice that Equation (95) includes three different parameters with $\lambda$-dependence. For simplicity, these parameters are assumed to change linearly and in the same way from a minimum (off) to a maximum (on) value as a function of $\lambda$. For example, $I_0(\lambda) = I_{off} + (I_{on} - I_{off}) \cdot \lambda$, where $I_{off}$ and $I_{on}$ represent the minimum (HRS) and maximum (LRS) current amplitude values, respectively. $\lambda$ runs from 0 to 1 and is called the memory state of the device (a state variable in the general memristor model). Its purpose is to control the transitions HRS $\leftrightarrow$ LRS. According to Equation (95), HRS depends exponentially on $V$, while LRS depends linearly on $V$ (because of the potential drop across the series resistances $R_i$ and $R_c$). $\lambda$ is described by a balance-type voltage-driven differential equation[164]

$$\frac{d\lambda}{dt} = \frac{1 - \lambda}{\tau_S(\lambda, V_{sb})} - \frac{\lambda}{\tau_R(\lambda, V_{sb})} \tag{96}$$

where $\tau_{S,R}$ [s], called TS and TR, respectively, in the model script in **Table 5**, are the characteristic times for the set and reset transitions. They are expressed as

$$\tau_S(V) = \exp[-\eta_S(V_{sb} - V_S)] \tag{97}$$

$$\tau_R(V) = \exp[\eta_R \lambda^\gamma (V_{sb} - V_R)] \tag{98}$$

where $\eta_{S,R}$ [V$^{-1}$], also called etas and etar, respectively, in the model script shown in Table 5, are the transition rates; $V_{S,R}$ [V] are the set and reset switching voltages, respectively. The snapback current ($i_{sb}$ in the model script) acts as a threshold current for the snapback effect and the snapforward parameter $\gamma \geq 0$ (called gam in the model script) controls the reset transition rate. H0 is the initial memory state. RPP in Table 5 is a parallel resistance required by the output current generator. Further details about the Dynamic Memdiode Model can be found in refs. [164,165].

The model script is presented in Table 5. The script contains the model parameters as well as the transport and memory equations. $+$ and $-$ are the conventional device terminals, while H is the output terminal for the memory state (not used in this work). For illustrative purposes, **Figure 30** shows a simulated curve with the DMM for a sinusoidal applied signal with amplitude 1.5 V and frequency 1 Hz. While Figure 30a shows the $I$–$V$ characteristic in logarithmic scale, Figure 30b shows the same $I$–$V$ curve (in black) in linear axis and the resulting curve after the snapback correction (in red). Notice the vertical current increase in the red curve occurring at $V_T$.

Before introducing variability in the script, it is essential to understand the model behavior and the role each parameter plays within it. In ref. [235], a one-way sensitivity analysis was performed in a simpler version of the memdiode model. In the

**Table 5.** LTSpice DMM script including the transport and memory equations. Variability is included in several key parameters. The Gaussian function is added in red for normal distributions and in blue for lognormal distributions. The LTSpice schematic is included for simulating 100 cycles under the voltage and timing considerations.

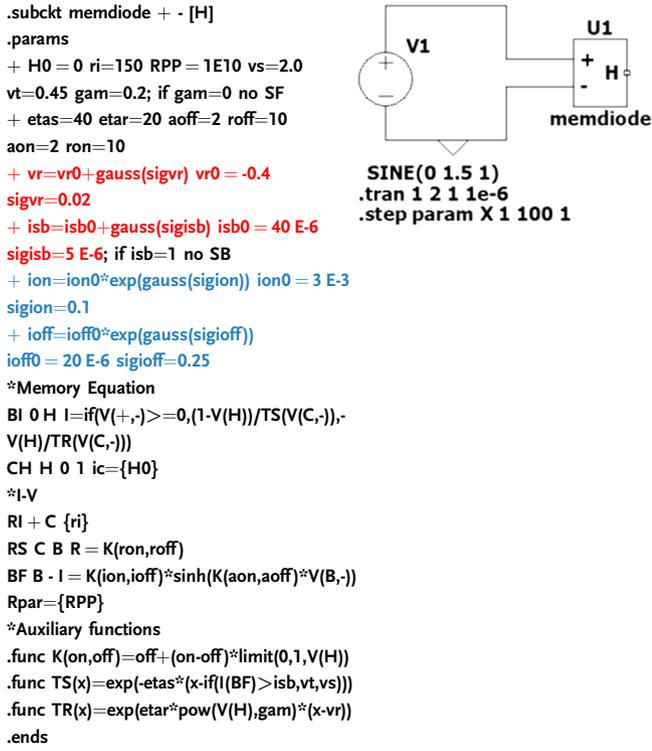

```
.subckt memdiode + - [H]
.params
+ H0 = 0 ri=150 RPP = 1E10 vs=2.0
vt=0.45 gam=0.2; if gam=0 no SF
+ etas=40 etar=20 aoff=2 roff=10
aon=2 ron=10
+ vr=vr0+gauss(sigvr) vr0 = -0.4
sigvr=0.02
+ isb=isb0+gauss(sigisb) isb0 = 40 E-6
sigisb=5 E-6; if isb=1 no SB
+ ion=ion0*exp(gauss(sigion)) ion0 = 3 E-3
sigion=0.1
+ ioff=ioff0*exp(gauss(sigioff))
ioff0 = 20 E-6 sigioff=0.25
*Memory Equation
BI 0 H I=if(V(+,-)>=0,(1-V(H))/TS(V(C,-)),-
V(H)/TR(V(C,-)))
CH H 0 1 ic={H0}
*I-V
RI + C {ri}
RS C B R = K(ron,roff)
BF B - I = K(ion,ioff)*sinh(K(aon,aoff)*V(B,-))
Rpar={RPP}
*Auxiliary functions
.func K(on,off)=off+(on-off)*limit(0,1,V(H))
.func TS(x)=exp(-etas*(x-if(I(BF)>isb,vt,vs)))
.func TR(x)=exp(etar*pow(V(H),gam)*(x-vr))
.ends
```

Schematic labels: V1, U1, + H +, memdiode, SINE(0 1.5 1), .tran 1 2 1 1e-6, .step param X 1 100 1

referred paper, the impact of the model parameters in the simulated $I$–$V$ characteristics was evaluated for four observables (HRS and LRS current magnitudes, SET and RESET voltage transitions) and the results reported in a summary table. Similarly, in this subsection, the effect of considering variability in the key model parameters is studied one at a time using the $\sigma$ (sig) values incorporated into the model script. The inclusion of C2C variability in the DMM is performed by adding a Gaussian random number generator to the model parameters

(gauss function in LTSpice). This function returns a zero centered normal distribution function with a standard deviation $\sigma$. This is indicated in Table 5: red for a normal distribution and blue for a lognormal distribution. This selection is consistent with the experimental study carried out in ref. [235]. **Figure 31** shows the results obtained from this analysis varying a) $I_{off}$, b) $I_{on}$, c) $I_{sb}$, and d) $v_r$ one at a time. The reported results correspond to 100 cycles.

Notice that $I_{sb}$ and $v_r$ changes only affect the transition regions as expected. Nevertheless, $I_{off}$ not only impacts the HRS current but also the set transition. Similarly, $I_{on}$ not only impacts the LRS current but also the reset transition. As it can be seen in Figure 31 as well as in our previous study,[235] a single random parameter can impact more than one region of the simulated curves. This is a consequence of the hysteresis effect and is referred to as variability propagation in the simulation process. The results reported in **Figure 32** indicate that an excess of model parameters with variability (concurrent variability) can introduce over-randomness in the simulated curves which in the context of circuit simulation can be hard to keep under control. This is further discussed next.

Once the impact of including variability in a single model parameter at a time has been analyzed, the second step consists in investigating how a combination of random model parameters affects the DMM output results. Here, for the sake of simplicity, variability was included in some of the key model parameters without considering cross-correlation effects among them. In ref. [235], the main features of C2C variability in a HfO2-based RRAM device were reproduced using an iterative approach for the selection of the model parameters but the process was applied to a simpler version of the memdiode model. Notice that by including variability in 4 parameters ($I_{off}$, $I_{on}$, $I_{sb}$, and $v_r$) simulations exhibiting C2C variability in all the curve regions can be achieved. As illustrated in the inset of Table 5, the procedure consists in generating a 100-cycle set of 1 s simulations, each one obtained after the application of the same input signal (sinusoidal voltage with frequency 1 Hz and amplitude 1.5 V). Importantly, each loop in the simulation is computed independently from the others ruling out correlated C2C. The results are obtained using the model script and the parameter values reported in Table 5. Figure 32 shows the simulated $I$–$V$ curves, a) in logscale and, b) in linear scale. The magnitude of the fluctuations can be changed using the $\sigma$ value associated with each model parameter. Variability in the rest of parameters: $r_i$, $eta_s$, $eta_r$, etc. is also possible but caution should be exercised since this can introduce over-randomness as discussed earlier.

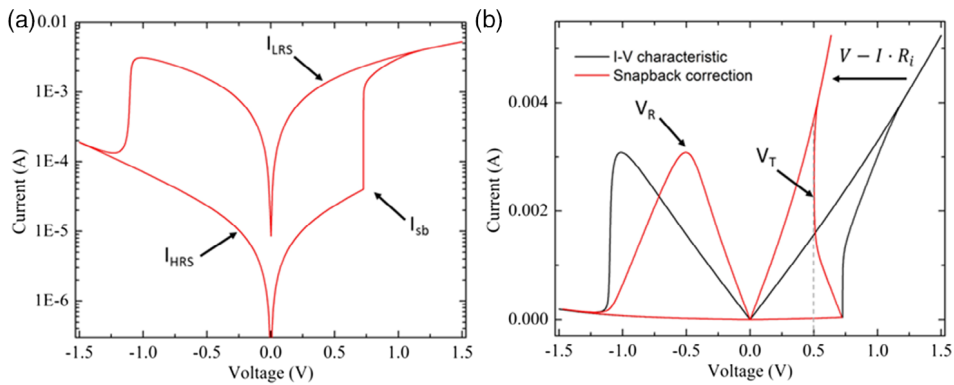

**Figure 30.** Simulated $I$–$V$ curve obtained with the script and parameter configuration in Table 5, without including variability. a) in logscale and b) in linear scale (black) presenting the snapback correction (red).

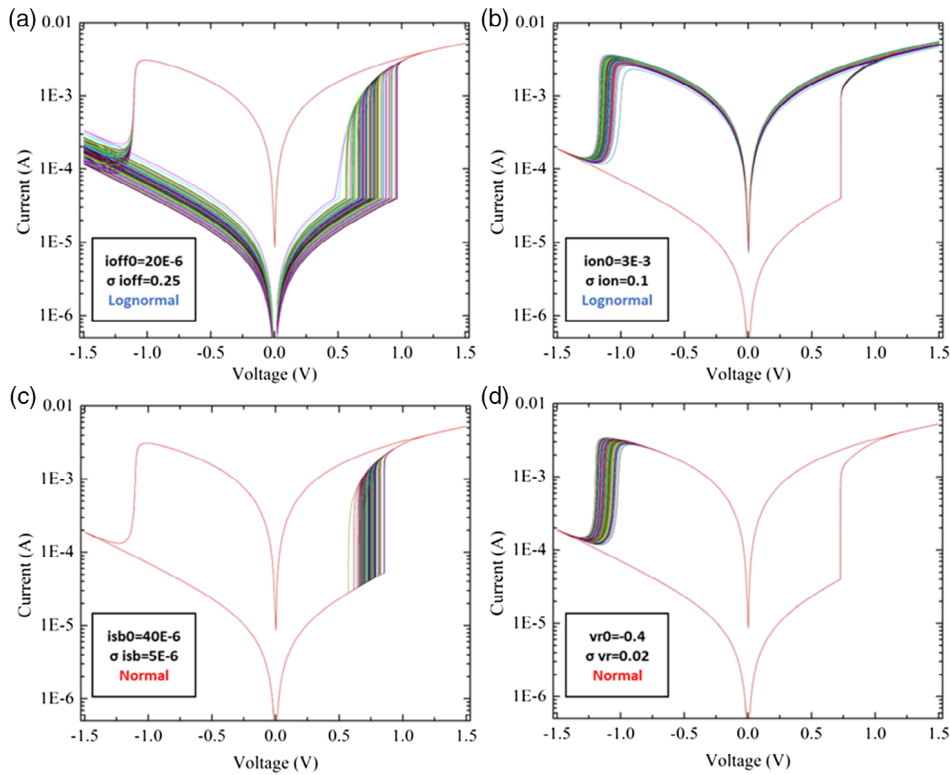

**Figure 31.** Impact in the *I–V* characteristics of including variability in model parameters one at a time: a) $I_{off}$, b) $I_{on}$, c) $I_{sb}$, and d) $\nu_r$.

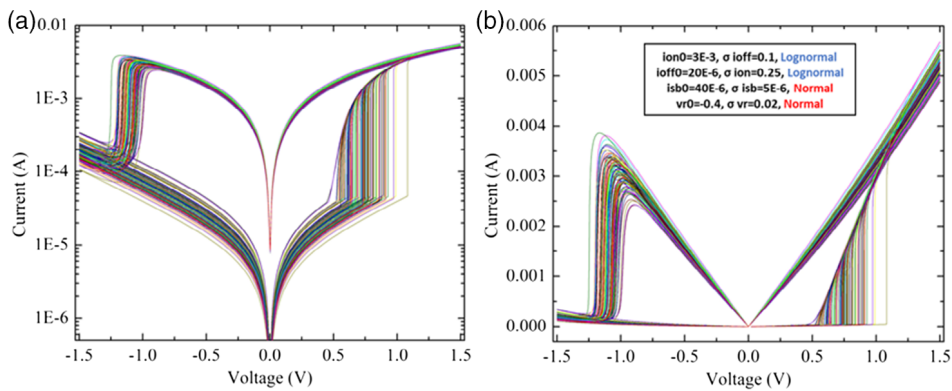

**Figure 32.** *I–V* curves from 100 cycles simulations including variability in the 4 specified parameters as presented in Table 5. a) in logscale and b) in linear scale.

After presenting the model response to concurrent variability, additional concerns about the variability propagation problem can arise. For instance, variability propagation can introduce an unexpected behavior in the HRS current: since for the example discussed here simulations start in the first quadrant and finish in the third quadrant, the initial and final spread of the resulting *I–V* curves (HRS) can be different if the experiment is not properly designed. However, this is a possible outcome that can occur under specific experimental circumstances. First, we have to take into account the impact of the applied voltage span, i.e., if this span is not large enough to produce a complete set or reset of the device, then variability in the simulated HRS curves will be different for positive and negative voltages. This aspect of the problem was studied in ref. [235], and it was referred to as voltage-induced variability.

Second, the simulation cycle number can also have influence on the symmetry of the *I–V* loop if cycles are not independent. A difference between the first and the subsequent cycles can be interpreted as a propagation variability issue. In this case, the first HRS cycle is strongly connected with the initial memory state parameter H0 which is sometimes unknown or simply put to zero. However, in the second cycle, if the cycles are not independent, the initial HRS current value is forced by the magnitude of the previous reset event. In **Figure 33**, a density plot comparing the initial and final HRS current values (extracted at $\pm 0.2$ V, respectively) are compared for the first and second cycles and the maximum applied voltage: a) 1st cycle and $V_{max} = 0.9$ V, b) 1st cycle and $V_{max} = 1.5$ V, c) 2nd cycle and $V_{max} = 0.9$ V, and d) 2nd cycle and $V_{max} = 1.5$ V. Figure 33b,d shows that the initial and final HRS current distributions coincide because the voltage span is high enough to induce the complete reset of the device.

However, Figure 33a,c shows a different behavior. In a), the densities are shifted, meaning that the initial and final HRS currents do not match as experimentally expected. This difference is voltage induced since 0.9 V is not high enough for the device to reach the full reset condition and therefore the final HRS current is ruled by this incomplete reset process. c) Corresponds to the second cycle but since the initial HRS current is ultimately determined by the previous incomplete reset process, both the initial and final HRS current distributions (measured at opposite voltages) match. If the reset process is not complete, the $I$–$V$ loop stabilizes only after one or more cycles.

In summary, this subsection reported a simple approach for including uncorrelated C2C variability in the DMM for RRAM devices. The model is able to generate random $I$–$V$ curves with the desired variability. A more realistic scheme not only should include self-correlation effects in the model parameters but also cross-correlation effects. This imperatively requires a detailed time-series analysis such as that reported in ref. [144]. In addition, the effects of variability propagation and the role played by the voltage sweep span were also discussed.

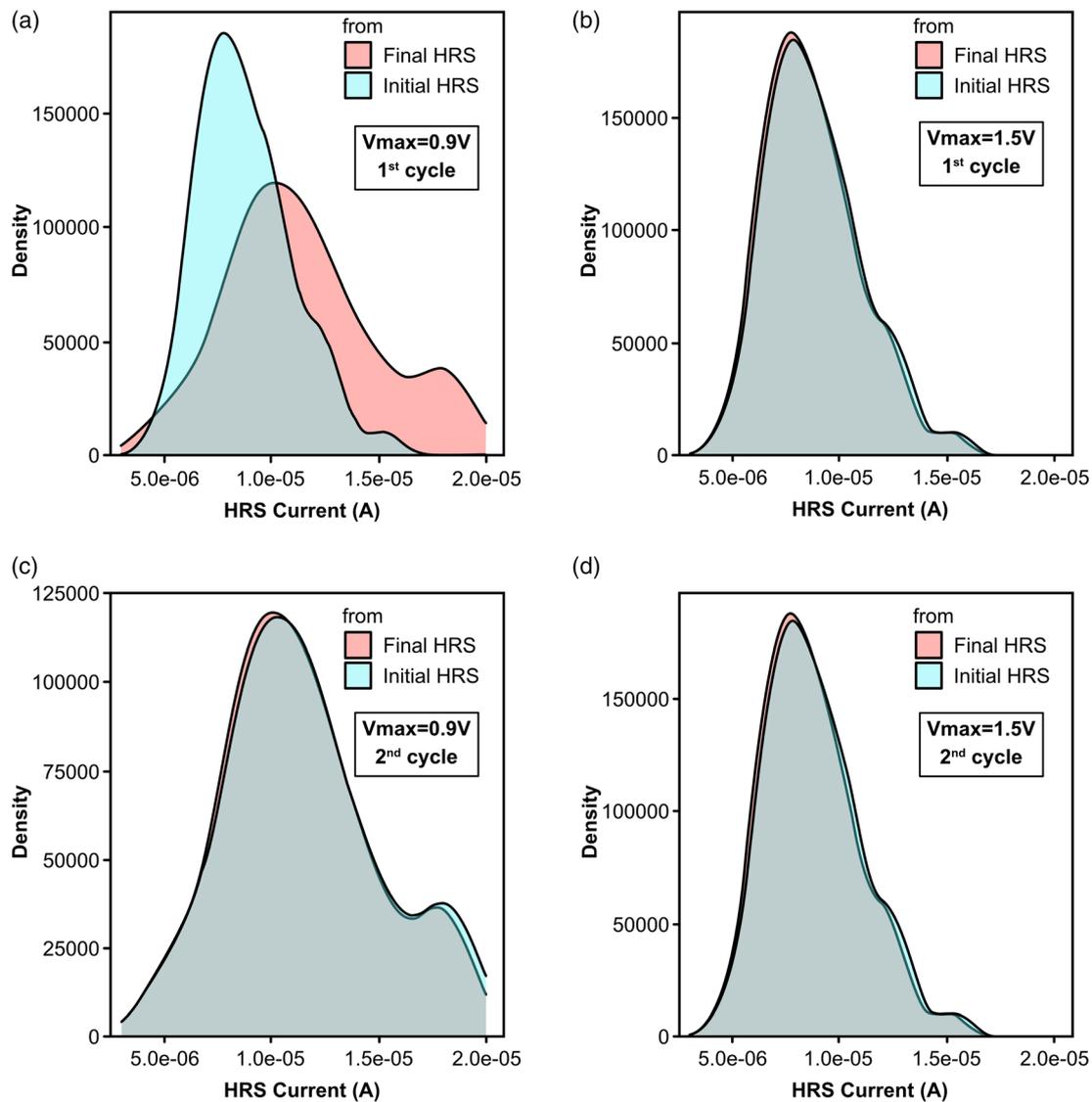

**Figure 33.** Comparison between the initial and final HRS current densities for different voltage and cycle situations: a) 1st cycle and $V_{max} = 0.9$ V, b) 1st cycle and $V_{max} = 1.5$ V, c) 2nd cycle and $V_{max} = 0.9$ V and d) 2nd cycle and $Vmax = 1.5$ V.

### 5.3. Time-Series Modeling of Cycle-To-Cycle Variability

In this final subsection, we will focus the attention on a completely different modeling approach: the times-series statistical analysis (TSSA). After the thorough experimental description of variability in Section 2, and the explanation of different modeling procedures (Section 3 and 4), we will develop expressions here that allow the calculation of RS parameters, such as the reset and set voltages, using previous values of these voltages, i.e., values corresponding to past cycles in a RS series.

From a purely statistical viewpoint, RRAM variability reflected in RS parameters has been analyzed using the Weibull distribution (WD).[4,236,237] This classical distribution is a reasonable option for filamentary RRAMs since it is a weakest-link type distribution (the failure of the whole is related to the weakest element). Reliability analyses with the WD assume that times to failure, in our case set voltage and/or reset voltage (we are considering devices under ramped voltage stress), are independent, although this assumption may not be valid in RRAMs, since successive observations (cycles) could be highly dependent (the CF is formed making use of broken parts of previous ones, as shown by kinetic Monte Carlo simulations[37,238]). Other approaches have been proposed to tackle variability statistical study; for instance, clustering statistical analysis, complementing the WD,[239] and convolution-based modeling.[240] Phase-type distribution functions have also been successfully employed for the description of RRAM variability.[32,241]

TSSA has been successfully applied in the fields of economics and sociology, and it has also been used in engineering and reliability of electronic devices.[242] TSSA was also valuable to analyze and model RRAM cycle-to-cycle variability.[144,243,244] This type of statistical technique is appropriate for modeling physical mechanisms that show "inertia" in some of their features. In RS cycling, for the devices we will consider below, a CF is created (set process) by using the remnants of the CF ruptured in the previous cycle (reset process). In this way, we study the relation between consecutive cycles and analyze the device "memory" in a long RS series, i.e., the parameters ($V_{set}$ or $V_{reset}$) that characterize consecutive cycles in a RS series are connected (therefore, the concept of autocorrelation comes up).

For the TSSA application, some mathematical conditions have to be fulfilled, for instance, time-series stationarity.[144,243,245,246] RS parameters can be forecasted by means of algebraic expressions that include the values of the RS parameters corresponding to previous cycles (in this manner we account for the recent history, or to put it in other words, the inertia in the RRAM RS operation). The number of previous cycles to consider in the models can be obtained through autocorrelation and partial autocorrelation analyses.[144,243–245]

To begin with, Equation (99) shows a univariate (only $V_{reset}$ values of different cycles are considered to model the current reset voltage) time-series model. The weights ($\varnothing_1...\varnothing_p$) of the autoregressive (AR) model have to be found. The determination of the order of the model (p) and weights to use (i.e., how many $V_{reset}$ values from previous cycles we need to forecast the current cycle in terms) is not straightforward. The order of the model depends on the physics governing RS; nevertheless, we do not deepen into this issue, assuming we have no knowledge about it. Instead, we will only use the information provided by the experimental data, assuming that the underlying physics and the technological details are "hidden" in the experimental RS data collected.

$$V_{reset_t} = \varnothing_1 V_{reset_{t-1}} + \varnothing_2 V_{reset_{t-2}} + \ldots + \varnothing_p V_{reset_{t-p}} + \varepsilon_t \quad (99)$$

The TSSA models using Equation (99), or the equivalent equation for $V_{set}$, are, therefore, empirical. Equation (99) could be modified to account for centered variables (e.g., $V_{reset}$-$\mu$, where $\mu$ is the $V_{reset}$ mean in the RS series), this formulation is equivalent to include a constant term $\Phi_0$ in the model.[144] The term $\varepsilon_t$ stands for a residual (in the TSSA terminology), it describes the model error, the difference between the measured and the modeled values. We can omit this term in the analytical expressions for the sake of brevity in an engineering context.

Equation (99) represents an AR model;[245] however, this type of model cannot be fitted satisfactorily to all data sets. If it does not work, we have to move towards more complex models, such as autoregressive moving average (ARMA) models. This alternative modeling approach includes AR and moving average (MA) parts combined.[245] MA models are made of a linear combination of past residuals,[245–247] see Equation (100) for the formulation of an ARMA(p,q) model for $V_{reset}$.

$$\begin{aligned} V_{reset_t} = {} &\varnothing_0 + \varnothing_1 V_{reset_{t-1}} + \ldots + \varnothing_p V_{reset_{t-p}} \\ &+ \theta_1 \varepsilon_{t-1} + \ldots + \theta_q \varepsilon_{t-q} \end{aligned} \quad (100)$$

where $V_{reset_t}$ is the reset voltage at current cycle and $V_{reset_{t-k}}$ are the reset voltages lagged k cycles, $\varepsilon_{t-k}$ are the errors (residuals) generated in earlier cycles, with $\varnothing_i$ (i = 1, ..., p) and $\theta_j$ (j = 1,..., q) being the unknown regression coefficients to be estimated in the modeling process. As highlighted above, in TSSA, the term $\varepsilon_t$ is usually included in the model, but in our case, we assume it. For the set voltage a parallel process could be performed.

As an example of the modeling capacity of the time-series approach, we used measured data from VCMs fabricated at the Institute of Microelectronics in Barcelona. The devices were fabricated with a W bottom electrode, a TiN/Ti stack for the top electrode and a dielectric made of a 5 nm HfO$_2$ layer on top of 2 nm Al$_2$O$_3$ layer.[33,248] Measurements were obtained using a Keysight B1500A Semiconductor Parameter Analyzer controlled by GPIB. Successive ramped voltage signals (0.08 V s$^{-1}$) were applied to the TiN/Ti electrode with respect to the grounded W electrode (see **Figure 34**). The reset voltage was determined as the value where the maximum current is obtained in a reset curve. The set voltage was obtained by finding the maximum separation of the measured curve to an imaginary straight line that joins the first and end points of the measured curve, in other words, finding the set curve knee (see Method 2 for Set voltage determination in ref. [127]). The reset and set voltages and currents series are shown in Figure 34.

The autocorrelation function (ACF) for these series of values gives us a useful measure of the degree of dependence among the data (reset or set voltages) of different cycles. It can be calculated as follows

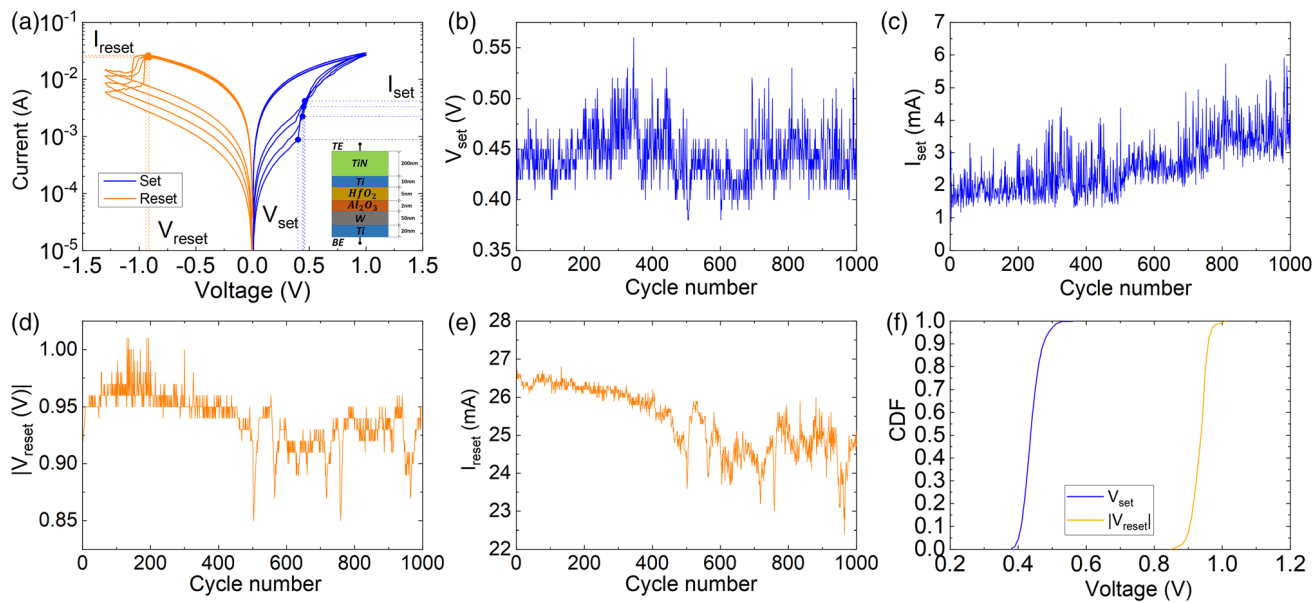

**Figure 34.** a) Experimental $I$–$V$ curves for several consecutive RS cycles showing the RS parameters $V_{set}$, $V_{reset}$, $I_{set}$, and $I_{reset}$ measured under ramped voltage stress. b) Set voltage, c) set current, d) absolute value of the reset voltage, e) reset current versus cycle number along the RS series measured. f) Cumulative distribution functions of the set and reset voltages.

$$\rho(k) = \text{Cor}(V_{reset_{t-k}}, V_{reset_t}) = \frac{\text{Cov}(V_{reset_{t-k}}, V_{reset_t})}{\sqrt{\text{Var}(V_{reset_t})\text{Var}(V_{reset_{t-k}})}} = \frac{\gamma(k)}{\gamma(0)} \tag{101}$$

where Cov and Var are the covariance and variance, respectively.[245] Notice that the ACF measures the correlation between two variables separated by k cycles. The ACF plot corresponding to our data is shown in **Figure 35**.

It can be seen that the correlation is important in both cases (the values are higher than the threshold bounds indicated by the dashed lines). This means that the past values of $V_{reset}$ and $V_{set}$ significantly affect the actual value. This effect might be related to the CF evolution from cycle to cycle, where the shape and remnants left in reset processes clearly affect the next values. The threshold bounds that mark the existence of correlation are important, their calculation can be refined, as explained in ref. [144], to account for the ACF values of previous lags, although we have simplified them in this work for the sake of clarity.

According to TSSA methodology, the next step is studying if the time series of $V_{set}$ or $V_{reset}$ values constitutes a stationary series[144] (the data mean, variance, and correlation do not depend on the cycle interval where they are computed), which is a necessary requirement for a correct modeling. If the series is not stationary (as it is our case here), different approaches can be used. For instance, differentiating the series (i.e., using the increments $V_{reset_t} - V_{reset_{t-1}}$ instead of $V_{reset_t}$). For our data set, this transformation generates new time series that is stationary. We calculate the ACF and partial ACF (PACF)[144] (the PACF measures the same correlation but eliminating the dependency due to the intermediated lags (1, 2, ..., $k-1$) of these differential time series to follow the modeling process, as shown in **Figure 36**.

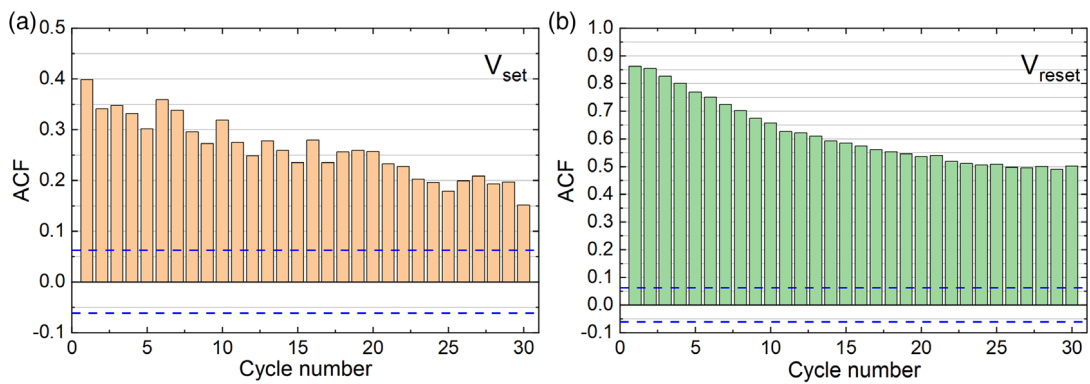

**Figure 35.** ACF versus cycle lag for a) $V_{set}$ and b) $V_{reset}$ series plotted in Figure 34. The ACF maximum and minimum threshold bounds are ±0.0619, as shown with blue dashed lines.

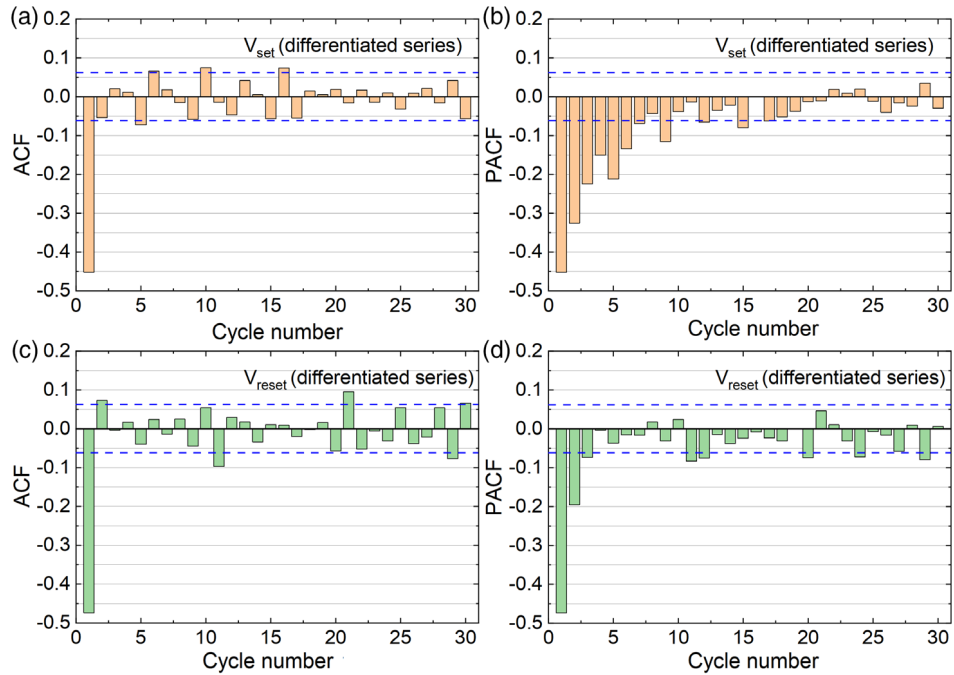

**Figure 36.** ACF and PACF versus cycle lag for the differentiated series. a) ACF and b) PACF for $V_{set_t} - V_{set_{t-1}}$, c) ACF and d) PACF for $V_{reset_t} - V_{reset_{t-1}}$. The ACF and PACF maximum and minimum threshold bounds are $\pm 0.0619$ shown with blue dashed lines.

In the case of the differential series for the set voltage, we have just one component, corresponding to the previous cycle (lag 1), which dominates over the rest. A comparison with respect to the threshold bounds is necessary. This result influences the modeling, since just this component will be considered. The analysis of the PACF (along with the ACF) suggests that some previous residuals have to be included in the model, see Equation (102). A similar analysis can be performed for the reset voltages (see Figure 36,c,d), although in this case more residual terms need to be incorporated.[245–247] In the modeling process, the simplest mathematical expression is chosen among the different alternatives that come up. This is known as a parsimonious model in the context of TSSA.[246]

The model selection procedure followed indicates that the best option is an ARIMA(0,1,1) for $V_{set}$ and an ARIMA(0,1,2) for $V_{reset}$. ARIMA models (autoregressive integrated moving average) are used when the series is differentiated.[246] After some algebra to isolate the current set and reset voltages the final models are obtained, see Equation (102) for the set voltage

$$V_{set_t} = V_{set_{t-1}} - 0.8634\varepsilon_{t-1} \tag{102}$$

and (103) for the reset voltage

$$V_{reset_t} = V_{reset_{t-1}} - 0.5811\varepsilon_{t-1} + 0.0924\varepsilon_{t-2} \tag{103}$$

These models can be used to predict the values of the series obtained in the laboratory for the devices under study, as shown in **Figure 37**. In this respect, in addition to generate the cycle-to-cycle variability randomly (accounting for the probability function and the fitting parameters), as it is done in circuit simulators such SPICE, this modeling approach would allow to reproduce the evolution of the RS parameters including the "memory" effect inherent to the RS series (see an implementation example in Verilog-A in ref. [145]).

The TSSA could be enhanced if a multivariate approach is followed as described in ref. [243]. In this case, for the modeling of $V_{reset}$, both, previous values of $V_{reset}$ and $V_{set}$ should be used. This would make sense after a causality study[243] to shed light on the goodness of a multivariate approach, i.e., an evaluation the improvement in the prediction capacity of the model including lag values of the set and reset voltages. For this purpose, the analysis of the Cross correlation function among the residual of the univariate models is useful. More details of this TSSA approach are given,[243] but this discussion exceeds the scope of this revision.

Some of the features and analyses connected to TSSA could be employed to study the stochasticity of RS in different materials. In this respect, a strong ACF in $V_{set}$ or $V_{reset}$ data implies that the CF or CFs involved in RS are stable (that is why they show a strong correlation between different cycles) and that could lead to CF formations along paths formed by grain boundaries in polycrystalline dielectrics or other features that fix the CF shape and size (in these cases, a connection between the grain boundaries or other material formations and the CF nature is present). In addition, a TSSA modeling approach can be used for circuit simulation.[145] Other materials show lower ACFs when analyzing $V_{set}$ or $V_{reset}$ data, this suggests different physical mechanisms behind RS operation with a lesser degree of correlation between consecutive cycles and, consequently, a reduced link to the dielectric morphology.[249]

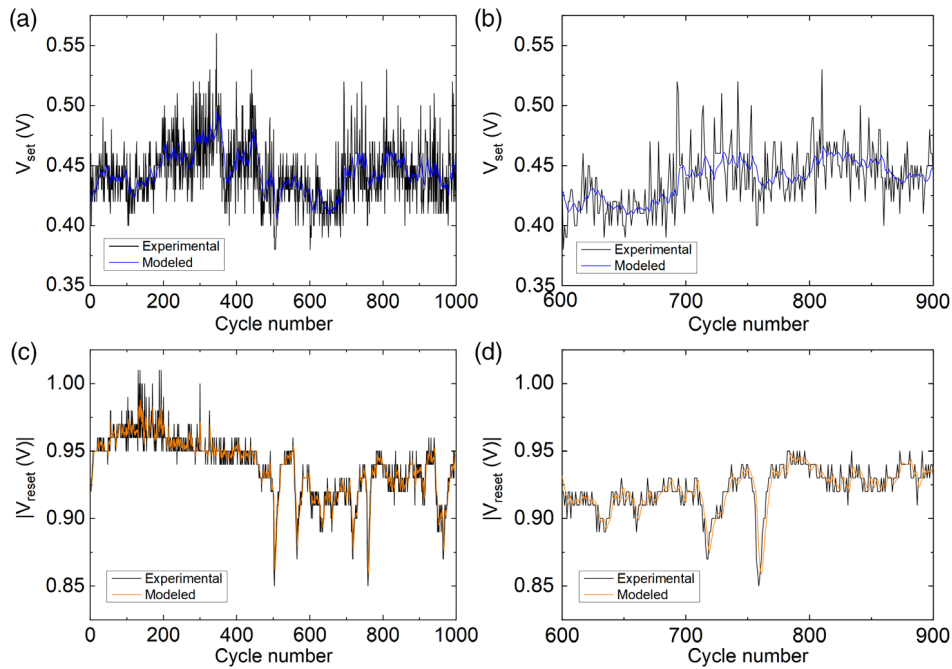

**Figure 37.** a) $V_{set}$ versus cycle number for the RS series under consideration. Measured values are shown by black lines and the modeled ones in blue, b) $V_{set}$ series detail for the cycle interval [600-900]. c) $V_{reset}$ versus cycle number for the RS series under consideration. Measured values are shown by black lines and the modelled ones in orange, d) $V_{reset}$ series detail for the cycle interval [600-900].

## 6. Discussion

We have shown in Section 2 that variability is an essential issue in resistive memories operation and, as discussed along this manuscript, this issue can be tackled in multiple and intricate manners. From physical to behavioral models, combining stochastic approaches as well as purely mathematical tools, all converge to point out the importance of incorporating this phenomenon into suitable EDA simulation tools which at the end of the day can aid to the actual device design. In this respect, working at the system level, where a large group of memory cells must operate together, the predictability of the behavior of all cells in a reasonable range is critical and this is something that we cannot control easily for the moment.

As a general principle, ICs are designed at many different levels: system, architecture, circuit, device, etc., being this activity one of the most interdisciplinary. Each of the involved tasks requires a different set of skills and, obviously, this is reflected in the different needs imposed on the EDA tools. For instance, at the device design level, it is most useful the capacity to simulate all the physical effects involved in the operation including possible quantum effects. This approach, while accurate, fails when a large circuit needs to be simulated since a highly descriptive level implies intensive computing. Thus, the device models have to be compact. These types of models fit the needs of the simulations they target. When thousands of devices are involved and the circuit size crosses the threshold of the system level, compact models (usually physically based) are replaced by simpler models (sometimes found in the behavioral realm), that takes abstraction a step further, with much simpler equations allowing faster simulations, but obscuring somehow the physical meaning of the model parameters.

It is difficult to establish clear IC design guidelines including RRAMs for the vast majority of the resistive memory technologies currently available, some of them still in its experimental phase, not only because of the complexity of the phenomenon and the many variables unknown but also because of the absence of well-defined design rules, even for the devices themselves. The way in which information is stored in RS devices inherently fluctuates from cycle to cycle, and this is something everybody agrees: circuit designers must cope with it. In some cases, during operation, devices will be subject to low-voltage pulsed signals so that the memory window and the HRS probability distribution are known. In general, LRS is much more stable and is not considered critical as HRS in terms of circuit design. This latter issue can be enlightened by the lumped compact model described in Section 4. The potential profile (Equation (65)) in the master equation (Equation (64)) presents a minimum for LRS (at $y = L$ during the set process at $V > 0$ in our formulation), but there is no minimum in this potential for HRS during the reset process at $V < 0$ (see also the comments after Equation (72)). Therefore, a dielectric layer that could be modeled in the HRS operation with a minimum in the effective potential $U_{eff}(y)$ would be of interest. Anyway, if only information related to the LRS and HRS is available, designers should be cautious. Models not only should include dispersion effects, often represented in terms of CDFs for the LRS and HRS, but also time correlation (time-series dependence) and possible drift (degradation) effects.

For the context described above, a complete knowledge of the electron transport mechanism might not be required, just the current value at a particular bias (read voltage), so that the variability problem simplifies a lot. However, introducing the voltage dependence of the randomness for all the operation regions is computationally demanding. For example, the Stanford (see Section 3.1) and the memdiode (see Section 5.2) models attempt to address this issue, but they assume no C2C correlation effects, only a univariate dispersion of the model parameters. Multivariate dispersion in the model parameters, and more complex noise activity such as Levy flights, are issues that deserve to be investigated in the future. See in Section 3.1 that different aspects of the Stanford model could be improved in connection to the variability description to account for experimental wide variations of the set and reset voltages, HRS and LRS currents, among other issues. This improvement could be implemented accounting for the fluctuation-dissipation relations described in Section 4.2.2. Another issue is related to Equation (3) that is nonlinear due to dependence $\gamma(g)$. This latter parameter accounts for the effect of electric field between the CF tip and the electrode. A more accurate description could be possible.

In connection with Section 3.2, it is important to highlight that the development of compact models for RS devices useful in the quantum regime is still in its infancy. In addition to the appropriate current–voltage relationship (quantum transport model) and the hysteresis mechanism in the memory state of the device (movement of ions/vacancies), the models must include randomness of different nature that leads to data-to-data, cycle-to-cycle, and device-to-device variability. This randomness depends on the particular current regime, which makes the problem even more difficult to solve in a general way. The model described in Section 3.2 should be considered as a physical framework based on the Büttiker–Landauer approach for the representation of the current fluctuations in the high resistance state (HRS) regime only. In this respect, this is not a complete model for fluctuations in RS devices, as expected for circuit simulators. Nevertheless, the proposed model can be easily incorporated into any circuit simulator because it simply consists in an expression for the current–voltage characteristic driven by Gaussian noise. This Gaussian noise in the context of a tunneling barrier height generates a lognormal distribution for the current magnitude. Although we focus on cycle-to-cycle variability, other kind of variability can be addressed making the appropriate changes in the generator of random processes. This is, for example, illustrated in Section 5.2, in which a particular version derived from the quantum model, the so-called memdiode model, is used to generate the C2C variability in the $I–V$ loops.

Concerning Section 5.1, it can be highlighted that a neuromorphic circuit could benefit from a model that can directly forecast the value of the device conductivity after N cycles of RS. This task is found to be particularly appropriate by means of the charge-flux paradigm. The description shown making use of only two parameters reproduces quite accurately the actual memristor behavior when a pulse train is applied. The corresponding Monte Carlo model for the implementation of variability has only two parameters ($n$ and $G_0$), and their variations are shown to be correlated, thus providing a variability model that can be applied even in hand calculations.

In the literature revised in this manuscript, we find different models developed in a wide variety of platforms (Python, MATLAB, PSpice, COMSOL, Ginestra, etc.) corresponding to different modeling approaches. It is clear therefore that a standard model as happens for other mature technologies is unavailable yet. In this respect, there is a long way to go in the modeling context. Some papers from the last few years point out that the mechanisms involved in RS phenomena are strongly linked to the device fabrication materials, as mentioned in Section 2.2. In this respect, existing and future models should account for material properties, together with the physical mechanisms behind RS. New and hard efforts should be focused on the device fabrication materials and architecture to unveil the variability physics and develop suitable models for their use in EDA tools.

Finally, for the sake of clarity, we include all the models presented in this manuscript in **Table 6** with their main characteristics highlighted.

## 7. Conclusions

Resistive memories have a simple structure and are compatible with CMOS technology; therefore, they can be integrated in the conventional processes of the current electronics industry. However, their huge potential in different applications could be limited by their inherent cycle-to-cycle variability. The variability issue is essential, and that is why we present here a revision manuscript to comprehensively study variability from the experimental and modeling viewpoints.

Variability is seen in experimental characterization under ramped and pulsed voltage signals. Its influence on the device operation depends on the fabrication materials and on the physics behind RS. In this work, current variations were analyzed accounting for their evolution in long RS series both for devices based on conventional transition metal oxides, such as $HfO_2$, and those with 2D materials dielectrics, such as h-BN. The modeling perspective to address variability is developed extensively throughout the manuscript, including the most representative approaches to analyze the variations in RS processes, with the potential to come up with analytical expressions for compact modeling purposes. The inclusion of variability in the Stanford model is described, studying the role of the model parameters. A quantum approach is also included to account for the flow of electrons in confining structures that might be created in the RS processes. In this approach, the device current is described using the Landauer's formulation for mesoscopic systems. Stochastic modeling is also presented. In doing so, the appropriateness of distributed and lumped models is discussed as well as the use of noise sources in the differential equations that are employed to describe variability. The estimation of stochastic model parameters is widely tackled. Behavioral models are also described from the perspective of the general memristor model formalism, the application of a dynamic memdiode model to calculate the device current and the use of time-series analysis to reproduce the correlation observed in measurements in long RS series.

**Table 6.** Summary of the models presented with the description of their main features.

| Model | Section | Modeling approach | SPICE/Verilog-A implementation | $R_{LRS}$ | $R_{HRS}$ | $V_{set}$ | $V_{reset}$ | $I–V$ curve shape | Fitting process |
|---|---|---|---|---|---|---|---|---|---|
| Stanford | 3.1 | Physical | Yes[132] | Yes | Yes | Yes | Yes | Yes | Yes |
| Mesoscopic | 3.2 | Physical | No | Yes | Yes | No | No | Partially | Yes |
| Stochastic distributed | 4.2.1 | Physical | No | Yes | Yes | Yes | Yes | Yes | Yes |
| Stochastic lumped | 4.2.1 | Physical | No | Yes | Yes | Yes | Yes | Yes | Yes |
| Stochastic lumped compact | 4.2.3 | Physical | No | Yes | Yes | Yes | Yes | Yes | Yes |
| Charge-flux approach | 5.1 | Behavioral | Yes[253,254] | Yes | Yes | Yes | Yes | Yes | Direct extraction |
| Memdiode | 5.2 | Behavioral | Yes[165] | Yes | Yes | Yes | Yes | Yes | Yes |
| Time series | 5.3 | Behavioral | No | No | No | No | No | No | Yes |

## Acknowledgements

This research was supported by project B-TIC-624-UGR20 funded by the Consejería de Conocimiento, Investigación y Universidad, Junta de Andalucía (Spain) and the FEDER program. F.J.A. acknowledges grant PGC2018-098860-B-I00 and PID2021-128077NB-I00 financed by MCIN/AEI/10.13039/501100011033/FEDER and A-FQM-66-UGR20 financed by the Consejería de Conocimiento, Investigación y Universidad, Junta de Andalucía (Spain) and the FEDER program. M.B.G. acknowledges the Ramón y Cajal Grant No. RYC2020-030150-I. M.L. and M.A.V. acknowledge generous support from the King Abdullah University of Science and Technology. A.N.M., N.V.A., A.A.D., M.N.K. and B.S. acknowledge the Government of the Russian Federation under Megagrant Program (agreement no. 074-02-2018-330 (2)) and the Ministry of Science and Higher Education of the Russian Federation under "Priority-2030" Academic Excellence Program of the Lobachevsky State University of Nizhny Novgorod (N-466-99_2021-2023). The authors thank D.O. Filatov, A.S. Novikov, and V.A. Shishmakova for their help in studying the dependence of MFPT on external voltage (Section 4). The devices in Section 4 were designed in the frame of the scientific program of the National Center for Physics and Mathematics (project "Artificial intelligence and big data in technical, industrial, natural and social systems") and fabricated at the facilities of Laboratory of memristor nanoelectronics (state assignment for the creation of new laboratories for electronics industry). E.M. acknowledges the support provided by the European project MEMQuD, code 20FUN06, which has received funding from the EMPIR programme co-financed by the Participating States and from the European Union's Horizon 2020 research and innovation programme.

## Conflict of Interest

The authors declare no conflict of interest.

## Keywords

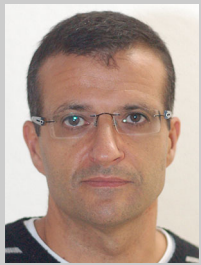

**Juan B. Roldán** received his degree in physics and Ph.D. degree from the University of Granada, Spain. His current research interests include characterization, physical simulation, and compact modeling of resistive switching memories. He has also been involved in physical simulation and modeling of multigate FETs, CMOS photodiodes, magnetic current sensors, and other types of electron devices. He has published over 190 papers in JCR journals on these subjects. He is currently a Full Professor with the Universidad de Granada.